\documentclass[11pt]{article}
\pdfoutput=1
\usepackage{caption}
\usepackage[titles]{tocloft}
\usepackage{jheppub}
\usepackage{makecell}
\usepackage[dvipsnames]{xcolor}
\usepackage{amsmath}
\usepackage{mathrsfs}

\usepackage{tikz}
\usetikzlibrary{decorations.pathmorphing,decorations.pathreplacing,calligraphy,calc,cd,external,arrows.meta,patterns}

\usepackage[export]{adjustbox}
\usepackage{hyperref}
\usepackage{bbm}
\usepackage{mathtools}
\usepackage{empheq}
\usepackage{nccmath}
\usepackage{blkarray}
\usepackage{esint}
\usepackage{microtype}
\usepackage{slashed}
\usepackage{physics}
\usepackage{amsfonts}
\usepackage{mathdots}
\usepackage{parskip}

\usepackage{csquotes}
\usepackage[compat=1.1.0]{tikz-feynman}
\usepackage{contour}

\usepackage[bb=boondox]{mathalfa}
\usepackage{amssymb}

\definecolor{maroon}{RGB}{0.51,0.02,0.02}
\definecolor{darkgreen}{rgb}{0.0, 0.4, 0.0}
\definecolor{thered}{rgb}{0.71,0.07,0.07}
\definecolor{theblue}{rgb}{0.27,0.27,0.73}
\colorlet{choral}{MidnightBlue}
\colorlet{darkred}{Maroon}

\hypersetup{
    pdfencoding=unicode,
	colorlinks=true,
	urlcolor=choral,
	linkcolor=choral,
	citecolor=Mahogany,
	pdftitle={Gravitational Radiation from Shockwave Scattering in the Self-Force EFT},
	pdfauthor={Emanuele Rosi},
	pdfdisplaydoctitle=true,
	pdfstartview=FitH,
	linktocpage=true
}

\usepackage[titles]{tocloft}
\usepackage{dirtytalk}
\usepackage{changepage}
\usepackage{subcaption}
\usepackage{float}
\usepackage{enumitem}  
\usepackage{stackengine}
\usepackage{bm}

\DeclarePairedDelimiterXPP\EV[1]{E}(){}{

#1}
\DeclarePairedDelimiterXPP\Var[1]{V}(){}{

#1}

\usepackage{cleveref}
\usepackage{nicematrix}
\newcommand{\gn}{G}
\newcommand{\vertx}{\mathcal{V}}
\newcommand{\vertp}{{\mathcal{V}}}
\newcommand{\propp}{{G}}
\newcommand{\propx}{{G}}
\newcommand{\ag}{\alpha_{_G}}
\newcommand{\bnab}{\bar{\nabla}}
\newcommand{\pd}{\partial}

\newcommand{\mh}{\mathrm{H}}
\newcommand{\ml}{\mathrm{L}}
\newcommand{\muir}{\mu_{\text{\tiny IR}}}
\newcommand{\muirt}{\tilde{\mu}_{\text{\tiny IR}}}
\newcommand{\wf}{\mathcal{W}}
\newcommand{\wfs}{\mathsf{WF}}

\newcommand{\pperp}{{\text{\tiny$\perp$}}}
\newcommand{\lo}{{\text{\tiny LO}}}

\newcommand{\pol}{\varepsilon}
\newcommand{\mpl}{M_{Pl}}

\makeatletter
    \newcommand*\bigcdot{\mathpalette\bigcdot@{1}}
    \newcommand*\smtimes{\mathpalette\smtimes@{.7}}
    \newcommand*\bigcdot@[2]{\mathbin{\vcenter{\hbox{\scalebox{#2}{$\m@th#1\bullet$}}}}}
    \newcommand*\smtimes@[2]{\mathbin{\vcenter{\hbox{\scalebox{#2}{$\m@th#1\times$}}}}}
\makeatother

\usetikzlibrary{fadings}
\tikzfading[name=fade right,
right color=transparent!0,
left color=transparent!100]

\newcommand{\middlearrow}[2]{
    \draw[-{Straight Barb[length=1.3mm,line width=1pt,black]}] ($(#1)!0.53!(#2)$) -- ($(#1)!0.54!(#2)$);
}
\newcommand{\middlearrowBig}[2]{
    \draw[-{Straight Barb[length=1.5mm,line width=1pt,black]}] ($(#1)!0.53!(#2)$) -- ($(#1)!0.54!(#2)$);
}

\usetikzlibrary{decorations.markings}
\tikzset{
    crossdot/.style={circle, draw, inner sep=0pt, minimum size=7pt, path picture={
            \draw[black]
                (path picture bounding box.north west) -- (path picture bounding box.south east)
                (path picture bounding box.north east) -- (path picture bounding box.south west);
}}}
\tikzset{
    partial ellipse/.style args={#1:#2:#3}{
        insert path={+ (#1:#3) arc (#1:#2:#3)}
}}
\tikzset{ shorten <>/.style={ shorten >=#1, shorten <=#1 } }
\tikzset{gradRtoB/.style={
    postaction={
        decorate,
        decoration={
            markings,
            mark=at position \pgfdecoratedpathlength-0.5pt with {\arrow[blue,line width=#1] {}; },
            mark=between positions 0 and \pgfdecoratedpathlength-0pt step 0.5pt with {
                \pgfmathsetmacro\myval{multiply(divide(
                    \pgfkeysvalueof{/pgf/decoration/mark info/distance from start}, \pgfdecoratedpathlength),100)};
                \pgfsetfillcolor{choral!\myval!Maroon!};
                \pgfpathcircle{\pgfpointorigin}{#1};
                \pgfusepath{fill};}
}}}}

\tikzset{
    vector/.style={
        decoration={snake, aspect=0.75, mirror, segment length=2mm},
        decorate
    },
    photon/.style={
        decorate,
        decoration={snake, amplitude=1pt, segment length=6pt}
    },
    graviton/.style={
        decorate,
        decoration={snake, amplitude=.4mm, segment length=1.5mm, pre length=.5mm, post length=.5mm},
        double
    }
}

\title{Gravitational Radiation from Shockwave Scattering in the Self-Force EFT}
\usepackage{orcidlink}
\definecolor{orcidlogocol}{named}{NavyBlue}
\author[\!a,b\,\orcidlink{0009-0009-2244-4511}]{Emanuele Rosi,}\emailAdd{emanuele.rosi@uniroma1.it}
\affiliation{$^a$ Dipartimento di Fisica, Universit\`a di Roma ``La Sapienza'', \\ Piazzale Aldo Moro, 2, 00185 Roma RM, Italia}
\affiliation{$^b$ Sezione INFN ``Laboratori Nazionali di Frascati'', \\ Via Enrico Fermi, 54, 00044 Frascati RM, Italia}

\abstract{
    We investigate the high-energy collision of two spinless bodies within the Self-Force Effective Field Theory (EFT), taking one of them to be massless and highly energetic, such that it generates a gravitational shockwave (GSW) background.
    The GSW is perturbed by the second body, either massless or massive, satisfying respectively $E_2/E_1\ll1$ or $m_2/E_1\ll1$.
    Although the Self-Force expansion for shockwaves is frame-dependent, we freely choose a frame in which this EFT manifests its power, obtaining results valid to all orders in Newton's constant that make the underlying resummation patterns evident.
    Specifically, we push the expansion to 1SF order and prove the all-order vanishing of the so-called recoil operators in light-cone gauge.
    We obtain the exact graviton propagator on the shockwave background, matching previous results, and use it to compute the on-shell waveform at 1SF order in momentum space, specialising to the massless-massless collision.
    The waveform is computed to all orders in the Post-Minkowskian, or weak-field, expansion.
    In the trans-Planckian regime, its resummed form is proved to yield a UV-finite angular spectrum of radiated energy.
}

\begin{document} 

\addtocontents{toc}{\protect\thispagestyle{empty}}

\maketitle

\thispagestyle{empty}

\setcounter{page}{1}
\allowdisplaybreaks
{\section{Introduction}\label{sec:introduction}}
The relativistic two-body problem has acquired direct phenomenological relevance with the birth of gravitational wave astronomy~\cite{LIGOScientific:2016aoc}, and will become even more pressing with the ongoing and upcoming LIGO-Virgo-KAGRA~\cite{LIGOScientific:2014qfs,LIGOScientific:2016aoc,VIRGO:2014yos,KAGRA:2020tym} observing runs and with the launch of next-generation detectors such as LISA~\cite{LISA:2017pwj}, all of which demand waveform templates of correspondingly higher accuracy.
Three complementary expansion schemes have been developed:
$(i)$ the Post-Newtonian expansion, valid for small velocities and large separation such that $v^2 \sim \gn M_{\rm tot}/r \ll 1$;
$(ii)$ the Post-Minkowskian (PM) expansion, still perturbative in $\gn$ but valid at any velocity;
$(iii)$ and the Self-Force (SF) expansion, organised as an expansion in the mass ratio of the two bodies, $q = M_\ml/M_\mh \ll 1$.
The study of the Self-Force expansion is strongly motivated by the prospective detection of extreme mass ratio inspirals (EMRIs) with LISA, in which a light compact object orbits around a much heavier one.
These systems are characterised by strong fields, requiring analytical results which are non-perturbative or resummed in $\gn$.
Moreover, EMRIs may feature large orbital eccentricity, for which even an expansion at low velocity is not viable.
It is therefore necessary to expand in the ratio $q$, treating the heavy body as the source of the background, and the light body as a perturbation.
This framework has recently been reformulated in the effective field theory language~\cite{Cheung:2023lnj,Cheung:2024byb} (see also~\cite{Kosmopoulos:2023bwc}), recasting the Self-Force expansion into effective operators, Feynman rules, propagators on curved backgrounds, order by order in $q$.
The Self-Force EFT can be applied, in principle, to any background sourced by the principal body.
However, already at first Self-Force order one runs into the computation of the Green function of the background.
For the Schwarzschild~\cite{Cheung:2024byb} and Kerr~\cite{Akpinar:2025huz} cases this has so far required a further expansion in Post-Minkowskian series, so as to avoid solving complicated equations such as the Regge-Wheeler-Zerilli or Teukolsky~\cite{Regge:1957td,Teukolsky:1973ha}.
In this paper we extend the Self-Force EFT construction to the Schwinger-Keldysh (in-in) path integral at $\mathcal{O}(q^2)$ (1SF) order.
Previous applications of the formalism only involved conservative quantities, for which the ordinary in-out path integral is sufficient.
Here, instead, we are interested in radiative contributions to the observables too, specifically the emitted waveform, which can only be correctly recovered within an in-in construction, following the strategy developed for other worldline EFTs~\cite{Galley:2013eba,Jakobsen:2022psy,Kalin:2022hph}.

On the other hand, the scattering of gravitational shockwaves (GSWs) - or, equivalently, of massless particles - has been, and remains, a very active area of theoretical research~\cite{Verlinde:1991iu,Gruzinov:2014moa,DEath:1992mef,DEath:1992nmz}.
First, it represents the simplest possible realisation of the two-body problem in General Relativity.
Second, it arises as the singular high-energy limit of the scattering of massive bodies~\cite{Aichelburg-Sexl} and the validity conditions of this limit are still to be completely understood.
Third, at sufficiently high energies it provides access to several long-standing open problems in modern physics, ranging from the finiteness of the radiated energy~\cite{Gruzinov:2014moa,Ciafaloni:2015xsr} to black hole formation~\cite{Amati:1988tn,Dvali:2021ooc}.
The trans-Planckian scattering regime $s\gg \mpl^2$ was first studied from the point of view of string amplitudes by Amati, Ciafaloni and Veneziano (ACV)~\cite{Amati:1987wq,Amati:1988tn,Amati:1990xe,Amati:1992zb}, in the field-theory limit and at various impact parameters, including the Post-Minkowskian regime $b\gg \gn\sqrt{s} \gg \ell_{\rm string}$.
The study of the classical dynamics of massless particle collision through amplitude methods has been largely motivated by the two-loop ACV result for the eikonal phase~\cite{Amati:1990xe} and was subsequently pursued - building on the Regge effective action of~\cite{Amati:1993tb} - in more recent works~\cite{Ciafaloni:2014esa,Ciafaloni:2015vsa,Ciafaloni:2015xsr,Ciafaloni:2017ort}, up to its matching onto the ultrarelativistic limit of the classical two-loop amplitudes for massive-body scattering~\cite{DiVecchia:2022nna,DiVecchia:2020ymx}.
Unlike QCD, where massless scattering is entirely dominated by $t$-channel evolution~\cite{Lipatov:1989bs,Lipatov:1993qn} and hence by the BFKL equation~\cite{Kuraev:1976ge,Kuraev:1977fs,Balitsky:1978ic}, massless scattering in General Relativity is instead dominated by eikonalisation~\cite{Amati:1987wq,Amati:1988tn,DiVecchia:2023frv,RajVenugopalan-universal-features}, which is an indirect consequence of the (negative) mass dimension of the coupling constant.
The classical corrections to the eikonal phase for the conservative scattering of two massless particles are captured by a sequence of classical H-diagrams, first introduced in~\cite{Amati:1993tb} as three-particle-irreducible diagrams dominated by Reggeon exchanges and soft exchanges.
The multi-H diagrams, which appear at even loop order, begin with the exchange of a single graviton in the Glauber region, continue at two loops with the 1990 ACV computation, and have recently been extended to four loops in~\cite{Alessio:2025isu,Alessio:2026bdi}.
There are however signatures of breakdown of the eikonal resummation, starting at three loops~\cite{DiVecchia:2019kta}, implying the need for alternative methods to obtain results resummed in $\gn$.
To this end, we work in a frame in which the Self-Force Effective Theory~\cite{Cheung:2023lnj,Cheung:2024byb} can be applied.
The two bodies travel along parallel, opposite directions (in light-cone coordinates)
\begin{equation}\label{eq:frame}
    p_1^\mu = E_1\xi^\mu = (2E_1,0,0,0) \,, \hspace{30pt} p_2^\mu = E_2\chi^\mu = (0,2E_2,0,0)\,,
\end{equation}
so that the trans-Planckian regime corresponds to $\gn s = 4 \gn E_1 E_2 \gg 1$, with $E_1$ large enough to generate an Aichelburg-Sexl shockwave background~\cite{Aichelburg-Sexl}, and $E_2 \ll E_1$ such that it sources a perturbation of order comparable to, or smaller than, $E_2/E_1$.\footnote{Equivalently, one may study the scattering of a shockwave of energy $E_1$ off a mass $m_2\ll E_1$, taken, for instance, to be initially at rest, as in~\cite{Fursaev:2026mvf}.}
Recent applications of this frame choice have termed this configuration the {\it dilute-dense} approximation~\cite{Blackstad:2026eoa}, in analogy with the QCD case.
Formulated in this configuration, the system can then be treated extensively using the Self-Force EFT~\cite{Cheung:2023lnj,Cheung:2024byb}, originally developed for the study of EMRI systems with a massive, quasi-static central body.

We now set aside all the considerations related to Newton's constant and the trans-Planckian regime, and expand the worldline action in the single small parameter $q=M_\ml/M_\mh\ll1$, with $M_\mh=E_1$ and $M_\ml=E_2$ (or, alternatively, we can consider $M_\ml=m_2$).
As noted above, already at first Self-Force order\footnote{Although the notion of Self-Force is itself frame-dependent~\cite{Galley:2013eba} for massless-particle scattering, we continue to use it here by a slight abuse of terminology.} we are led to compute the graviton propagator on the shockwave background, together with the geodesics of this metric.
The latter are straightforward to obtain and have been available in the literature for almost forty years~\cite{Dray:1984ha,FerrariVeneziano}.
The solution for the propagator was recently derived in the eikonal approximation by Raj and Venugopalan~\cite{Raj:2024xsi,Raj:2025hse}, and a study carried out in parallel to the present work~\cite{Bohnenblust:2026ujk} confirmed that the result of~\cite{Raj:2024xsi} is, in fact, exact rather than approximate.
Working in the light-cone gauge, we show that the recoil of the primary body, with momentum $p_1$, is null and thus it does not contribute to the propagator.
We obtain a Born-like sequence~\cite{Correia:2024jgr,Caron-Huot:2025tlq} in the Post-Minkowskian expansion, which resums exponentially in position space, demonstrating the existence of an underlying $s$-channel ladder structure.
The result is calculated using a light-cone regulator - originally introduced in the context of SCET~\cite{Rothstein:2016bsq} - to cure the UV divergences that arise in this calculation and are not handled by dimensional regularisation.

Finally, we use the graviton propagator to solve the equation of motion for the gravitational perturbation at 1SF order in the massless-particle scattering.
A resummed waveform, valid to all orders in Newton's constant and computed in the centre-of-mass frame, was previously obtained in~\cite{Gruzinov:2014moa} using purely general-relativistic methods, and subsequently in~\cite{Ciafaloni:2015xsr} by resumming the eikonal ladder contributions to all orders.
Here, by combining the graviton propagator with the graviton source - obtained from the geodesic motion of the body with energy $E_2$ on the background - we obtain all the {\it rescattering} contributions~\cite{Amati:1993tb,Ciafaloni:2014esa,Blackstad:2026eoa} of the emitted graviton off the worldline of the more energetic body, schematically given by:
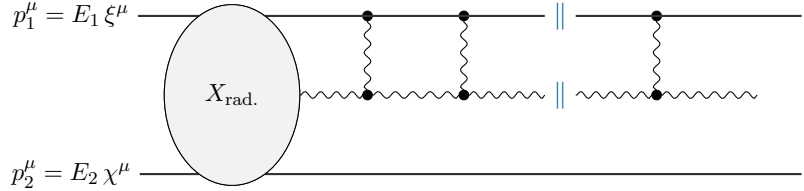
\vspace{10pt}
\begin{center}
    \begin{tikzpicture}[baseline=(blob.center),scale=0.85,transform shape]
        \centering
        \begin{feynman}
            \vertex[dot, style={fill=gray!10, shape=ellipse, minimum width=60pt, minimum height=80pt=80pt}] (blob) {};
            \vertex[dot, right=60pt of blob] (int0) {};
            \vertex[dot, right=1.5cm of int0] (int1) {};
            \vertex[choral, font=\large, right=1.5cm of int1] (int2) {$\Vert$};
            \vertex[dot, right=1.5cm of int2] (int3) {};
            \vertex[right=1.7cm of int3] (out) {};

            \vertex[dot, above=35pt of int0] (a0) {};
            \vertex[dot, above=35pt of int1] (a1) {};
            \vertex[choral, font=\large, above=35pt of int2] (a2) {$\Vert$};
            \vertex[dot, above=35pt of int3] (a3) {};
            
            \vertex[above=35pt of blob] (intUp) {};
            \vertex[left=2.5cm of intUp] (inUp) {$p_1^\mu=E_1\,\xi^\mu$};
            \vertex[right=9cm of intUp] (outUp) {};

            \vertex[below=35pt of blob] (intLow) {};
            \vertex[left=2.5cm of intLow] (inLow) {$p_2^\mu=E_2\,\chi^\mu$};
            \vertex[right=9cm of intLow] (outLow) {};
        
            \diagram{
                {   (blob) -- [edges=photon] (int0) -- [edges=photon] (int1) -- [edges=photon] (int2) -- [edges=photon] (int3) -- [edges=photon] (out),
                    (inUp) -- [black!80, thick] (intUp) -- [black!80, thick] (a2) -- [black!80, thick] (outUp),
                    (inLow) -- [black!80, thick] (intLow) -- [black!80, thick] (outLow),
                    (a0) -- [edges=photon] (int0),
                    (a1) -- [edges=photon] (int1),
                    (a3) -- [edges=photon] (int3)
                },
            };
            \vertex[dot, style={fill=gray!10, shape=ellipse, minimum width=60pt, minimum height=80pt}, right=0cm of blob] (blob2) {$X_{\rm rad.}$};
        \end{feynman}
    \end{tikzpicture}
\end{center}
\vspace{10pt}
where $X_{\rm rad.}$ encodes the geodesic motion of $E_2$ on the background of $E_1$, together with the emission of a graviton.
The diagram, implicitly equipped with retarded causality flow, shows how taking the limit $E_1\gg E_2$ suppresses a large set of diagrams involving the interaction of the emitted graviton with the secondary worldline.
The resulting waveform is obtained to all orders of the Post-Minkowskian series.
A resummation can be obtained by ``interchanging'' the PM series with the soft ($\omega\to0$) one, using the results of Appendix~\ref{app:integrals}.

The paper is organised as follows.
In Section~\ref{sec:sfeft-section} we review the basics of the Self-Force Effective Field Theory, expanding the action up to $\mathcal{O}(q^2)$ and extending it to the in-in formalism;
then we specialise the EFT to the physical system discussed above, introducing the gravitational shockwave metric, its geodesics, and fixing the gauge.
Section~\ref{sec:propagators} is entirely devoted to the study of the graviton's retarded Green function on the GSW background.
Finally, in Section~\ref{sec:waveform} we compute the waveform at leading order in $q$ (1SF), its Post-Minkowskian and soft expansions, and derive the properties of its asymptotic form at large frequencies and beamed emission around the trajectory of the secondary shockwave.
Conclusions are in Section~\ref{sec:conclusions}.

\paragraph{Conventions}
We work in Minkowski space in the mostly minus signature and dimensional regularisation with $D=4-2\epsilon$ dimensions and infrared regularisation scale $\muir$.
For convenience we also define $\muirt^2 = \pi\muir^2 e^{\gamma_{\rm E}}$.
We use light-cone coordinates $(x^+,x^-,\vec{x})$, where $x^{\pm} = t \pm x^\parallel$ and $x_\pperp^\mu = (0,0,\vec{x})$, and choose $\xi_{\mu}$ and $\chi_{\mu}$ as a basis such that $\xi \cdot x = x^-$ and $\chi \cdot x = x^+$, while $\chi\cdot\chi=\xi\cdot\xi=0$ and $\chi\cdot\xi=2$.
Consequently, $d=D-2=2-2\epsilon$ is the dimensionality of the transverse space.
Greek indices are reserved for $D$-dimensional vectors, while Latin characters are used for transverse ones.
We keep the $\perp$ symbol implicit when working with Latin indices so that $p^i = (0,0,\vec{p})$ and $p_i = (0,0,-\vec{p})$.
We stress that $p_\pperp^2 = - \vec{p}^{\,2} \le 0$.
The only exception is the impact parameter between the two scattered bodies, for which we use the convention $b^2 = \vec{b}^{\,2} = -b_\pperp^2 > 0$.
The gravitational coupling and Newton's constant are related by $\kappa^2 = 32\pi\gn$.
We use $\eta_{\alpha\beta}$ for the Minkowski metric, $\bar{g}_{\alpha\beta}$ for the unperturbed background, and $h_{\alpha\beta}$ for the perturbations, such that the full metric is $g_{\alpha\beta} = \bar{g}_{\alpha\beta} + \kappa h_{\alpha\beta}$.
A hat over volume forms and Dirac delta functions implies a normalisation factor
\begin{equation}
    \hat{d}^Dk = \frac{d^Dk}{(2\pi)^D}\,, \qquad \hat{\delta}^{(D)}(k) = (2\pi)^D {\delta}^{(D)}(k)\,,
\end{equation}
and integration measures implicitly contain the metric determinant
\begin{equation}
    \int_x = \int d^4x\,\sqrt{-\eta}\,, \qquad \int_p = \int \hat{d}^4p\,\sqrt{-\eta}\,,
\end{equation}
which is equal to $\sqrt{-\eta}=1/2$ in light-cone coordinates.
When working in $D$ dimensions, a factor $(\muir^{2}e^{\gamma_\mathrm{E}})^\epsilon$ is implicit in the momentum integrals and its inverse in spacetime integrals. 
The conventions for the Fourier transform and its inverse are
\begin{equation}
    \tilde{f}(p) = \mathrm{FT}\big[f\big](p) = \int_x f(x) \, e^{ip\cdot x}\,, \qquad \mathrm{FT}^{-1}\big[\,\tilde{f}\,\big](x) = \int_p \tilde{f}(p)\,e^{-ip\cdot x}\,.
\end{equation}
As usual, the Heaviside step function is defined at the origin as $\Theta(0)=1/2$.
\section{Self-Force EFT on GSW background}\label{sec:sfeft-section}
In this section we set up the Self-Force EFT framework used throughout the paper.
We first review, in Section~\ref{sec:SFEFT}, the generic Self-Force Effective Field Theory for two interacting spinless bodies, following~\cite{Cheung:2023lnj,Cheung:2024byb}.
In Section~\ref{sec:schwinger-keldysh} we extend this construction to the Schwinger-Keldysh (in-in) path integral, in order to obtain causal observables that correctly capture radiation-reaction effects.
Finally, in~\ref{sec:2.3}, we specialise the formalism to the case in which the first body\footnote{The primary body is usually called {\it heavy body} within the SF-EFT. We improperly keep using this name, also for massless sources.} is massless and sources a gravitational shockwave background.
We review the shockwave solution, its light-like and time-like geodesics, and finally we introduce the light-cone gauge and its properties.

\subsection{Self-Force Effective Field Theory}\label{sec:SFEFT}
We review the Self-Force Effective Field Theory developed in~\cite{Cheung:2023lnj,Cheung:2024byb} for generic backgrounds and sources.
The two interacting bodies with masses $m_a$ and worldline coordinates $x_a^\alpha(\tau_a)$ are described by the action
\begin{equation}\label{eq:wl-act}
    S_{\mathrm{matter}} = -\sum_{a=1,2} m_a \int d\tau_a \sqrt{g_{\alpha\beta}(x_a(\tau_a)) \dot{x}_a^\alpha(\tau_a) \dot{x}_a^\beta(\tau_a)} \,,
\end{equation}
which, after applying the Brink-Di Vecchia-Howe reparametrisation~\cite{Brink:1976sc}, is expressed in terms of two auxiliary einbein fields $e_a(\tau_a)$
\begin{equation}\label{eq:brink-trick}
    S_{\mathrm{matter}} = -\sum_{a=1,2} \frac{m_a}{2} \int d\tau_a \left(\frac{g_{\alpha\beta}(x_a(\tau_a)) \dot{x}_a^\alpha(\tau_a) \dot{x}_a^\beta(\tau_a)}{e_a(\tau_a)} + e_a(\tau_a)\right).
\end{equation}
By varying the action with respect to one of the two einbeins one gets the constraint $e_a^2 =\dot{x}_a^2$, reproducing the original action~\eqref{eq:wl-act}.
For massive particles we can choose $e_a = 1$, named ``proper-time gauge'' as $\tau_a$ is now identified with the proper time according to the velocity normalisation $\dot{x}_a^2 = 1$.
Instead we approach the massless limit using $e_a = m_a/E_a$, with $E_a$ their initial energy in a scattering framework.
This definition allows a smooth ultrarelativistic limit for the mass-shell constraint $\dot{x}_a^2 = m_a^2/E_a^2\to0$.
In both massive and massless cases we have chosen a dimensionless einbein, consequently the worldline parameters $\tau_a$ have dimensions of a length.
Moreover, the second term in the round brackets in~\eqref{eq:brink-trick} is discarded for both $a=1,2$ since it is non-dynamical once the einbein is fixed.
Then we write down a generic action that holds for both massive and massless particles, introducing two quantities $M_a$ that could be either an energy or a mass, {\it i.e.} $M_a\in\{m_a,E_a\}$.
Relabelling $\tau_a = \tau$, we get
\begin{equation}\label{eq:matter-action}
    S_{\mathrm{matter}} = -\sum_{a=1,2} \frac{M_a}{2} \int d\tau \; g_{\alpha\beta}(x_a(\tau)) \dot{x}_a^\alpha(\tau) \dot{x}_a^\beta(\tau)\,.
\end{equation}
The worldlines naturally define the stress-energy tensors of the two bodies
\begin{equation}\label{eq:stress-energy-definition}
    T^{\alpha\beta}_a(x) = M_a\int d\tau \, \frac{\delta^{(D)}(x-x_a(\tau))}{\sqrt{-g(x)}}\, \dot{x}_a^\alpha \, \dot{x}_a^\beta \,,
\end{equation}
such that
\begin{equation}
    S_{\mathrm{matter}} = -\frac{1}{2} \sum_{a=1,2} \int_x T_a^{\alpha\beta}(x) g_{\alpha\beta}(x) \,.
\end{equation}
The full action is the sum of the matter action and the bulk action, consisting of Einstein-Hilbert gravity and a gauge-fixing term
\begin{equation}\label{eq:matter-plus-bulk}
    S = S_{\mathrm{matter}} + S_{\mathrm{bulk}} = -\frac{1}{2} \sum_{a=1,2} \int_x T_a^{\alpha\beta}(x) g_{\alpha\beta}(x) - \frac{2}{\kappa^2} \int_x R[g] + S_{\mathrm{g.f.}}
\end{equation}
We now proceed to expand the above action in Self-Force (SF) orders in terms of the small ratio $q=M_\ml/M_\mh\ll1$, where $M_\mh = M_1$ and $M_\ml = M_2$ are the {\it heavy} and {\it light} masses or energies of the two bodies respectively,
\begin{equation}
    S=\sum_{n=0}^\infty S^{(n)} \,, \quad\quad S^{(n)} \propto q^n \,.
\end{equation}
To this purpose, the gravitational field and the trajectories are expressed as their unperturbed solutions plus a perturbation
\begin{equation}
    g_{\alpha\beta} = \bar{g}_{\alpha\beta} + \kappa h_{\alpha\beta} \,, \hspace{30pt} x_a^\mu = \bar{x}_a^\mu + \delta x_a^\mu \,,
\end{equation}
with $h_{\alpha\beta}$ and $\delta x_a^\mu$ being Self-Force perturbations at least of order $\mathcal{O}(q)$.
All barred variables refer to quantities calculated on the unperturbed background $\bar{g}_{\alpha\beta}$.
Within this EFT, we will treat the unknown variables $\delta x_a^\mu$ classically: we solve order by order in $q$ the equations of motion for $\delta x_a^\mu$ in terms of the unknown metric perturbation $h_{\alpha\beta}$, place back these solutions into the action and thus obtain an EFT featuring classical sources and effective operators involving just one quantum field $h_{\alpha\beta}$.
Observables are then obtained as the expectation value of some operators $\mathcal{O}$ on the $h_{\alpha\beta}$ field
\begin{equation}\label{eq:operator-exp-value-conservative}
    \big\langle \mathcal{O} \big\rangle = \frac{\int \mathcal{D}h_{\alpha\beta} \, \mathcal{O}[h] \,e^{iS_\mathrm{EFT}[h]}}{\int \mathcal{D}h_{\alpha\beta} \,e^{iS_\mathrm{EFT}[h]}} = \sum_n q^n \, \big\langle\mathcal{O}\big\rangle^{(n)}\,,
\end{equation}
expanded in the small parameter $q$.
In the next paragraph we extend this formulation to the in-in path integral to include radiation-reaction effects~\cite{Kalin:2022hph,Jakobsen:2022psy}.
Now we proceed to analyse the SF sectors of the action up to $\mathcal{O}(q^2)$, namely the 1SF order.
We remark that the gauge-fixing action in~\eqref{eq:matter-plus-bulk} is quadratic in $h_{\alpha\beta}$ and thus $\mathcal{O}(q^2)$.

At 0SF, which includes both $\mathcal{O}(q^0)$ and $\mathcal{O}(q)$, the presence of the secondary source does not generate any perturbation of the metric field. Instead, it follows the geodesic motion on the background generated by the first body.
The first two orders of the action are
\begin{align}
    & S^{(0)} = -\frac{1}{2}\int_x \left(\bar{T}_\mh^{\alpha\beta}\bar{g}_{\alpha\beta} + \frac{4}{\kappa^2} R[\bar{g}]\right), \label{eq:S0}\\
    & S^{(1)} = -\frac{1}{2}\int_x \left(\kappa h_{\alpha\beta} \left(\bar{T}_{\mh}^{\alpha\beta} - \frac{4}{\kappa^2}\bar{G}^{\alpha\beta}\right) + \bar{g}_{\alpha\beta}\left(\bar{T}_{\ml}^{\alpha\beta}+\delta T_\mh^{\alpha\beta}\right)\right). \label{eq:S1}
\end{align}
Here $S^{(0)}$ is a constant while the only surviving term in $S^{(1)}$ is the one containing the light stress-energy tensor $\bar{T}_\ml^{\alpha\beta}$.
Indeed the first two terms in~\eqref{eq:S1} cancel because they satisfy the unperturbed Einstein equations given below in~\eqref{eq:einstein-equations}, while the piece involving $\delta T_\mh^{\alpha\beta}$ is reabsorbed in the (re)normalisation of the metric evaluated on the position of the heavy source $\bar{x}_\mh$.
It is proved in~\cite{Cheung:2024byb} that such renormalisation is needed to handle divergences arising from self-energy diagrams.
These contributions are reabsorbed by appropriate counterterms, whose effect is to redefine the metric evaluated on the worldline of the source $\bar{g}_{\alpha\beta}(\bar{x}_\mh) = \eta_{\alpha\beta}$.\footnote{Consequently, all the curvature symbols computed on $\bar{x}_\mh$ vanish $\bar{\Gamma}^\lambda_{\alpha\beta}(\bar{x}_\mh) = \bar{R}^{\gamma}_{\,\,\alpha\beta\delta}(\bar{x}_\mh) = 0$.}
We thus use $S^{(1)}$ to get the 0SF equations of motion
\begin{subequations}
\begin{align}
    & \bar{G}_{\alpha\beta} = 8\pi\gn \bar{T}_{\mh\,\alpha\beta} \,, \label{eq:einstein-equations} \\
    & \ddot{\bar{x}}_\ml^\mu + \bar{\Gamma}_{\,\rho\sigma}^\mu (\bar{x}_\ml)\,\dot{\bar{x}}_\ml^\rho \dot{\bar{x}}_\ml^\sigma = 0 \,, \label{eq:geodesic-equations}
\end{align}
\end{subequations}
that are detailed in the next paragraphs specialising to the gravitational shockwave background allowing the light body to be either massive or massless.

At 1SF order, the action reads
\begin{equation}
    \begin{aligned}
        S^{(2)} = & -\frac{M_\mh}{2} \int d\tau\left(\delta\dot{x}_\mh^2 - 2\delta x_\mh^\rho \dot{\bar{x}}_\mh^\alpha \dot{\bar{x}}_\mh^\beta \delta\Gamma_{\rho\alpha\beta}(\bar{x}_\mh)\right) - \frac{\kappa}{2}M_\ml\int d\tau \, \dot{\bar{x}}_\ml^\alpha \dot{\bar{x}}_\ml^\beta h_{\alpha\beta} (\bar{x}_\ml) \\
            & + \int d^Dx \sqrt{-\bar{g}}\, h^{\alpha\beta}\bar{\mathcal{D}}_{\alpha\beta}^{\;\;\;\;\gamma\delta}h_{\gamma\delta} + S_{\mathrm{g.f.}}
    \end{aligned}
\end{equation}
with $\bar{\mathcal{D}}$ a second-order differential operator~\cite{Andersson:2020gsj}
\begin{equation}\label{eq:diff-operator-graviton}
    \begin{aligned}
        \bar{\mathcal{D}}_{\alpha\beta}^{\;\;\;\;\gamma\delta} [\bar g] = & - \delta_\alpha^\gamma \delta_\beta^\delta \bnab_\lambda \bnab^\lambda + \bar g_{\alpha\beta} \bar g^{\gamma\delta} \bnab_\lambda \bnab^\lambda - \bar g^{\gamma\delta} \bnab_\alpha \bnab_\beta + \\
            & - \bar g_{\alpha\beta} \bnab^\gamma \bnab^\delta + \delta_\beta^\delta \bnab^\gamma \bnab_\alpha + \delta_\alpha^\delta \bnab^\gamma \bnab_\beta \,,
    \end{aligned}
\end{equation}
and
\begin{equation}
    \delta \Gamma^\omega_{\alpha\beta} = \frac{\kappa}{2} \bar{g}^{\omega\rho} \left(\bnab_\alpha h_{\rho\beta} + \bnab_\beta h_{\alpha\rho} - \bnab_\rho h_{\alpha\beta}\right)
\end{equation}
that is a difference of connections and thus a tensor.
From the first term of $S^{(2)}$ we derive the equations of motion for $\delta x_\mh^\mu$
\begin{equation}\label{eq:recoil-diffeq}
    \delta \ddot{x}_\mh^\mu + \delta\Gamma_{\alpha\beta}^\mu (\bar{x}_\mh) \dot{\bar{x}}_\mh^\alpha \dot{\bar{x}}_\mh^\beta = 0 \,,
\end{equation}
that are solved classically and put back into the action, obtaining a 1SF recoil effective operator,\footnote{The double integration $1/\partial_\tau^2$ is to be understood in analogy with solving the classical equations of motion~\eqref{eq:recoil-diffeq}, and it is accordingly regularised in frequency space ($\tau\leftrightarrow\omega$) via the retarded prescription $(\omega+i0^+)^{-2}$, as explained in~\cite{Jakobsen:2022psy}.}
\begin{equation}\label{eq:recoil1SF}
    S^{(2)}_{\mathrm{recoil}} = -\frac{M_\mh}{2}\int d\tau \, \dot{\bar{x}}_\mh^\alpha \dot{\bar{x}}_\mh^\beta \delta\Gamma^{\mu}_{\alpha\beta}(\bar{x}_\mh) \frac{1}{\partial_\tau^2} \dot{\bar{x}}_\mh^\gamma \dot{\bar{x}}_\mh^\delta \delta\Gamma_{\mu\gamma\delta}(\bar{x}_\mh) \,,
\end{equation}
that is quadratic in the gravitational perturbation.
Indeed it will be convenient to recast it as a quadratic differential operator $\bar{\mathcal{R}}$ such that it contributes explicitly to the $\mathcal{O}(q^2)$ action as follows
\begin{equation}\label{eq:S2}
        S^{(2)}_{\rm EFT} = - \frac{\kappa}{2}M_\ml \int d\tau \, \dot{\bar{x}}_\ml^\alpha \dot{\bar{x}}_\ml^\beta h_{\alpha\beta} (\bar{x}_\ml) + \int d^Dx \sqrt{-\bar{g}}\, h^{\alpha\beta}\left( \bar{\mathcal{D}}_{\alpha\beta}^{\;\;\;\;\gamma\delta} + \bar{\mathcal{R}}_{\alpha\beta}^{\;\;\;\;\gamma\delta} \right)h_{\gamma\delta} + S_{\mathrm{g.f.}}
\end{equation}
where the recoil differential operator is symbolically\footnote{This expression is purely symbolic, as performing the $1/\partial_\tau^2$ integrals requires the explicit form of $\bar{x}_\mh(\tau)$. Nevertheless, we find it useful to retain this form as it will prove useful in the next section.}
\begin{equation}
    \begin{aligned}
        & \bar{\mathcal{R}}_{\alpha\beta}^{\;\;\;\;\gamma\delta} = \frac{\kappa^2}{2}M_\mh\int d\tau\, \frac{\delta^{(D)}(x-\bar{x}_\mh(\tau))}{\sqrt{-\bar{g}}} \, \bar{\gamma}^{\mu}_{\;\;\alpha\beta} \, \frac{1}{\partial_\tau^2} \, \bar{\gamma}_\mu^{\;\;\gamma\delta}\,, \\
        & \bar{\gamma}^{\mu\,\alpha\beta} = \frac{1}{2}\bar{g}^{\mu\omega} \left(\left(\delta_\omega^\alpha \dot{\bar{x}}_\mh^\beta + \delta_\omega^\beta \dot{\bar{x}}_\mh^\alpha\right)\dot{\bar{x}}_\mh^\rho \bnab_\rho - \dot{\bar{x}}_\mh^\alpha \dot{\bar{x}}_\mh^\beta \bnab_\omega \right) \,.
    \end{aligned}
\end{equation}
The 1SF recoil operator $S^{(2)}_\mathrm{recoil}$ -- or equivalently $\bar{\mathcal{R}}_{\alpha\beta}^{\;\;\;\;\gamma\delta}$ -- describes the first-order fluctuation of the heavy body about its straight-line trajectory due to the presence of gravitational perturbations, which are sourced by the current $\bar{T}_\ml^{\alpha\beta}$.
The rest of the paper focuses on solving the 0SF sector~\eqref{eq:einstein-equations} and~\eqref{eq:geodesic-equations} classically and on the derivations and applications of the Feynman rules associated with the 1SF action~\eqref{eq:S2}.
The 1SF Feynman rules displayed in Figure\,\ref{fig:1SF-diagrams} are the diagrammatic representation, in terms of external graviton legs, of the 1SF action $S^{(2)}$.

\begin{figure}[t!]
    \begin{center}
        \begin{tikzpicture}[baseline=(int1.base)]
            \centering
            \begin{feynman}
                \vertex[dot, style={fill=NavyBlue!40, minimum size=10pt}] (int1) {};
                \vertex[left=1cm of int1] (wl1) {};
                \vertex[right=1cm of int1] (wl2) {};
                \vertex[above=1.8cm of int1] (em) {\(h_{\alpha\beta}\)};
                \diagram{
                    {[edges=graviton] (int1) -- (em)},
                };
            \end{feynman}
        \end{tikzpicture}
        \hspace{1cm}
        \begin{tikzpicture}[baseline=(int1.base)]
            \centering
            \begin{feynman}
                \vertex[] (int1) {};
                \vertex[left=1.5cm of int1] (wl1) {\(h_{\alpha\beta}\)};
                \vertex[right=1.5cm of int1] (wl2) {\(h_{\gamma\delta}\)};
                \vertex[dot, style={fill=gray!5, pattern=north east lines, minimum size=15pt}, above=1.2cm of int1] (int2) {};
                \diagram{
                    {(wl1) -- [edges=graviton, bend left=30] (int2) -- [edges=graviton, bend left=30] (wl2)},
                };
            \end{feynman}
        \end{tikzpicture}
        \hspace{1cm}
        \begin{tikzpicture}[baseline=(int1.base)]
            \centering
            \begin{feynman}
                \vertex[] (int1) {};
                \vertex[left=1.5cm of int1] (wl1) {\(h_{\alpha\beta}\)};
                \vertex[right=1.5cm of int1] (wl2){\(h_{\gamma\delta}\)};
                \vertex[dot, style={fill=thered!40, minimum size=15pt}, above=1.2cm of int1] (em) {};
                \diagram{
                    {(wl1) -- [edges=graviton, bend left=30] (em) -- [edges=graviton, bend left=30] (wl2)},
                };
            \end{feynman}
        \end{tikzpicture}
        \caption{The three types of interactions at 1SF order: the first one consists of the graviton field sourced by the geodesic motion of the light body, while the second and third are the two different contributions to the quadratic part of the action. 
        We use blue for graviton interactions with the light trajectory scaling with $M_\ml$, red for interactions with the heavy one scaling with $M_\mh$ and white dashed for the background. We adopt an all-ingoing convention.}
        \label{fig:1SF-diagrams}
    \end{center}
\end{figure}

The 2SF expansion has been derived in~\cite{Cheung:2024byb} and features the presence of deviations $\delta x_\ml^\alpha$ from the geodesic motion, together with the three-graviton vertex, which is separated into a background part and a 2SF recoil operator.
Pictorially, they resemble the second and third diagrams of Figure~\ref{fig:1SF-diagrams} with one more external leg.
We do not detail the 2SF order as it goes beyond the scope of the present work.

\subsection{Self-Force EFT and Schwinger-Keldysh formalism}\label{sec:schwinger-keldysh}
The Self-Force EFT introduced in Section~\ref{sec:SFEFT} is formulated, so far, as an ordinary in-out quantum field theory predicting time-ordered (Feynman) correlators of operators $\mathcal{O}[h]$ as in equation~\eqref{eq:operator-exp-value-conservative}, the natural output being S-matrix-like objects.
However, the classical equations we are interested in are naturally described in terms of Cauchy conditions, thus requiring the knowledge of initial state only.
Within the QFT formalism, this amounts to formulating an in-in (or Schwinger--Keldysh) path integral~\cite{Schwinger:1960qe} instead of the standard in-out one: the classical equations of motion and their retarded solutions are recovered exactly, order by order, without ever having to discard by hand the advanced or anti-time-ordered pieces of the in-out correlators.
This strategy was first introduced for effective actions of non-conservative classical systems in~\cite{Galley:2012hx}, and has become the standard tool to extract classical, radiation-reacted dynamics from worldline effective field theories~\cite{Kalin:2022hph,Jakobsen:2022psy,Caron-Huot:2023vxl}.
We adapt this construction to the Self-Force EFT introduced in the previous paragraph.

Starting from the EFT action obtained by integrating out the worldline perturbations, we promote the gravitational field to a pair of independent fields $h_A^{\alpha\beta}$ living on two copies $(A=1,2)$ of the original path integral contour, and assign to their actions a relative sign
\begin{equation}
    S_{\rm EFT}[h_1,h_2] = S_{\rm EFT}[h_1] - S_{\rm EFT}[h_2] \,,
\end{equation}
which has the effect of evolving the two fields forward or backward in time.
Truncating the action at the 1SF order we get 
\begin{equation}\label{eq:1sf-doubled-action}
    S^{(2)}_{\rm EFT}[h_1,h_2] = \int_x \left[h_{1}^{\alpha\beta}\left(\bar{\mathcal{D}}+\bar{\mathcal{R}}\right)_{\alpha\beta}^{\;\;\;\;\gamma\delta} h_{1\,\gamma\delta}
    - h_{2}^{\alpha\beta}\left(\bar{\mathcal{D}}+\bar{\mathcal{R}}\right)_{\alpha\beta}^{\;\;\;\;\gamma\delta} h_{2\,\gamma\delta}
    - J_{1}^{\alpha\beta}h_{1\,\alpha\beta} + J_{2}^{\alpha\beta}h_{2\,\alpha\beta}\right] ,
\end{equation}
with $J_A^{\alpha\beta}$ the external sources of gravitons.
The doubled 1SF generating functional is defined by
\begin{equation}
    Z^{(2)}[J_1,J_2] = \int \left[\mathcal{D}h_{1}\right] \, \left[\mathcal{D}h_{2}\right] \, \text{exp}\left(iS^{(2)}_{\rm EFT}[h_1,h_2]\right) \,.
\end{equation}
We stress that the background data -- the unperturbed metric $\bar g_{\alpha\beta}$ and the heavy trajectory $\bar x_\mh^\alpha$ -- are not doubled: they are external, fixed classical input common to both branches, rather than dynamical variables that are path-integrated over.
The path integral is then supplemented by the boundary conditions
\begin{equation}
    \begin{aligned}
        & \text{(i)}\quad \lim_{t\to-\infty}h_1^{\alpha\beta}(x) = \lim_{t\to-\infty} h_2^{\alpha\beta}(x) = 0\,,\\
        & \text{(ii)}\quad \lim_{t\to+\infty}h_1^{\alpha\beta}(x) = \lim_{t\to+\infty}h_2^{\alpha\beta}(x)\,.
    \end{aligned}
\end{equation}
Condition (ii) is the ``folding'' condition of the closed-time-path contour~\cite{Galley:2012hx,Caron-Huot:2023vxl}: since the two actions have opposite signs, the fields $h_1$ and $h_2$ evolve forward and backward in time respectively, and the folding condition forces them to agree in the asymptotic future, so that they describe a single physical field there; going backwards in time, they split into the two independent branches $h_1$ and $h_2$.
Again, in the asymptotic past they are required to be equal and they are trivially equal to zero because of condition (i).
To make the physical field manifest, we rotate to the Keldysh basis denoted by the indices $a,b=+,-$, with $h^+$ the physical field and $h^-$ its quantum fluctuations
\begin{equation}\label{eq:Keldyshrot}
    h^{+\,\alpha\beta} \equiv \frac{1}{2}\big(h_1^{\alpha\beta}+h_2^{\alpha\beta}\big)\,,
    \qquad
    h^{-\,\alpha\beta} \equiv h_1^{\alpha\beta}-h_2^{\alpha\beta}\,,
\end{equation}
so that $h_1 = h^+ + \tfrac12 h^-$ and $h_2 = h^+ - \tfrac12 h^-$; an analogous definition holds for the currents $J^{\pm}$.
Since the 1SF action $S_{\rm EFT}^{(2)}[h]$ is a linear and bilinear functional in $h_{\alpha\beta}$, the doubled action \eqref{eq:1sf-doubled-action} truncates exactly at linear order in $h^-$
\begin{equation}
    S^{(2)}_{\rm EFT}[h^+,h^-] = \int_x \left[
    2 h^{-\,\alpha\beta} \left(\bar{\mathcal{D}}+\bar{\mathcal{R}}\right)_{\alpha\beta}^{\;\;\;\;\gamma\delta} h^{+}_{\gamma\delta}
    - J^{+\,\alpha\beta} h^{-}_{\alpha\beta} - J^{-\,\alpha\beta} h^{+}_{\alpha\beta} \right]\,,
\end{equation}
without actually needing to expand in powers of $h^-$.
Then the 1SF generating functional $Z^{(2)}[J^+,J^-]$ is computed by integrating out the $h^-$ field first and the classical field $h^+$ afterwards.
The first integration gives
\begin{equation}
    Z^{(2)}[J^+,J^-] = \mathcal{N} \int \left[\mathcal{D}h^+\right] \text{exp}\left\{-i\int_x J^{-\,\alpha\beta}h^+_{\alpha\beta} \right\} \delta \big[ 2\left(\bar{\mathcal{D}}+\bar{\mathcal{R}}\right)_{\alpha\beta}^{\;\;\;\;\gamma\delta} h^{+}_{\gamma\delta} - J^{+}_{\alpha\beta} \big] \,,
\end{equation}
with an unknown normalisation $\mathcal{N}$.
The delta function coming from the $h^-$ integration provides the classical solution for the causal field $h^+_{\alpha\beta}$
\begin{equation}\label{eq:classical-eom-hplus}
    \big\langle h^+_{\alpha\beta}(x) \big\rangle = \frac{1}{2} \int_{x^\prime} \mathbf{G}^{\rm ret.}_{\alpha\beta\,\gamma\delta}(x,x^\prime) J^{+\,\gamma\delta}(x^\prime)\,,
\end{equation}
where $\mathbf{G}^{\rm ret.}$ is the inverse of the operator $\bar{\mathcal{D}} + \bar{\mathcal{R}}$ calculated using the retarded prescription.
Further details on the propagator are given in section~\ref{sec:propagators}.
Then the 1SF generating functional is
\begin{equation}\label{eq:generating-functional-2}
    Z^{(2)}[J^+,J^-] = Z_0 \; \text{exp}\left\{ -\frac{i}{2} \int_{x,x^\prime} J^{-\,\alpha\beta}(x) \mathbf{G}^{\rm ret.}_{\alpha\beta\,\gamma\delta}(x,x^\prime) J^{+\,\gamma\delta}(x^\prime) \right\} \,.
\end{equation}
with $Z_0$ a normalisation constant.
In Section~\ref{sec:propagators} we compute the retarded Green function on the GSW background, while in Section~\ref{sec:waveform} we use it to compute the gravitational waveform from equation~\eqref{eq:classical-eom-hplus}.
As explained in~\cite{Jakobsen:2022psy}, the effect of using the Schwinger-Keldysh folding conditions consists in considering {\it causal diagrams} in which we have a definite causality flow dictated by the retarded conditions on the propagators; specifically, each diagram has only one outgoing field $h^+$ and all ingoing contributions $h^-$ sourced by the currents $J^+$.
To give an example on flat space $(\bar{g}_{\alpha\beta}=\eta_{\alpha\beta} \text{ and } \bar{\mathcal{R}}_{\alpha\beta}^{\;\;\;\;\gamma\delta}=0)$, the retarded graviton propagator is~\cite{Kalin:2022hph}
\begin{equation}
    i\propx_{\alpha\beta\gamma\delta}^{(0)}(x_1,x_2) = 2\,\frac{i\delta}{\delta J^{-\,\alpha\beta}(x_1)} \, \frac{i\delta}{\delta J^{+\,\gamma\delta}(x_2)}\bigg|_{J^\pm=0} = \big\langle h^{+}_{\alpha\beta}(x_1) h^{-}_{\gamma\delta}(x_2) \big\rangle \,,
\end{equation}
which is in momentum space
\begin{equation}\label{eq:flat-propagator-2}
    i\propx_{\alpha\beta\gamma\delta}^{(0)}(k_1,k_2) = 
    \begin{tikzpicture}[baseline=(cntr.base)]
        \centering
        \begin{feynman}
            \vertex[] (cntr) {\(\text{\small$\gamma\delta(-)$}\)};
            \vertex[right=2.2cm of cntr] (phi) {\(\text{\small$(+)\alpha\beta$}\)};
            \diagram* {
                {[edges=photon]
                (cntr) -- (phi)
                }
            };
        \end{feynman}
        \middlearrow{cntr}{phi}
    \end{tikzpicture}
    = \frac{iP_{\alpha\beta\gamma\delta}(k_1)}{k_1^2+i0^+(k_1\cdot\xi)}\frac{\hat{\delta}^{(D)}(k_1+k_2)}{\sqrt{-\eta}}\,,
\end{equation}
with $P_{\alpha\beta\gamma\delta}$ the helicity projector specified below in paragraph~\ref{sec:gauge-choice}.
We now focus on the denominator: the retarded prescription in terms of light-cone coordinates is
\begin{equation}
    k^2+i k^0\, 0^+ = k^2+i0^+\left(k\cdot\chi+k\cdot\xi\right)\simeq\left(k\cdot\chi+i0^+\right)\left(k\cdot\xi+i0^+\right) + k_\pperp^2\,.
\end{equation}
After applying a complex rotation to the minus integration variable $k\cdot\xi\to k\cdot\xi-i0^+$ we get exactly the Mandelstam-Leibbrandt (ML) prescription~\cite{Leibbrandt:1983pj,Leibbrandt:1987qv,Mandelstam:1982cb} of equation~\eqref{eq:flat-propagator-2}.
More precisely, equation~\eqref{eq:flat-propagator-2} coincides with the ML propagator for positive $k\cdot\xi>0$ only, condition that is always satisfied for on-shell momenta in our computations.
We checked that all the computations that we present are either invariant under the change of variable $k\cdot\xi\to k\cdot\xi-i0^+$ or properly regularised by this shift.
In fact, when using the axial-gauge projector $P_{\alpha\beta\gamma\delta}(k)$ (introduced later in this Section) we find additional singularities $(k\cdot\xi)^{-n}$ produced by the gauge choice -- see {\it e.g.} equations~\eqref{eq:pijkl-projected} and~\eqref{eq:pijkl-projected2} -- whose poles are mapped into
\begin{equation}
    \text{axial gauge:} \quad \frac{1}{(k\cdot\xi)^n} \hspace{3pt} \longmapsto \hspace{3pt} \frac{1}{(k\cdot\xi-i0^+)^n} \,, 
\end{equation}
thanks to the rotation mentioned above, making the retarded prescription manifest as a regularisation of the axial gauge singularities.
Nevertheless, the computations appearing in this paper do not need this kind of regularisation; thus from now on we ignore this subtlety and we just use $k^2+i0^+(k\cdot\xi)$ as a propagator.

\subsection{Self-Force EFT with Gravitational Shockwave background}\label{sec:2.3}
In this subsection we introduce the gravitational shockwave (GSW), or Aichelburg-Sexl~\cite{Aichelburg-Sexl}, solution of Einstein's equations, sourced by a massless, point-like particle with energy $E_1$ and uniform velocity $\xi^\alpha$.
We then review the light-like and time-like geodesics of this metric, connecting them to the operators that source gravitons in the 1SF action $S_{\rm EFT}^{(2)}$.
Finally, we covariantly fix the gauge of the perturbation $h_{\alpha\beta}$ and outline many properties and simplifications that derive from it.

\subsubsection{The Gravitational Shockwave solution}\label{sec:GSW}
We specialise to the case where the highly energetic particle is massless and generates a gravitational shockwave (GSW) background.
This solution was first derived by Aichelburg and Sexl~\cite{Aichelburg-Sexl} by taking the ultrarelativistic limit of a boosted Schwarzschild metric.
Equivalently, it is obtained as an exact solution of the Einstein equations~\eqref{eq:einstein-equations} sourced by a massless particle with velocity $\dot{\bar{x}}_\mh^\mu=\xi^\mu=(2,0,0,0)$ and energy $E_1$, corresponding to the unperturbed stress tensor of the ``heavy'' particle
\begin{equation}
    \bar T_{\mh}^{\alpha\beta}(x) = E_1 \delta(\xi\cdot x) \delta^{(d)}(x_\pperp) \xi^\alpha \xi^\beta \,.
\end{equation}
We recall that $\xi^\alpha$ and $\chi^\alpha$ are two vectors such that $x^- = x\cdot\xi, x^+=x\cdot\chi$, $\xi^2 = \chi^2 = 0$, and $\xi\cdot\chi=2$.
The velocity $\xi$ is a Killing vector of the metric, which is expressed in Kerr-Schild form
\begin{equation}\label{eq:sw-metric}
    \bar g_{\alpha\beta} = \eta_{\alpha\beta} + \Phi(x)\xi_\alpha\xi_\beta\,.
\end{equation}
Thanks to the null nature of $\xi$, metric indices are raised and lowered by $\eta$.
The function $\Phi$, which satisfies $\xi^\mu \bnab_\mu \Phi= 0$, describes the transverse shape of the GSW wavefront localised on $x^{-}=\xi\cdot x$
\begin{equation}\label{eq:function-phi}
    \Phi(x) = \ag \delta\left(\xi\cdot x\right) G_d(r)\,,
\end{equation}
with $\ag = 16\pi\gn E_1$, $r=|x_\pperp|$ and $G_d$ the Green function of the Laplace operator in $d=D-2$ Euclidean dimensions~\cite{FerrariVeneziano}
\begin{subequations}
    \begin{align}
        & G_d (r) = - \frac{r^{4-D}}{(D-4)\Omega_{D-2}}\,, & \hspace{-50pt} D > 4\,, \label{eq:Gd-dimGen} \\
        & G_d (r) = \frac{\log(r\muirt)}{2\pi}\,, & \hspace{-50pt} D = 4\,, \label{eq:Gd-dim4}
    \end{align}
\end{subequations}
where $\Omega_{D-2} = \Omega_d = {2\pi^{d/2}}/{\Gamma(d/2)}$ is the solid angle of the $S^{d-1}$ sphere, while $\muirt$ is a mass scale introduced to make the argument of the logarithm dimensionless, that we fix for convenience to be related to the dimensional regularisation scale $\muir$ through $\muirt^2 = \pi \muir^2 e^{\gamma_{\rm E}}$.
In that way we there is a smooth transition from the $D$-dimensional function~\eqref{eq:Gd-dimGen} to the four-dimensional one~\eqref{eq:Gd-dim4} through the replacement of the pole $1/\epsilon\to2\log\muir$.
From now on we investigate the perturbations of the GSW metric induced by the presence of a secondary body with much smaller mass or energy $q=M_2/E_1\ll1$.

\subsubsection{Geodesic motion and graviton sources}\label{sec:graviton-sources}
At 1SF order the gravitons are sourced by the geodesic motion of the light body through the first term in~\eqref{eq:S2},
\begin{equation}
    S^{(2)}_{h-\mathrm{source}} = - \frac{\kappa}{2} \int_x \bar{T}_\ml^{\alpha\beta}(x) \, h_{\alpha\beta}(x)\,,
\end{equation}
which is pictured in blue in Figure\,\ref{fig:1SF-diagrams} and where the unperturbed stress-energy tensor is
\begin{equation}\label{eq:tmunu-ft-def}
    \bar{T}^{\alpha\beta}_{\ml}(k) = M_\ml \int d\tau \, e^{ik\cdot\bar{x}_\ml(\tau)} \, \dot{\bar{x}}_\ml^\alpha(\tau)\, \dot{\bar{x}}_\ml^\beta(\tau) \,.
\end{equation}
The corresponding Feynman rule for graviton emission in momentum space is
\begin{equation}\label{eq:grav-sources}
    \begin{tikzpicture}[baseline=(bs.base)]
        \centering
        \begin{feynman}
            \vertex[dot, style={fill=NavyBlue!40, minimum size=10pt}] (int1) {};
            \vertex[left=1cm of int1] (wl1) {};
            \vertex[right=1cm of int1] (wl2) {};
            \vertex[above=1.6cm of int1] (em) {\(h_{\alpha\beta}\)};
            \vertex[above=0.6cm of int1] (bs) {};
            \diagram{
                {[edges=graviton] (int1) -- (em)},
            };
        \end{feynman}
    \end{tikzpicture}
    \hspace{-15pt}
    = \mathrm{FT}\left[ i\frac{\delta S^{(2)}_{h-\mathrm{source}}}{\delta h_{\alpha\beta}} \right] (k) = -i\frac{\kappa}{2} \, \bar{T}_\ml^{\alpha\beta} (k) \,,
\end{equation}
To compute it, we need the solutions $\bar{x}_\ml^\alpha(\tau)$ of the geodesic equations~\eqref{eq:geodesic-equations} in the presence of the GSW background, previously derived by Ferrari, Pendenza and Veneziano~\cite{FerrariVeneziano} and by Dray and 't Hooft~\cite{Dray:1984ha}.
In Appendix~\ref{app:geodesic-motion} we re-compute them within two distinct frameworks: $(i)$ a massless body initially travelling in the light-cone direction opposite to the shockwave, with energy $E_2 \ll E_1$ - namely with initial momentum $p_2^\mu = E_2\chi^\mu$ - and $(ii)$ a massive particle initially at rest with a small mass $m_2 \ll E_1$, both taken at an impact parameter $b$ from the GSW, which is purely transverse, $b = |b_\pperp|$.
We first focus on the former.
Choosing $\bar{x}_\ml^- = s$ as the affine parameter, the massless geodesics read
\begin{subequations}\label{eq:geo-m0}
    \begin{align}
        & \bar{x}^{-}_\ml(s) = s, \label{eq:geo1}\\
        & \bar{x}_{\ml,\pperp}^{\alpha}(s) = b_\pperp^\alpha + s \Theta(s) f_2 \,\hat{b}_\pperp^\alpha, \label{eq:geo2}\\
        & \bar{x}^{+}_\ml(s) = \Theta(s) f_1 + s\Theta(s) f_3\,,
    \end{align}
\end{subequations}
where the constants $f_i$ are related to the GSW transverse shape function $G_d(r)$ through
\begin{equation}\label{eq:f1-f2-f3}
    f_1 = - \ag G_d(b) \,, \qquad f_2 = \frac{1}{2}\frac{df_1}{db} = -\frac{\ag}{2}G_d'(b) \,, \qquad f_3 = f_2^2 = \frac{\ag^2}{4}\left(G_d'(b)\right)^2 \,,
\end{equation}
and represent the instantaneous changes in position and velocity of the projectile as it crosses the wavefront.
Indeed in $D=4$ dimensions they reduce to
\begin{subequations}\label{eq:deflections-geodesics}
    \begin{align}
        & \Delta \bar{x}_{\ml}^{+} = f_1\big|_{D=4} = -8\gn E_1\log\left(b\muirt\right) \,,\\
        & \Delta \dot{\bar{x}}_{\ml}^{+} = f_3\big|_{D=4} = \frac{16G^2E_1^2}{b^2} \,,\\
        & \Delta \dot{\bar{x}}_{\ml,\perp}^\alpha = f_2\,\hat{b}_\pperp^\alpha\big|_{D=4} = - \frac{4GE_1\hat{b}_\perp^\alpha}{b} \,,
    \end{align}
\end{subequations}
in agreement with previous computations~\cite{FerrariVeneziano,Dray:1984ha,Bohnenblust:2026ujk}.
Within the massive-particle framework, we solve the geodesic equations for the mass $m_2$ initially at rest, finding that they yield the same shifts $f_1, f_2$ and $f_3$ and the same $\bar{x}_{\ml,\pperp}, \bar{x}_{\ml}^{-}$ but a different plus trajectory
\begin{equation}\label{eq:xplus-geo-massive}
    \bar{x}^{+}_{\ml,m\ne0}(s) = s + \Theta(s) f_1 + s\Theta(s) f_3\,,
\end{equation}
which leads to the same deflections as in equation~\eqref{eq:deflections-geodesics}.
All the computational details are provided in Appendix~\ref{app:geodesic-motion}.
In order for the Self-Force expansion to be valid, the ratio of the energies of the two bodies must be small also for the outgoing particles.
This implies that the velocity kicks are much smaller than unity, forcing the impact parameter to be large, $b\gg\gn E_1$.

The Feynman rule for the graviton source in~(\ref{eq:grav-sources}\,-\,\ref{eq:tmunu-ft-def}) is then computed using the geodesics derived above, with affine parameter $\tau = s/2$, and with the help of the identity
\begin{equation}\label{eq:theta-integral}
    \int_\mathbb{R} dx \, g(x) \, e^{f(x)\Theta(x)} = \int_\mathbb{R} dx \, g(x) + \int_{\mathbb{R}^+} dx \, g(x) \, \left(e^{f(x)} - 1 \right) \,.
\end{equation}
In performing the computation, we need to introduce an $i0^+$ regulator to make the Fourier exponential $e^{ik\cdot\bar{x}_\ml(\tau)}$ vanish at infinity, so that the integral converges,
obtaining
\begin{subequations}\label{eq:graviton-sources-massless}
    \begin{align}
        & \bar{T}^{--}_{\ml}(k) = 4iE_2 \, e^{-i\vec{k}\cdot\vec{b}} \left(\psi(k) - \frac{1}{(k\cdot\chi)-i0^+}\right)\,, \label{eq:separation0}\\
        & \bar{T}^{-i}_{\ml}(k) = 4iE_2 f_2 \frac{b^i}{b}\, e^{-i\vec{k}\cdot\vec{b}} \psi(k)\,, \\
        & \bar{T}^{ij}_{\ml}(k) = 4iE_2 f_2^2 \frac{b^i b^j}{b^2} \, e^{-i\vec{k}\cdot\vec{b}} \psi(k)\,,
    \end{align}
\end{subequations}
where $\psi(k)$ is defined as
\begin{equation}
    \psi(k) = \frac{e^{i(k\cdot\xi)f_1/2}}{(k\cdot\chi) + f_3(k\cdot\xi) - 2 f_2\,\vec{k}\cdot\hat{b} +i0^+}\,.
\end{equation}
This function describes the instantaneous deflection of the second body after its passage through the wavefront of the shockwave (see Figure~\ref{fig:pict-trajectory}).
Accordingly, we define two phases of the trajectory: its {\it early} component ${}^e\bar{T}_\ml$ -- which is simply a straight line in the $\chi^\alpha$ direction with energy $E_2$ and impact parameter $b$ -- and its {\it late} component ${}^\ell\bar{T}_\ml$.
In particular, the early component is obtained from equations~\eqref{eq:graviton-sources-massless} by setting $\psi(k) = 0$, while ${}^\ell\bar{T}_\ml$ is given by the terms proportional to $\psi(k)$
\begin{equation}\label{eq:separation}
    \bar{T}^{\alpha\beta}_\mathrm{L}(k) = {}^{e}\bar{T}^{\alpha\beta}_\mathrm{L}(k) + {}^{\ell}\bar{T}^{\alpha\beta}_\mathrm{L}(k)\,.
\end{equation}
We anticipate that, thanks to the gauge choice for the graviton propagator outlined in Sec.\,\ref{sec:gauge-choice}, we will never need the plus components of the light stress-energy tensor $\bar{T}_\ml^{+\alpha}$, which we have therefore not shown.

The momentum-space stress-energy tensor for the massive particle initially at rest is computed analogously from the geodesic motion~(\ref{eq:geo1}\,-\,\ref{eq:geo2}\,-\,\ref{eq:xplus-geo-massive}), obtaining
\begin{equation}
    {}^e\bar{T}_{\ml,m\ne0}^{--}(k) = -\frac{2im_2e^{-i\vec{k}\cdot\vec{b}}}{(k\cdot\chi)+(k\cdot\xi)-i0^+}\,,
\end{equation}
for the early trajectory and
\begin{equation}
    {}^\ell\bar{T}_{\ml,m\ne0}^{\alpha\beta}(k) = {}^\ell\bar{T}_{\ml}^{\alpha\beta}(k) \Big|_{E_2\to m_2/2,\,f_3\to f_3+1}\,,
\end{equation}
for the late component.

\begin{figure}[t]
    \begin{center}
        \begin{tikzpicture}
            \begin{feynman}
                \vertex[crossdot, thered] (cntr) {};
                \vertex[left=7cm of cntr] (E1in) {};
                \vertex[right=3cm of cntr] (E1out) {};
            
                \vertex[above=2.5cm of cntr] (frontUp) {};
                \vertex[below=1.5cm of cntr] (frontLo) {};
            
                \vertex[above=1.5cm of cntr, style={fill=gray!5, pattern=north east lines, minimum size=15pt}] (int) {};
                \vertex[right=6cm of int] (earlyGeo) {};
                \vertex[left=5cm of int] (GeoVar) {};
                \vertex[above=1.3cm of GeoVar] (lateGeo) {};
            
                \vertex[left=1.5cm of earlyGeo] (e2tail) {};
                \vertex[left=0.5cm of earlyGeo] (e2arr) {};
                \vertex[right=1.5cm of E1in] (e1tail) {};
                \vertex[right=0.5cm of E1in] (e1arr) {};
                
                \diagram* {
                    (E1in) -- [thick, style={dash dot}] (cntr) -- [thick, style={dash dot}] (E1out),
                    (frontUp) -- [thered!80, ultra thick] (cntr) -- [thered!80, ultra thick] (frontLo),
                    (earlyGeo) -- [NavyBlue, thick, style={dash dot}, edge label=${}^{e}\bar{T}^{\alpha\beta}_\ml$] (int) -- [NavyBlue, thick, style={dash dot}, edge label=${}^{\ell}\bar{T}^{\alpha\beta}_\ml \propto \psi$] (lateGeo),
                    (e1arr) -- [style={-{Latex}}, edge label=$E_1\xi^\mu$] (e1tail),
                    (e2arr) -- [style={-{Latex}}, edge label=$E_2\chi^\mu$] (e2tail)
                };
            \end{feynman}
        \end{tikzpicture}
    \end{center}
    \caption{Schematic picture of the massless two-body collision in GR. The black trajectory generates the shockwave front (red) and does not recoil according to the vanishing of~\eqref{eq:recoil1SF}. 
    The less energetic particle follows the blue trajectory derived in App.\,\ref{app:geodesic-motion}, encoded in the stress energy tensor, which we separate in an early contribution ${}^{e}\bar{T}_\ml^{\alpha\beta}$ and a late (or {\it deflection}) contribution ${}^{\ell}\bar{T}_\ml^{\alpha\beta}$, as in equation~\eqref{eq:separation}.}
    \label{fig:pict-trajectory}
\end{figure}

\subsubsection{Light-cone gauge and its properties}\label{sec:gauge-choice}
Before turning to the diagrammatics, we fix the gauge for $h_{\alpha\beta}$ and derive some gauge-dependent features that dramatically simplify the computations.
Keeping in mind that $\xi$ is a uniform Killing field of the GSW metric~\eqref{eq:sw-metric}, we pick an axial gauge in which the reference vector is $\xi$, referred to as the {\it light-cone gauge}.
It is defined by the constraint $\xi^\alpha h_{\alpha\beta} = 0$, which holds at the level of equations of motion, while the explicit expression for the polarisation tensors is obtained by double-copying the Yang-Mills light-cone gauge~\cite{Bern:2019prr,Raj:2024xsi,DelDuca:1995zy,Alessio:2026bdi}
\begin{equation}
    \pol_{\alpha\beta}^{(\sigma)}(k) = \pol_{\alpha}^{(\sigma)}(k) \, \pol_{\beta}^{(\sigma)}(k)\,,
\end{equation}
with $\sigma = \oplus,\ominus$, $\left(\pol_{\oplus}\right)^{*} = \pol_\ominus$ and
\begin{equation}\label{eq:polarisation}
    \pol^{\alpha}_\oplus(k) = \frac{1}{\sqrt2} \left(\frac{2\zeta_k}{(\xi\cdot k)},0;1,i\right) .
\end{equation}
We have defined the complex transverse component $\zeta_k = k^x +ik^y$.
The Yang-Mills polarisations satisfy $k\cdot\pol(k)=0$, as well as $\pol^{(\sigma)}_{\alpha}(k_1)\pol^{\alpha\,(\sigma^\prime)}(k_2) = \pol^{(\sigma)}_{i}(k_1)\pol^{i\,(\sigma^\prime)}(k_2) = \delta_{\sigma\sigma^\prime}-1$, and
\begin{equation}\label{eq:sum-polarisations-ym}
    P_{\alpha\beta}(k) = \sum_{\sigma} \pol_\alpha^{(\sigma)} (k) \pol^{(\sigma)\,*}_\beta (k) = -\eta_{\alpha\beta} + \frac{\xi_\alpha k_\beta + k_\alpha \xi_\beta}{k\cdot\xi}
\end{equation}
on-shell.
The projector $P_{\alpha\beta}$ satisfies $\xi^\alpha P_{\alpha\beta}(k) = 0$ also for off-shell momenta, and serves as a building block for the sum over the graviton helicities~\cite{Raj:2024xsi}
\begin{equation}\label{eq:sum-polarisations}
    P_{\alpha\beta\gamma\delta}(k) = \sum_{\sigma} \pol_{\alpha\beta}^{(\sigma)}(k)\, \pol^{(\sigma)\,*}_{\gamma\delta}(k) = \frac{1}{2}P_{\alpha\gamma}P_{\beta\delta} + \frac{1}{2}P_{\alpha\delta}P_{\beta\gamma} - \frac{1}{D-2}P_{\alpha\beta}P_{\gamma\delta}\,.
\end{equation}
For computational purposes, the two projectors~\eqref{eq:sum-polarisations-ym} and~\eqref{eq:sum-polarisations} are decomposed in light-cone coordinates in Appendix~\ref{app:projectors}.

Although the Self-Force EFT allows for non-perturbative computations in $\gn$, it is in most cases necessary to expand around Minkowski space and treat the background field as an external perturbation, see {\it e.g.}~\cite{Cheung:2023lnj,Cheung:2024byb} and Section~\ref{sec:propagators} of this paper.
However, the polarisations that we define in Minkowski space preserve their properties when used on $\bar{g}$, since their indices can be equivalently raised and lowered with $\eta_{\alpha\beta}$ or $\bar{g}_{\alpha\beta}$, owing to the Kerr-Schild form of the metric.
The graviton propagator on the GSW background -- defined in equation~\eqref{eq:two-point-bra} below -- is obtained perturbatively by inserting infinite corrections proportional to $\bar{g}_{\alpha\beta}-\eta_{\alpha\beta}$ and its resummed form~\eqref{eq:vert-prop-2} features a tensor structure similar to $P_{\alpha\beta\gamma\delta}$, which is valid in flat space.
The only difference with $P_{\alpha\beta\gamma\delta}$ lies in the two-momentum sum, since energy and momentum are not conserved in a generic curved background,
\begin{equation}\label{eq:projector-P2}
    \mathcal{P}_{\alpha\beta\gamma\delta}(k_1,k_2) \coloneq \sum_{\sigma} \pol_{\alpha\beta}^{(\sigma)}(k_1) \,\pol_{\gamma\delta}^{(\sigma)\,*}(k_2) \,.
\end{equation}
which reduces to $P_{\alpha\beta\gamma\delta}$ for $k_1=k_2$.
Therefore, combining the Kerr-Schild form of the metric and the gauge choice allows to bootstrap the tensor structure $\mathcal{P}_{\alpha\beta\gamma\delta}$ of the graviton's Green function, that will be confirmed by an explicit computation in the next section.

As a consequence of the properties of the light-cone gauge, any interaction diagram containing at least one $\xi^\alpha$ vanishes automatically when contracted with a graviton propagator, since this contains a factor $P_{\alpha\beta\gamma\delta}$.
This allowed us to neglect $\bar{T}^{\alpha+}_\ml$ in~\eqref{eq:graviton-sources-massless}, and likewise to set $\xi^\alpha h_{\alpha\beta} = 0$ in the action~\eqref{eq:S2}, the latter proved in the next section.
Finally, since $\dot{\bar{x}}_\mh^\alpha = \xi^\alpha$, the 1SF heavy recoil operator~\eqref{eq:recoil1SF} gives a null contribution to the diagrams, as does the 2SF recoil operator derived in equation~(124) of~\cite{Cheung:2024byb}, which we write here for completeness
\begin{equation}
    \begin{aligned}
        S^{(3)}_{\rm recoil}= - E_1 \int d\tau \Big[& \frac{1}{2}\dot{\bar{x}}_\mh^{\alpha} \dot{\bar{x}}_\mh^{\beta} \delta \Gamma^{\mu}_{\alpha \beta}\frac{1}{\overleftarrow{\partial_\tau}}\delta g_{\mu \nu}\frac{1}{\overrightarrow{\partial_{\tau}}}\dot{\bar{x}}_\mh^{\gamma} \dot{\bar{x}}_\mh^{\delta} \delta \Gamma^{\mu}_{\gamma \delta}\\
        & + \dot{\bar{x}}_\mh^{\alpha} \dot{\bar{x}}_\mh^{\beta} \delta \Gamma^{\mu}_{\alpha \beta}\frac{1}{\overleftarrow{\partial_\tau}}\dot{\bar{x}}_\mh^{\rho} \delta \Gamma_{\mu \nu \rho}\frac{1}{\overrightarrow{\partial_{\tau}^{2}}}\dot{\bar{x}}_\mh^{\gamma} \dot{\bar{x}}_\mh^{\delta} \delta \Gamma^{\mu}_{\gamma \delta}\\
        & + \frac{1}{2} \dot{\bar{x}}_\mh^{\alpha} \dot{\bar{x}}_\mh^{\beta} \delta \Gamma^{\mu}_{\alpha \beta}\frac{1}{\overleftarrow{\partial_{\tau}^{2}}}\dot{\bar{x}}_\mh^{\rho} \dot{\bar{x}}_\mh^{\sigma} \delta R_{\rho \mu \nu \sigma}\frac{1}{\overrightarrow{\partial_{\tau}^{2}}}\dot{\bar{x}}_\mh^{\gamma} \dot{\bar{x}}_\mh^{\delta} \delta \Gamma^{\mu}_{\gamma \delta}\Big] \Big|_{\,\dot{\bar{x}}_\mh^\alpha = \xi^\alpha}\,.
    \end{aligned}
\end{equation}
We therefore state that {\it there are no shockwave recoil operators at any Self-Force order in the light-cone gauge}.
This statement can easily be proved.
Imposing the gauge condition $\xi^\alpha h_{\alpha\beta} = 0$ on the perturbation, we obtain $\xi_\alpha = g_{\alpha\beta} \xi^\beta$ also for the full metric, and thus $\Gamma^{\mu}_{\alpha\beta}\xi^\alpha\xi^\beta = 0$.
Therefore, the geodesic equation for the null trajectory $x_\mh^\alpha$ on the perturbed background is identical to the equation for $\bar{x}_\mh^\alpha$ on the unperturbed metric, {\it i.e.},
\begin{equation}
    \begin{cases}
        \ddot{x}_\mh^\alpha(\tau) = \ddot{\bar{x}}_\mh^\alpha(\tau) = 0\,, \\
        \dot{x}_\mh^\alpha(-\infty) = \dot{\bar{x}}_\mh^\alpha(-\infty) = \xi^\alpha\,,
    \end{cases}
\end{equation}
with the solution $x_\mh^\alpha(\tau) = \bar{x}_\mh^\alpha(\tau) = \xi^\alpha\tau$ that is unique once the transverse component is fixed to $x_{\mh\,\pperp}^\alpha(\tau) = 0$.
Consequently, each retarded solution for the variation of motion $\delta x_\mh^\alpha$ vanishes identically at any SF order, which proves the statement
\begin{equation}
    S_{\rm recoil}^{(n)} = 0 \hspace{10pt} \forall n \ge 2\,.
\end{equation}
The choice of the light-cone gauge makes the recoil of the first body vanish, implying that there is no deflection.
The whole physical recoil is entirely encoded in the gravitational field.
{\section{Graviton propagator on GSW background}\label{sec:propagators}}
The last ingredient required for computing observables in the Self-Force expansion -- also one of the principal results of this paper -- is the Green's function of the gravitational perturbation on the GSW background, {\it aka} the graviton propagator.
It is the inverse of the quadratic operator $\bar{\mathcal{D}}_{\alpha\beta}^{\;\;\;\gamma\delta}+\bar{\mathcal{R}}_{\alpha\beta}^{\;\;\;\gamma\delta}$ in~\eqref{eq:S2}, depicted in Figure~\ref{fig:propagator}.
\begin{figure}[t]
    \begin{center}
        \begin{tikzpicture}[baseline=(int1.base)]
            \centering
            \begin{feynman}
                \vertex[] (wl1) {\(h_{\gamma\delta}\,(-)\)};
                \vertex[right=3cm of wl1] (wl2) {\((+)\,h_{\alpha\beta}\)};
                \diagram{
                    {(wl1) -- [edges=graviton] (wl2)},
                };
            \end{feynman}
            \middlearrowBig{wl1}{wl2}
        \end{tikzpicture}
        $=\Bigg($
        \begin{tikzpicture}[baseline=(int1.base)]
            \centering
            \begin{feynman}
                \vertex[dot, style={fill=gray!5, pattern=north east lines, minimum size=15pt}] (int1) {};
                \vertex[left=1.5cm of int1] (wl1) {\(h_{\alpha\beta}\)};
                \vertex[right=1.5cm of int1] (wl2) {\(h_{\gamma\delta}\)};
                \diagram{
                    {(wl1) -- [edges=graviton] (int1) -- [edges=graviton] (wl2)},
                };
            \end{feynman}
        \end{tikzpicture}
        +
        \begin{tikzpicture}[baseline=(int1.base)]
            \centering
            \begin{feynman}
                \vertex[dot, style={fill=thered!40, minimum size=15pt}] (int1) {};
                \vertex[left=1.5cm of int1] (wl1) {\(h_{\alpha\beta}\)};
                \vertex[right=1.5cm of int1] (wl2){\(h_{\gamma\delta}\)};
                \diagram{
                    {(wl1) -- [edges=graviton] (int1) -- [edges=graviton] (wl2)},
                };
            \end{feynman}
        \end{tikzpicture}
        $\Bigg)^{-1}_{\rm ret.}$
        \caption{The graviton propagator obtained, as in~\eqref{eq:two-point-bra}, by inverting the quadratic part of the 1SF action~\eqref{eq:S2}, given by the sum of the background contribution and the recoil effective operator. The subscript ``ret.'' stands for {\it retarded} prescription.}
        \label{fig:propagator}
    \end{center}
\end{figure}
Within a path-integral formulation, it corresponds to the two-point function
\begin{equation}\label{eq:two-point-bra}
    \big\langle h_{\alpha\beta}^{+}(x) h_{\gamma\delta}^{-}(x') \big\rangle = i\mathbf{G}^{\rm ret.}_{\alpha\beta\gamma\delta}(x,x') = \frac{2}{Z_0}\frac{i\delta}{\delta J^{-\,\alpha\beta}(x)}\frac{i\delta}{\delta J^{+\,\gamma\delta}(x')}\bigg|_{J^\pm=0} Z_c^{(2)}[J^+,J^-] \,,
\end{equation}
where $Z_c$ stands for the connected part of the generating functional in equation~\eqref{eq:generating-functional-2}.
We anticipated in Section~\ref{sec:gauge-choice} that it is convenient to work in both Self-Force and Post-Minkowskian expansions, thus expanding the generating functional~\eqref{eq:generating-functional-2} in powers of $\kappa$.
After expanding the differential operator $\bar{\mathcal{D}}_{\alpha\beta}^{\;\;\;\gamma\delta}+\bar{\mathcal{R}}_{\alpha\beta}^{\;\;\;\gamma\delta}$, the terms proportional to the gravitational coupling are factored out of the generating functional
\begin{equation}\label{eq:pert-path-integral}
    Z[J] = Z_0\exp \left\{2i\int_x \left(\frac{i\delta}{\delta J^{\alpha\beta}_{+}}\right) \left(\bar{\mathcal{D}}_{(\Phi)}^{\alpha\beta\gamma\delta} + \bar{\mathcal{R}}^{\alpha\beta\gamma\delta}\right) \left(\frac{i\delta}{\delta J^{\gamma\delta}_{-}}\right)\right\} \text{exp}\left\{-\frac{i}{2}\int_{x,x'} J^{-\,\lambda\omega} \propx^{(0)}_{\lambda\omega\rho\sigma} J^{+\,\rho\sigma}\right\},
\end{equation}
where we separated the quadratic operator $\bar{\mathcal{D}}$ plus the gauge fixing term in a flat and a shockwave contribution
\begin{equation}
    \bar{\mathcal{D}}^{\alpha\beta\gamma\delta} + \bar{\mathcal{D}}^{\alpha\beta\gamma\delta}_{\mathrm{g.f.}} = {\mathcal{D}}_{(0)}^{\alpha\beta\gamma\delta} + \bar{\mathcal{D}}_{(\Phi)}^{\alpha\beta\gamma\delta}\,,
\end{equation}
and where $\propx^{(0)}_{\alpha\beta\gamma\delta} = {\mathcal{D}}^{-1}_{(0)\,\alpha\beta\gamma\delta}$ is the flat-space graviton propagator in light-cone gauge introduced in equation~\eqref{eq:flat-propagator-2}
\begin{equation}\label{eq:flat-propagator}
    i\propx_{\alpha\beta\gamma\delta}^{(0)}(k_1,k_2) = \frac{iP_{\alpha\beta\gamma\delta}(k_1)}{k_1^2+i0^+(k_1\cdot\xi)} \,\frac{\hat{\delta}^{(D)}(k_1+k_2)}{\sqrt{-\eta}}\,,
\end{equation}
while $\bar{\mathcal{D}}_{(\Phi)}^{\alpha\beta\gamma\delta}$ contains at least one power of $\kappa$.
It is usual to simplify the notation in flat space by omitting the second argument $k_2$ and the normalised delta function.
We stress here the presence of the retarded (ML) prescription introduced in Section~\ref{sec:schwinger-keldysh}.

Having expressed the 1SF generating functional as~\eqref{eq:pert-path-integral}, we prove that the perturbation $\bar{\mathcal{D}}_{(\Phi)} + \bar{\mathcal{R}}$ results very simplified in the light-cone gauge.
First, we decompose these operators in light-cone coordinates through the projectors in equation~\eqref{eq:projectors}.
Whenever a product $\xi^\alpha G^{(0)}_{\alpha\beta\gamma\delta}$ appears, it vanishes identically as a consequence of the gauge properties.
This makes the action of the operator $\bar{\mathcal{R}}$ vanish, while $\bar{\mathcal{D}}_{(\Phi)}$ is reduced from hundreds of terms to
\begin{subequations}\label{eq:intv-Dbar}
    \begin{gather}
        \bar{\mathcal{D}}_{(\Phi,\,\mathrm{red.})}^{\alpha\beta\gamma\delta} = \Phi(x) X^{\alpha\beta\gamma\delta} \xi^\mu \xi^\nu \partial_\mu \partial_\nu\,, \label{eq:diffop-reduced} \\[3pt]
        X^{\alpha\beta\gamma\delta} = \frac{1}{2}\eta^{\alpha\gamma} \eta^{\beta\delta} + \frac{1}{2}\eta^{\alpha\delta} \eta^{\beta\gamma} - \eta^{\alpha\beta} \eta^{\gamma\delta} \,. \label{eq:Xtensor}
    \end{gather}
\end{subequations}
The interaction is now linear in $\Phi$ and therefore in Newton's constant, so that we can construct by iteration an integral equation for the full propagator~\eqref{eq:two-point-bra}.
Indeed the generating functional~\eqref{eq:pert-path-integral} reduces to
\begin{equation}\label{eq:pert-path-integral-active}
    Z[J] = Z_0\exp \left\{2i\int_x \left(\frac{i\delta}{\delta J^{+\,\alpha\beta}}\right) \bar{\mathcal{D}}_{(\Phi\,,\mathrm{red.})}^{\alpha\beta\gamma\delta} \left(\frac{i\delta}{\delta J^{-\,\gamma\delta}}\right)\right\} \text{exp}\left\{-\frac{i}{2}\int_{x,x'} J^{-\,\lambda\omega} \propx^{(0)}_{\lambda\omega\rho\sigma} J^{+\,\rho\sigma}\right\},
\end{equation}
that contains the flat propagator and a vertex $\vertx^{\alpha\beta\gamma\delta}(x) = \vertx(x)\cdot X^{\alpha\beta\gamma\delta}$, whose Feynman rule -- derived below -- is associated to the operator $\bar{\mathcal{D}}^{\alpha\beta\gamma\delta}_{(\Phi,{\rm red.})}$.
The vertex is proportional to $\ag$, hence to $\gn$, and we focus here just on its tensor structure $X$.
Working in momentum space and applying the functional derivatives in~\eqref{eq:pert-path-integral-active} order by order in $\ag$, one can prove that the connected two-point function is
\begin{equation}\label{eq:vert-prop}
    i\mathbf{G}_{\alpha\beta\gamma\delta} (k_1,k_2) = iG^{(0)}_{\alpha\beta\gamma\delta}(k_1,k_2) + iG^{(0)}_{\alpha\beta\mu\nu}(k_1) \, i\mathbf{V}^{\mu\nu\rho\sigma}(k_1,k_2)\, i G^{(0)}_{\rho\sigma\gamma\delta}(-k_2) \,,
\end{equation}
where the non-perturbative vertex admits the Post-Minkowskian expansion
\begin{equation}\label{eq:full-vertex-as-sum}
    \mathbf{V}^{\mu\nu\rho\sigma}(k_1,k_2) = \sum_{n=1}^\infty \ag^n \vertx_{(n)}^{\mu\nu\rho\sigma}(k_1,k_2)\,,
\end{equation}
The content of equation~\eqref{eq:vert-prop}, for which we adopt an all ingoing convention, is depicted in Figure~\ref{fig:series}.
Each term of the sequence~\eqref{eq:full-vertex-as-sum} is obtained from the previous one by iterating it with a propagator and one further vertex.
\begin{figure}[t]
    \begin{center}
    \hspace{-60pt}
    \begin{tikzpicture}[baseline=(int1.base)]
        \centering
        \begin{feynman}
            \vertex[] (int1) {\(h_{\alpha\beta}\)};
            \vertex[right=3cm of int1] (wl1) {\(h_{\gamma\delta}\)};
            \diagram{
                (wl1) -- [edges=graviton] (int1)
            };
        \end{feynman}
        \middlearrowBig{int1}{wl1}
    \end{tikzpicture}
    =
    \begin{tikzpicture}[baseline=(cntr.base)]
        \centering
        \begin{feynman}
            \vertex[] (cntr) {\(\text{\small$\alpha\beta$}\)};
            \vertex[right=2.2cm of cntr] (phi) {\(\text{\small$\gamma\delta$}\)};
            \diagram* {
                {[edges=photon]
                (cntr) -- (phi)
                }
            };
        \end{feynman}
        \middlearrow{cntr}{phi}
    \end{tikzpicture}
    +
    \begin{tikzpicture}[baseline=(wl1.base)]
        \centering
        \begin{feynman}
            \vertex[style={fill=gray!5, pattern=north east lines, minimum size=15pt}, label={above:$\mathbf{V}$}] (int1) {};
            \vertex[left=1.9cm of int1] (wl1) {\(h_{\alpha\beta}\)};
            \vertex[right=1.9cm of int1] (wl2) {\(h_{\gamma\delta}\)};
            \diagram{
                {(wl1) -- [edges=photon] (int1) -- [edges=photon] (wl2)},
            };
        \end{feynman}
        \middlearrowBig{wl1}{int1}
        \middlearrowBig{int1}{wl2}
    \end{tikzpicture}
    \\
    \vspace{10pt}\hspace{40pt}=
    \begin{tikzpicture}[baseline=(cntr.base)]
        \centering
        \begin{feynman}
            \vertex[] (cntr) {\(\text{\small$\alpha\beta$}\)};
            \vertex[right=2.2cm of cntr] (phi) {\(\text{\small$\gamma\delta$}\)};
            \diagram* {
                {[edges=photon]
                (cntr) -- (phi)
                }
            };
        \end{feynman}
        \middlearrow{cntr}{phi}
    \end{tikzpicture}
    +
    \begin{tikzpicture}[baseline=(bg.base)]
        \centering
        \begin{feynman}
            \vertex[dot, style={fill=gray!5, pattern=north east lines, minimum size=10pt}, label={above:$\vertx$}] (bg) {\(\)};
            \vertex[left=1.3cm of bg] (h1) {\(\text{\small$\alpha\beta$}\)};
            \vertex[right=1.3cm of bg] (h2) {\(\text{\small$\gamma\delta$}\)};
            \diagram* {
                {[edges=photon]
                (h1) -- (bg) -- (h2)
                }
            };
        \end{feynman}
        \middlearrow{h1}{bg}
        \middlearrow{bg}{h2}
    \end{tikzpicture}
    +
    \begin{tikzpicture}[baseline=(bg1.base)]
        \centering
        \begin{feynman}
            \vertex[dot, style={fill=gray!5, pattern=north east lines, minimum size=10pt}, label={above:$\vertx$}] (bg1) {\(\)};
            \vertex[left=1.3cm of bg1] (h1) {\(\text{\small$\alpha\beta$}\)};
            \vertex[right=1.2cm of bg1] (int2);
            \vertex[dot, style={fill=gray!5, pattern=north east lines, minimum size=10pt}, right=1cm of bg1,label={above:$\vertx$}] (bg2) {\(\)};
            \vertex[right=1.3cm of bg2] (h2) {\(\text{\small$\gamma\delta$}\)};

            \diagram* {
                {[edges=photon]
                (h1) -- (bg1) -- (bg2) -- (h2)
                }
            };
        \end{feynman}
        \middlearrow{h1}{bg1}
        \middlearrow{bg1}{bg2}
        \middlearrow{bg2}{h2}
    \end{tikzpicture}
    $+\;\dots$
    \caption{The Neumann series defining the non-perturbative in $\gn$ equation for the graviton propagator in GSW background. A double line stands for a graviton on a non-trivial background $\bar{g}$, while the single line for the flat-space propagator.
    We recognise this sequence to resemble the diagrams of the Born series appearing in~\cite{Correia:2024jgr,Caron-Huot:2025tlq}.}
    \label{fig:series}
    \end{center}
\end{figure}
Each iteration involves contractions to the left and to the right with propagators on flat space, so that only the transverse components of the tensor $X^{\mu\nu\rho\sigma}$ are effective, as proved by explicit computation.
Moreover, one can prove that $X_{ij}^{\;\;rs}P_{\pperp\,rskl} = P_{\pperp\,ijkl}$; then, after the first iteration, the tensor structure $X$ is replaced by $P_\pperp$, given in equation~\eqref{eq:pijkl-projected}, and it remains the same at every order.
Nevertheless, the difference between $X_\pperp^{ijkl}$ and $P_{\pperp}^{ijkl}$ is proportional to a trace structure $\eta_\pperp^{ij}\eta_\pperp^{kl}$, and we know that $\eta_\pperp^{kl} \pol_{kl}^{(\sigma)} = 0$.
Therefore, it makes no difference to replace $X_\pperp$ with $P_\pperp$ also at the leading order, which justifies restricting the computation of the following scalar vertex $\mathbf{V}$ solely
\begin{equation}\label{eq:vert-tens}
    \mathbf{V}^{ijkl}(k_1,k_2) = \mathbf{V}(k_1,k_2) P_\pperp^{ijkl}\,.
\end{equation}
The Post-Minkowskian expansion~\eqref{eq:full-vertex-as-sum} of the (scalar) vertex is characterised by the following contributions
\begin{equation}\label{eq:Vn-definition}
   i\vertp_{(n)}(k_1,k_2) = \int \left(\prod_{r=1}^{n-1} \hat{d}^D\ell_r\sqrt{-\eta}\right) i\vertp(\ell_{0}, -\ell_{1}) \prod_{j=1}^{n-1} i\propp^{(0)}(\ell_{j}) \, i\vertp(\ell_j,-\ell_{j+1}) \,,
\end{equation}
with $\ell_0 = k_1$ and $\ell_n = -k_2$ and
\begin{equation}
    iG^{(0)}(\ell) = \frac{i}{\ell^2 + i0^+(\ell\cdot\xi)}\,,
\end{equation}
which is the ML propagator of a scalar field.
Applying $i\vertx\otimes iG^{(0)}\otimes$ to the above equation and integrating one more time, a recursive structure for~\eqref{eq:full-vertex-as-sum} emerges naturally.
It takes the form of a Fredholm type II integral equation
\begin{equation}\label{eq:integral-eqn}
    i\mathbf{V}(k_1,k_2) = i\ag\vertp(k_1,k_2) + \ag \int_\ell i\vertp(k_1,-\ell) \, i\propp^{(0)}(\ell) \, i\mathbf{V}(\ell,k_2)\,,
\end{equation}
whose solution is given by a Neumann series governed by the parameter $\ag$.

One then substitutes the solution of~\eqref{eq:integral-eqn} for the scalar vertex into the definition of $\mathbf{V}^{\alpha\beta\gamma\delta}$ and subsequently into~\eqref{eq:vert-prop}, obtaining then the Green function $\mathbf{G}_{\alpha\beta\gamma\delta}$.
With the help of the property $\pol_{ij}^{(\sigma)}(k_1) \pol^{ij\,(\sigma')}(k_2) = 1-\delta_{\sigma,\sigma'}$, we evaluate the final tensor product
\begin{equation}
    P_{\alpha\beta\mu\nu}(k_1) P^{\mu\nu\rho\sigma}_\pperp P_{\rho\sigma\gamma\delta}(k_2) = \sum_{\sigma} \pol_{\alpha\beta}^{(\sigma)}(k_1) \,\pol_{\gamma\delta}^{(\sigma)\,*}(k_2) = \mathcal{P}_{\alpha\beta\gamma\delta}(k_1,k_2)\,.
\end{equation}
Therefore the full propagator~\eqref{eq:vert-prop} is
\begin{equation}\label{eq:vert-prop-2}
    \begin{aligned}
        &i\mathbf{G}^{\rm ret.}_{\alpha\beta\gamma\delta} (k_1,k_2) = \\
        &\hspace{40pt} = \mathcal{P}_{\alpha\beta\gamma\delta}(k_1,k_2) \left(i G^{(0)}(k_1) \frac{\hat{\delta}^{(D)}(k_1+k_2)}{\sqrt{-\eta}} + iG^{(0)}(k_1)\,i\mathbf{V}(k_1,k_2)\,iG^{(0)}(-k_2)\right)\,,
    \end{aligned}
\end{equation}
which has the same tensor structure that we have predicted in the previous section without explicit proof.
The sign in front of $k_2$ in the last scalar propagator serves to reverse the sign of its pole, according to the all ingoing convention for $\mathbf{G}$.
In this paragraph we have rephrased the complicated problem of determining the graviton propagator on the GSW background into an easier one, that consists in solving the integral equation for the scalar vertex $\mathbf{V}$ through the Post-Minkowskian series~\eqref{eq:full-vertex-as-sum}.
The coefficients of that series are defined in~\eqref{eq:Vn-definition} and we explicitly compute them in subsections~\ref{sec:vertex-regularisator} and~\ref{sec:series-coefficients} below.
For the readers not interested in technical details, we suggest skipping these paragraphs.

\subsection{Interaction vertex and shockwave regulators\label{sec:vertex-regularisator}}
In order to compute all of the PM terms $\vertx_{(n)}^{\mu\nu\rho\sigma}$ in~\eqref{eq:full-vertex-as-sum}, two ingredients are required.
The first is the flat-space propagator $G^{(0)}(k)$ that we have already discussed, the other is the PM interaction vertex in equation~\eqref{eq:intv-Dbar} and~\eqref{eq:pert-path-integral-active}.
The latter is computed by integrating by parts the differential operator~\eqref{eq:diffop-reduced} and, after the Fourier transform it reads
\begin{equation}
    i\mathcal{V}(k_1,k_2) = i (\xi\cdot k_1)(\xi\cdot k_2)\hat{\delta}(\xi\cdot k_1+\xi\cdot k_2)\,\frac{1}{(k_1+k_2)_\pperp^2}\,,
\end{equation}
which, not surprisingly, contains a propagator which resembles the Glauber interaction with a highly energetic Wilson line within the SCET framework~\cite{Rothstein:2016bsq,Rothstein:2024nlq,Saavedra:2026ttk}.
However, when trying to compute $\mathcal{V}_{(n)}$ with $n>1$, we immediately get divergent integrals.
These are intuitively understood as a physical issue related to the delta function governing the wavefront of the background metric.
To overcome this, we need to introduce a regulator.
Among the possible choices, we found two of them that are very convenient, one for its physical interpretation, the other for its simplicity.

The first regulator that we consider has been introduced in~\cite{Bohnenblust:2026ujk} and regularises the delta function $\delta(\xi\cdot x)$ using one of its parametric representation, in particular a symmetric exponential function
\begin{equation}\label{eq:regul-berlin}
    \delta (\xi\cdot x) = \lim_{\Lambda\to\infty} \rho_\Lambda (\xi\cdot x) = \lim_{\Lambda\to\infty} \frac{\Lambda}{2}\,e^{-\Lambda\,\mid\xi\cdot x\mid}\,.
\end{equation}
Instead of focusing the whole shockwave of energy $E_1$ and direction $\xi$ on the hyperplane $\xi\cdot x=0$, it sets a continuous spectrum of waves, not self-interacting and each one shifted with respect to the other according to the distribution $\rho_\Lambda$.
In momentum space this amounts to regularising the PM interaction vertex as
\begin{equation}
    i\mathcal{V}_\Lambda(k_1,k_2) = i (\xi\cdot k_1)(\xi\cdot k_2) \hat{\delta}(\xi\cdot k_1+\xi\cdot k_2)\frac{1}{(k_1+k_2)_\pperp^2} \frac{\Lambda^2}{(k_1\cdot\chi+k_2\cdot\chi)^2+\Lambda^2}\,,
\end{equation}
therefore introducing a dependence on the plus components $k_{1,2}^+$ that makes the integrals~\eqref{eq:Vn-definition} ultraviolet finite.
This dependence disappears when taking the $\Lambda\to\infty$ limit, according to $\mathrm{FT}[\rho_\Lambda]\to1$ for large $\Lambda$.

The second regulator that we consider is a simplified version of the rapidity regulators used within the SCET~\cite{Chiu:2012ir}, reviewed in Section 5.2 of~\cite{Rothstein:2016bsq}.
In our case it is sufficient to introduce a parameter $\lambda \to 0^+$ to regularise the integrals.
In particular we correct the vertex with a power law
\begin{equation}\label{eq:regul-scet-0}
    i\mathcal{V}_\lambda(k_1,k_2) = i (\xi\cdot k_1)(\xi\cdot k_2) \hat{\delta}(\xi\cdot k_1+\xi\cdot k_2) \, \frac{|2(k_1\cdot\chi + k_2\cdot\chi)|^{-\lambda}}{(k_1+k_2)_\pperp^2} \,,
\end{equation}
and we compute loops using the following Fourier transforms~\cite{Rothstein:2016bsq}
\begin{subequations}\label{eq:regul-scet}
    \begin{align}
        & |2\ell^+|^{-\lambda} = \frac{\lambda}{2} \, e_\lambda \int_\mathbb{R} dx\, |x|^{\lambda-1}\, e^{-ix\ell^+} \,, \label{eq:regIZR-eq1} \\
        & \frac{1}{\ell^{+} + \Delta + i0^{+}} = -i\int_\mathbb{R} ds\, \Theta(s) \, e^{is\ell^++is\Delta} \,,
    \end{align}
\end{subequations}
where
\begin{equation}
    e_\lambda = 2^{-\lambda} \Gamma(1-\lambda) \frac{\sin(\pi\lambda/2)}{\pi\lambda/2} = 1 + \mathcal{O}(\lambda)\,.
\end{equation}
We prove in the next section that what has been computed within the WQFT approach with the regulator~\eqref{eq:regul-berlin}, is obtained here using the SCET regulator~\eqref{eq:regul-scet-0} that we found convenient for this computation.

\subsection{Solution for the integral equation at any loop order\label{sec:series-coefficients}}
We start by computing the first iteration $\vertp_{(2)}$ as a first step, then we move to the  generic $n$-th order.
Afterwards we give an expression for the resummed form~\eqref{eq:vert-tens}.

\paragraph{First iteration}
The first iteration of the scalar vertex, that we show as the third diagram of the second line of Fig.~\ref{fig:series}, involves one loop integration.
We prove a factorisation into collinear and transverse integrals after applying the regularisation limit $\lambda\to0$. The loop expression is
\begin{equation}\label{eq:v2-first-iteration}
    \begin{aligned}
        i\vertp_{(2)}(k_1,k_2) & = \int_\ell \, i\vertp_\lambda(k_1,-\ell) \, i\propp^{(0)}(\ell) \, i\vertp_\lambda(\ell,k_2) \\
            & = \frac{i}{2} (\xi\cdot k_1)^2 (\xi\cdot k_2) \hat{\delta}(\xi\cdot k_1 + \xi\cdot k_2) \int \frac{\muir^{2\epsilon}e^{\epsilon\gamma_\mathrm{E}}\hat{d}^d\ell_\pperp}{(k_1-\ell)_\pperp^2(k_2+\ell)_\pperp^2} \cdot C_2 (\lambda)
    \end{aligned}
\end{equation}
with
\begin{equation}
    C_2(\lambda) = \int \hat{d}\ell^+ \frac{\mid\!2(k_1^+-\ell^+)\!\mid^{-\lambda}\,\mid\!2(k_2^++\ell^+)\!\mid^{-\lambda}}{\ell^+ + \ell_\pperp^2/k_1^- + i0^+}\,.
\end{equation}
The integral $C_2$ is computed using the transforms~\eqref{eq:regul-scet} as in~\cite{Rothstein:2016bsq}.
Relabelling $\ell_\pperp^2/\ell^- = \Delta$ we get
\begin{equation}
    \begin{aligned}
        C_2(\lambda) = -i & \left(\frac{\lambda}{2}\,e_\lambda\right)^2 \int dx_1 \, dx_2 \mid \!x_1 x_2\! \mid^{\lambda-1} \, e^{-i(x_1 k_1^+ + x_2k_2^+)} \int ds \,\Theta(s) \, e^{is\Delta} \\
            & \cdot \int_{\ell^+} \text{exp}\left(is\ell^+ + ix_1\ell^+ -ix_2\ell^+\right) \,.
    \end{aligned}
\end{equation}
The last integral gives a delta function constraining $s = x_2-x_1$ that we use to remove the variable $s$, obtaining a $\Theta(x_2-x_1)$.
The two remaining integrals in $x_1$ and $x_2$ have integrands of the same form and are entangled by the Heaviside function.
Since the prefactor is proportional to $\lambda^2$, we need the leading divergence $1/\lambda^2$ of the integral, that is independent of $\Delta,k_1^+$ and $k_2^+$, thus we symmetrise it as
\begin{equation}\label{eq:one-two-fact}
    \Theta(x_2-x_1) \hspace{5pt}\to\hspace{5pt} \frac{1}{2}\left(\Theta(x_1-x_2) + \Theta(x_2-x_1)\right)\,,
\end{equation}
that gives
\begin{equation}
    C_2(\lambda) = -\frac{i}{2}\left(\frac{\lambda}{2}\,e_\lambda\right)^2 \int dx_1 \mid\! x_1 \!\mid^{\lambda-1}\,e^{-ix_1 F_1} \int dx_2 \mid\! x_2 \!\mid^{\lambda-1} e^{-ix_2 F_2} \,,
\end{equation}
for some $F_{1,2}$. Inverting~\eqref{eq:regIZR-eq1} and keeping the finite order only, we get
\begin{equation}
    C_2(\lambda) = -\frac{i}{2}\Big(1+\mathcal{O}(\lambda)\Big) \,.
\end{equation}
This proves that the finite part $C_2$ does not depend on any variable, which is in agreement with the high-energy factorisation~\cite{Rothstein:2024nlq,Rothstein:2016bsq,Melville:2013qca,White:2019ggo}.
Below we prove it at any loop order $C_n$.
Therefore, we integrate the transverse dynamics of~\eqref{eq:v2-first-iteration} using Euclidean massless bubble integrals~\cite{Kotikov:2018wxe} for which we follow the conventions used in Appendix B of~\cite{Alessio:2026bdi}
\begin{equation}
    i\vertp_{(2)}(k_1,k_2) = \frac{i}{2} C_2 (\xi\cdot k_1)^2 (\xi\cdot k_2) \hat{\delta}(\xi\cdot k_1 + \xi\cdot k_2)\,\mathcal{B}_{1,1}^{(1)}(k_{1\pperp}+k_{2\pperp})\,,
\end{equation}
with
\begin{equation}
    \mathcal{B}_{1,1}^{(1)}(\vec{q}) = \int_{\vec{\ell}} \left[\ell_\pperp^2(q-\ell)_\pperp^2\right]^{-1} = \left(\frac{4\pi\muir^2 e^{\gamma_\mathrm{E}}}{\vec{q}^{\,2}}\right)^\epsilon \frac{1}{4\pi \vec{q}^{\,2}} \frac{\Gamma(-\epsilon)^2\Gamma(1+\epsilon)}{\Gamma(-2\epsilon)} \,.
\end{equation}

\paragraph{Generic order}
Repeating the same steps done before, we obtain the following form for the $n-1$ loop order, namely the $n-$th term of the series
\begin{equation}
    \begin{aligned}
        i\vertp_{(n)}(k_1,k_2) = &\frac{i}{2^{n-1}} (\xi\cdot k_1)^{n} (\xi\cdot k_2) \hat{\delta}(\xi\cdot k_1 + \xi\cdot k_2)\,\text{\tiny$\times$} \\
            & \left(\muir^2 e^{\gamma_\mathrm{E}}\right)^{(n-1)\epsilon} \int \left(\prod_{j=1}^{n} \frac{\hat{d}^{d}\ell_{j\pperp}}{\ell_{j\pperp}^2}\right) \hat{\delta}^{(d)}\big(k_{1\pperp}+k_{2\pperp}-{\textstyle\sum}_{i=1}^{n}\ell_{i\pperp}\big) C_n(\lambda)\,,
    \end{aligned}
\end{equation}
in which $C_n(\lambda)$ is defined as before
\begin{equation}
    C_n(\lambda) = \int \prod_{j=1}^{n-1} \frac{\hat{d}\ell_j^+}{\ell_j^+ + \Delta_j + i0^+} \prod_{i=0}^{n-1} \mid\! 2\left(\ell_{i}-\ell_{i+1}\right) \!\mid^{-\lambda} \,,
\end{equation}
with $\Delta_j$ depending on $\xi\cdot k_1$ and on the transverse loop momenta, while $\ell_0 = k_1$ and $\ell_n = -k_2$.
Using again the representation~\eqref{eq:regul-scet} and performing the trivial integrals we are left with
\begin{equation}
    C_n(\lambda) = (-i)^{n-1} \left(\frac{\lambda}{2}\,e_\lambda\right)^n \int_\mathbb{R} d^nx \mid\! x_1 \cdots x_n \!\mid^{\lambda-1} \, e^{-i\sum_{r} x_r F_r} \, \Theta(x_1 > \cdots > x_n) \,.
\end{equation}
Then the integrals are symmetrised as before, generating the $1/n!$ factor
\begin{equation}
    \Theta(x_1 > \cdots > x_n) \hspace{5pt}\rightarrow\hspace{5pt} \frac{1}{n!} \sum_{g\in S_n}\Theta(x_{g_1} > \cdots > x_{g_n}) \,,
\end{equation}
which gives
\begin{equation}
    C_n (\lambda) = \frac{(-i)^{n-1}}{n!} + \mathcal{O}(\lambda) \,,
\end{equation}
in agreement with the result predicted in~\cite{Bohnenblust:2026ujk} using the regulator $\rho_\Lambda (\xi\cdot x)$.
Similarly, in~\cite{Rothstein:2016bsq} this coefficient is used to resum the infinite series of Glauber exchanges between two highly energetic Wilson lines in quantum chromodynamics.
The vertex at the $n$-th PM order is
\begin{equation}\label{eq:vert-n-computed}
    i\vertp_{(n)}(k_1,k_2) = - \frac{2}{n!} \left(\frac{1}{2i}\right)^n (\xi\cdot k_1)^n (\xi \cdot k_2) \hat{\delta}(\xi\cdot k_1+\xi\cdot k_2)\,\mathcal{B}_{1,1}^{(n-1)}(k_{1\pperp}+k_{2\pperp})\,,
\end{equation}
where
\begin{equation}\label{eq:iterated-bubbles}
    \begin{aligned}
        \mathcal{B}_{1,1}^{(n)}(\vec{q}) =\int_{\ell_{1}^\pperp,\dots,\ell_{n+1}^\pperp} \frac{\hat{\delta}^{(d)}\big(q_\pperp-{\textstyle\sum}_{i=1}^{n+1}\ell_{i\pperp}\big)}{\prod_{j=1}^{n+1} \ell_{j\pperp}^2} = \left(\frac{4\pi\muir^{2}e^{\gamma_\mathrm{E}}}{\vec{q}^{\,2}}\right)^{n\epsilon}\frac{(-1)^{n+1}}{(4\pi)^{n} \vec{q}^{\,2}} \frac{\Gamma(1+n\epsilon)\Gamma\left(-\epsilon\right)^{n+1}}{\Gamma\left(-\epsilon(n+1)\right)}\,.
    \end{aligned}
\end{equation}

\subsection{Result: resummed propagator and Compton amplitude\label{sec:results-two-pts}}
Once the terms of the series have been computed, we investigate whether their sum $\mathbf{V}^{\mu\nu\rho\sigma}$
admits a closed-form expression.
In order to work it out, it is necessary to use a special version of the Fourier space, namely momentum space for the collinear dynamics $k^\pm$ and position space for the transverse components $k_\pperp \to x_\pperp$.
This space has recently been used to resum high-energy contributions to the momentum kick in massive spinless gravitational scattering, as it allows one to manage the infinite series of exchanges that dominates the eikonal function in the Glauber region while keeping advantage of the momentum space computations for the collinear components~\cite{Saavedra:2026ttk}.
Indeed the iterated bubble integrals $\mathcal{B}_{1,1}^{(n)}$ in~\eqref{eq:iterated-bubbles} are nothing but a convolution of transverse propagators, which we know to factorise in position space thanks to the properties of the Fourier transform
\begin{equation}
    \int \hat{d}^dq_\pperp \, e^{i \vec{x}\cdot\vec{q}}\,\mathcal{B}_{1,1}^{(n-1)}(\vec{q}\,) =  \left(-\frac{\Gamma(-\epsilon)}{4\pi} \left(\muirt^2\vec{x}^{\,2}\right)^{\epsilon} \right)^n .
\end{equation}
where $\muirt^2=\pi\muir^2e^{\gamma_{\rm E}}$.
The $n$-th order vertex
\begin{equation}\label{eq:vert-n-momenutm-space}
    i\vertp_{(n)}(k_1^{\pm},k_2^\pm;\vec{x}\,) = - \frac{2}{n!} (\xi \cdot k_2) \hat{\delta}(\xi\cdot k_1+\xi\cdot k_2)\, \left(i(\xi\cdot k_1)\frac{\Gamma(-\epsilon)}{8\pi} \left(\muirt^2\vec{x}^{\,2}\right)^{\epsilon} \right)^n,
\end{equation}
is recognised to belong to an exponential series that resums to
\begin{equation}\label{eq:resummed-V-xspace}
    i\mathbf{V}(k_1^\pm,k_2^\pm,\vec{x}\,) = - 2 (\xi\cdot k_2) \hat{\delta}(\xi\cdot k_1 +\xi\cdot k_2) \left(e^{2i\gn E_1(\xi\cdot k_1)\Gamma(-\epsilon)\left(\muirt^2\vec{x}^{\,2}\right)^{\epsilon}}-1\right) ,
\end{equation}
which is one of the main results of this paper.
The resummed vertex in the full momentum space is easily obtained as a series for $\epsilon\to0$.
Using the expanded version of the exponentiated factor
\begin{equation}\label{eq:expansion-eps}
    2i\gn E_1 \left(\xi \cdot k_1\right) \left( - \frac{1}{\epsilon} - \log\left(\muirt^2 \vec{x}^{\,2} e^{\gamma_\mathrm{E}}\right) + \mathcal{O}(\epsilon) \right),
\end{equation}
and truncating the exponent at $\mathcal{O}(\epsilon)$ one finds
\begin{equation}\label{eq:resummed-V-qspace}
    i\mathbf{V}(k_1,k_2) = - 2 (\xi\cdot k_2) \hat{\delta}(\xi\cdot q) \left( \frac{4\pi iW}{\vec{q}^{\,2}} e^{-iW/\epsilon} \frac{\Gamma(1-iW)}{\Gamma(1+iW)}\left(\frac{\vec{q}^{\,2}}{4\pi\muir^2 e^{2\gamma_\mathrm{E}}}\right)^{iW} - \hat{\delta}^{(d)}(q_\pperp)\right),
\end{equation}
with $q=k_1+k_2$ and $W = 2\gn E_1 (\xi\cdot k_1)$ the Weinberg factor responsible for the exponentiation of the infrared divergences~\cite{Weinberg:1965nx} that we discuss below.
Therefore, the full graviton propagator on a GSW background is
\begin{equation}\label{eq:grav-prop-resummed}
    \begin{aligned}
        &i\mathbf{G}^{{\rm ret.}}_{\alpha\beta\gamma\delta} (k_1,k_2)\big|_{D=4} = i{G}^{(0)}_{\alpha\beta\gamma\delta}(k_1,k_2)\,- iG^{(0)}(k_1)\,\mathcal{P}_{\alpha\beta\gamma\delta}(k_1,k_2)\,iG^{(0)}(-k_2) \, \text{\tiny$\times$} \\
        &\hspace{40pt} \text{\tiny$\times$}\,2(\xi\cdot k_2) \hat{\delta}(\xi\cdot q) \left( \frac{4\pi iW}{\vec{q}^{\,2}} \frac{\Gamma(1-iW)}{\Gamma(1+iW)}\left(\frac{\vec{q}^{\,2}}{4\pi Q^2 e^{2\gamma_\mathrm{E}}}\right)^{iW} - \hat{\delta}^{(2)}(q_\pperp)\right) ,
    \end{aligned}
\end{equation}
where we used $1/\epsilon\to2\log(Q/\muir)$ to rephrase the Weinberg divergence in $D=4$ dimensions and remove the dependence on $\muir$, while $Q$ is an arbitrary mass scale.
A few comments are needed.
First of all, we recognise in~\eqref{eq:grav-prop-resummed} echoes of the early 't Hooft amplitude computation~\cite{tHooft:1987vrq}.
Secondly, we find agreement with the WQFT response theory computation~\cite{Bohnenblust:2026ujk} and with the shockwave propagator computed by Raj and Venugopalan in the eikonal approximation~\cite{Raj:2024xsi}, a result that is surprisingly exact beyond the approximation.
The only difference between our result and the two-point function in response theory~\cite{Bohnenblust:2026ujk} lies in {\it how} we got the result.
The authors of~\cite{Bohnenblust:2026ujk} obtain the two-point function as the sum of a background contribution and an additional ``remainder'' part, which corresponds to the contributions from the 1SF recoil operator $\bar{\mathcal{R}}_{\alpha\beta}^{\;\;\;\gamma\delta}$ that here is zero due to our gauge choice.
Instead of considering a recoil operator and background contribution, we use the light-cone gauge to make the former vanish and move its contribution into the latter.
Indeed, in the present case we have proved that the contribution from $S_\mathrm{recoil}$ to the graviton propagator is exactly zero at any Self-Force order.

Lastly, we can use the results of this section to give an expression for the classical gravitational Compton amplitude\footnote{Since we used retarded propagators, we should name it {\it generalised amplitude} as in~\cite{Caron-Huot:2023vxl}, or {\it response function} as in~\cite{Bohnenblust:2026ujk}.}, namely for the process
\begin{equation}\label{eq:compton-process}
    p_1 + k_1^{\sigma_1} \;\rightarrow\; p_1^\prime + k_2^{\sigma_2}\,,
\end{equation}
where $k_{1,2}^\mu$ are graviton momenta and $p_{1}^\mu,p_1^{\prime\mu}$ refer to an ultrarelativistic spinless body with initial momentum $p_1^\mu = E_1\xi^\mu$.
The Compton amplitude is given by the on-shell two-point function, which is obtained from the off-shell Green's function by removing the external flat-space propagators and contracting it with the polarisations of the external gravitons
\begin{equation}
    i\mathcal{A}_{\mathrm{Compton}}(k_1^{\sigma_1},k_2^{\sigma_2}) = \pol_{\alpha\beta}^{(\sigma_1)}(k_1)\,i\mathbf{V}^{\alpha\beta\gamma\delta}(k_1,-k_2)\,\pol_{\gamma\delta}^{(\sigma_2)*}(k_2) \big|_{k_1^2=k_2^2=0}\,.
\end{equation}
The tensor structure yields the helicity conservation expected in high-energy scattering~\cite{Chen:2022clh,DelDuca:1996nom,Bogdan:2002sr}
\begin{equation}\label{eq:compton-amplitude}
    i\mathcal{A}_{\mathrm{Compton}}(k_1^{\sigma_1},k_2^{\sigma_2}) = \frac{8\pi iW}{\vec{q}^{\,2}} \delta_{\sigma_1,\sigma_2} (\xi\cdot k_2) \hat{\delta}(\xi\cdot q) e^{-iW/\epsilon} \frac{\Gamma(1-iW)}{\Gamma(1+iW)}\left(\frac{\vec{q}^{\,2}}{4\pi\muir^2 e^{2\gamma_\mathrm{E}}}\right)^{iW}
\end{equation}
from which we removed the term $\hat{\delta}^{(d)}(q_\pperp)$ in $\mathbf{V}$, since it corresponds to a diagram in which the two particles do not interact.
Given the equation above, it is clear that the infrared Weinberg factor $W$ in the scalar vertex $\mathbf{V}(k_1,k_2)$ refers to the Compton process depicted in equation~\eqref{eq:compton-process}.
We re-express the IR factor as
\begin{equation}\label{eq:weinberg-IR-Compton}
    -\frac{iW}{\epsilon} = -\frac{iG}{\epsilon} \left(p_1\cdot k_1 + p_1^\prime \cdot k_2\right) = -\frac{2iG}{\epsilon} \, p_1\cdot k_1 \,,
\end{equation}
which matches the infrared-divergent phase of the one-loop gravitational Compton amplitude~\cite{Bjerrum-Bohr:2025bqg,Akpinar:2025byi}, predicting an IR factor equal to $-2i\gn M \omega/\epsilon$ in the frame where the graviton frequency is $\omega$ and the classical body is initially at rest with mass $M$.
Therefore, $M\omega$ is the scalar product between the momentum of the graviton and the momentum of the source, which in the present case reads $E_1(\xi\cdot k_1)$, reproducing equation~\eqref{eq:weinberg-IR-Compton}.
More rigorously, we can directly compute the Weinberg factor from known equations, see {\it e.g.}~\cite{Dunbar:1995ed,Brandhuber:2023hhy,Georgoudis:2023eke}.
For massless particles, one has
\begin{equation}
    -\frac{iW}{\epsilon} = \frac{G}{\pi\epsilon} \sum_{a<b=1}^{4} \eta_a\eta_b(p_a\cdot p_b)\log\big(-\eta_a\eta_b (p_a\cdot p_b)/\mu^2 - i0^+ \big)\,,
\end{equation}
with $\eta_a = +1$ for ingoing particles and $-1$ for outgoing ones, and with $\mu$ an arbitrary reference scale.
By expressing $\log(-x - i0^+) = \log(x)-i\pi$ for $x>0$, the imaginary part of the expression above gives the result~\eqref{eq:weinberg-IR-Compton}, while the real part is
\begin{equation}\label{eq:real-weinberg}
    \mathsf{Re}\left(-\frac{iW}{\epsilon}\right) = \frac{G}{\pi\epsilon}\left(s\log\left(\frac{s}{\mu^2}\right)+u\log\left(-\frac{u}{\mu^2}\right)+t\log\left(-\frac{t}{\mu^2}\right)\right),
\end{equation}
where $s,t,u$ are the standard Mandelstam invariants for a four-particle process.
The result above matches the predictions of~\cite{Dunbar:1995ed}.
One then has to take the classical limit of eq.~\eqref{eq:real-weinberg}, which amounts to using $t\propto\hbar^2$, so that $s\simeq-u$.
This makes the equation above vanish up to corrections of order $\mathcal{O}(\hbar^2)$.
For this reason, we do not find any real part in the IR-exponentiated factor of the classical Compton amplitude.
The expression for $i\mathcal{A}_{\mathrm{Compton}}$ is obtained by resumming the ladder of rescatterings of the incoming graviton on the GSW background, as shown in Figure~\ref{fig:series}.
On the other hand, the Compton amplitude involving gravitons and a massive body initially at rest has recently been calculated in $\mathcal{N}=8$ supergravity, showing a similar ladder structure order by order in perturbation theory~\cite{n8compton}.
Given the dominance of graviton exchanges at leading order in the high-energy expansion of any theory~\cite{tHooft:1987vrq,DiVecchia:2020ymx}, known as {\it graviton dominance}, we assume that the ultrarelativistic limit of the result in $\mathcal{N}=8$ matches the present computation, provided that the massless limit of the classical body is sufficiently smooth.

{\section{Waveform from massless-particle collision at leading order in $q$}\label{sec:waveform}}
In this section we compute the variation of the metric perturbation $h_{\alpha\beta}$ for a massless-massless particle collision at first Self-Force order, for which the (frame-dependent) expansion parameter is $q=E_2/E_1$.
After deriving the waveform from the Schwinger-Keldysh generating functional, we introduce the on-shell waveform in momentum space as the sum of two pieces, the {\it free propagating waveform} and the {\it interacting waveform}, depicted in Figure~\ref{fig:wf}.
While the former is computed exactly at 1SF, the latter is treated by expanding it in a Post-Minkowskian series, which we compute to all orders without resumming.
Then we check that the given waveform satisfies several non-trivial consistency checks:
$(i)$ the leading order waveform in the PM expansion agrees with the tree-level KMOC~\cite{Kosower:2018adc,Cristofoli:2021vyo} computation, that we calculated in Appendix~\ref{app:lowf};
$(ii)$ the waveform, expanded for small frequency of the emitted graviton, is matched against the classical soft theorems of Saha, Sahoo and Sen~\cite{Saha:2019tub,Sahoo:2021ctw} (see also~\cite{Alessio:2024onn}) at orders $1/\omega,\log\omega$ and $\omega\log^2\omega$.
Finally, in paragraph~\ref{sec:trans-planckian} we approximate the PM-resummed 1SF waveform in the strong-field regime, which allows us to demonstrate the UV convergence of the angular spectrum of emitted energy in massless-massless particle scattering, partially addressing the problem of the logarithmic UV divergence of the total energy~\cite{DiVecchia:2022nna,Ciafaloni:2015xsr,RajVenugopalan-universal-features,Gruzinov:2014moa}.
We stress that such convergence property is already known in literature.

Using the Post-Minkowskian resummed 1SF generating functional~\eqref{eq:generating-functional-2}, we derive the expectation value of the perturbation field in the presence of a source
\begin{equation}
    \big\langle h_{\alpha\beta}^+ (x) \big\rangle = \frac{1}{Z_0} \frac{i\delta Z_c[J]}{\delta J^{-\,\alpha\beta}(x)} \bigg|_{J^-=0,\,J^+=\frac{\kappa}{2}\bar{T}_{\ml}} = \frac{\kappa}{4} \int_{x'} \mathbf{G}^{\rm ret.}_{\alpha\beta\gamma\delta}(x,x') \, \bar{T}_\ml^{\gamma\delta} (x')\,,
\end{equation}
obtaining the known classical equation of motion.
Working out its Fourier transform, we define the off-shell momentum-space waveform
\begin{equation}\label{eq:wf-definition}
    \wfs_{\alpha\beta}(k) = -\frac{i\kappa}{4} \int_{r} i\mathbf{G}^{\rm ret.}_{\alpha\beta\gamma\delta}(-k,r) \, \bar{T}_\ml^{\gamma\delta} (r)\,,
\end{equation}
which is given by the single diagram in Figure~\ref{fig:wf}, made up of the non-perturbative building blocks derived in the previous sections, $\mathbf{G}^{\rm ret.}$ and $\bar{T}_\ml$.
\begin{figure}[t]
    \begin{center}
        \begin{tikzpicture}[baseline=(base.base)]
            \centering
            \begin{feynman}
                \vertex[] (int1) {};
                \vertex[above=.5cm of int1] (base) {};
                \vertex[left=1.3cm of int1, dot, style={fill=NavyBlue!40, minimum size=10pt}] (wl1) {};
                \vertex[right=2cm of int1] (wl3) {};
                \vertex[above=1cm of wl3] (wl2){\(h_{\alpha\beta}\)};
                \vertex[above=.6cm of wl1] (wl1prime) {};
                \vertex[above=.63cm of wl2] (wl2prime) {};
                \diagram{
                    {(wl1) -- [edges=graviton, bend left=30] (wl2)},
                };
            \end{feynman}
            \middlearrowBig{wl1prime}{wl2prime};
        \end{tikzpicture}
        \hspace{5pt}
        =
        \hspace{5pt}
        \begin{tikzpicture}[baseline=(base.base)]
            \centering
            \begin{feynman}
                \vertex[] (int1) {};
                \vertex[above=.5cm of int1] (base) {};
                \vertex[left=1.3cm of int1, dot, style={fill=NavyBlue!40, minimum size=10pt}] (wl1) {};
                \vertex[right=1.8cm of int1] (wl3) {};
                \vertex[above=1cm of wl3] (wl2){\(h_{\alpha\beta}\)};
                \vertex[above=.6cm of wl1] (wl1prime) {};
                \vertex[above=.58cm of wl2] (wl2prime) {};
                \diagram{
                    {(wl1) -- [edges=photon, bend left=30] (wl2)},
                };
            \end{feynman}
            \middlearrowBig{wl1prime}{wl2prime};
        \end{tikzpicture}
        \hspace{5pt}
        +
        \begin{tikzpicture}[baseline=(base.base)]
            \centering
            \begin{feynman}
                \vertex[] (int1) {};
                \vertex[above=.5cm of int1] (base) {};
                \vertex[left=1.6cm of int1, dot, style={fill=NavyBlue!40, minimum size=10pt}] (wl1) {};
                \vertex[right=1.8cm of int1] (wl3) {};
                \vertex[above=1cm of wl3] (wl2){\(h_{\alpha\beta}\)};
                \vertex[style={fill=gray!5, pattern=north east lines, minimum size=15pt}, above=1.5cm of int1] (em) {};
                \vertex[above=.4cm of wl1] (wl1ri) {};
                \vertex[left=.4cm of wl1ri] (wl1prime) {};
                \vertex[above=.3cm of em] (emprime) {};
                \vertex[above=.1cm of em] (emsec) {};
                \diagram{
                    {(wl1) -- [edges=photon, bend left=30] (em) -- [edges=photon, bend left=5] (wl2)},
                };
            \end{feynman}
            \middlearrowBig{wl1prime}{emprime};
            \middlearrowBig{emsec}{wl2};
        \end{tikzpicture}
    \end{center}
    \caption{The two contributions to the 1SF waveform: $(i)$ emission of a graviton from the unperturbed geodesic motion of the light body, $\wf^\mathrm{free}_{\sigma}$, and $(ii)$ the non-trivial contribution from the interaction of the emitted graviton with the wavefront of the GSW, $\wf^\mathrm{int}_{\sigma}$.}
    \label{fig:wf}
\end{figure}
Similarly, the on-shell waveform is obtained from the off-shell one as~\cite{Mougiakakos:2021ckm,Bohnenblust:2025gir}
\begin{equation}\label{eq:on-shell-wf-def}
    \wf_\sigma(k) = -\frac{ik^2}{\sqrt{-\bar{g}}} \, \pol_{\alpha\beta}^{(\sigma)*}(k) \, \wfs^{\alpha\beta}(k) \big|_{k^2=0}\,,
\end{equation}
which we simply call the {\it waveform amplitude} because of its analogy to the (FT of the) KMOC amplitude~\cite{Cristofoli:2021vyo,Caron-Huot:2023vxl}.\footnote{We remark that the waveform amplitude is not simply the five-point amplitude of QFT in the classical limit; rather, it is the generalised amplitude introduced in~\cite{Caron-Huot:2023vxl}, namely the in-in expectation value $\text{Exp}_k$ of the graviton annihilation operator in the future.}
Bearing in mind the decomposition of the graviton propagator into a free propagator and an interacting part in~\eqref{eq:vert-prop-2}, and using the definition of the projector $\mathcal{P}_{\alpha\beta\gamma\delta}(r,-k)$, we express the amplitude as
\begin{equation}\label{eq:wf-def-2}
    \wf_{\sigma}(k) = \wf_{\sigma}^{\rm free}(k) + \wf_{\sigma}^{\rm int}(k) \,,
\end{equation}
with
\begin{subequations}
    \begin{align}
        & \wf_{\sigma}^{\rm free}(k) = -\frac{i\kappa}{2}\, \pol_{\alpha\beta}^{(\sigma)*}(k) \, \bar{T}_\ml^{\alpha\beta}(k) \,, \label{eq:wf-def-2-free} \\
        & \wf_{\sigma}^{\rm int}(k) =  -\frac{i\kappa}{2} \int_{r} \pol_{\alpha\beta}^{(\sigma)*}(r) \, \bar{T}_\ml^{\alpha\beta}(r) \, i\propp^{(0)}(r) i\mathbf{V}(r,-k) \label{eq:wf-def-2-int}\,,
    \end{align}
\end{subequations}
which has a simple interpretation in terms of the diagrams in Figure~\ref{fig:wf}.
The first term is referred to as the {\it free propagating} waveform, and the second as the {\it interacting} waveform - though the labels are not to be taken too literally, since they are simply two distinct contributions to the same 1SF observable.
The free part represents the contribution to the waveform from the geodesic emission $\bar{T}_\ml^{\alpha\beta}$ alone, in which the graviton propagates freely to the observer without any interaction with the background.
The second term $\wf_{\sigma}^{\mathrm{int}}$ describes the non-trivial crossing of the emitted graviton through the shockwave front, {\it i.e.} the second diagram of Fig.~\ref{fig:wf}.

We now compute the two terms in equation~\eqref{eq:wf-def-2} that contribute to the momentum-space waveform amplitude.
The free waveform is obtained from equation~\eqref{eq:wf-def-2-free}, for which we use the decomposition of the stress-energy tensor into an early and a late part, $\bar{T}_{\ml}^{\alpha\beta} = {}^e\bar{T}_{\ml}^{\alpha\beta} + {}^\ell\bar{T}_{\ml}^{\alpha\beta}$, together with the two contractions
\begin{subequations}
    \begin{align}
        {}^{e}\bar{T}_\ml^{\alpha\beta}(k)\pol_{\alpha\beta}^{(\sigma)*}(k) & = -4i E_2 \frac{{k}^i {k}^j \pol_{ij}^{(\sigma)*}}{(\xi\cdot k)^2 (\chi\cdot k-i0^+)} \, e^{-i\vec{k}\cdot\vec{b}} \,, \label{eq:contraction-1}\\
        {}^{\ell}\bar{T}_\ml^{\alpha\beta}(k)\pol_{\alpha\beta}^{(\sigma)*}(k) & = 4i E_2 \,\frac{\psi(k) \, k_s^i \, k_s^j \, \pol_{ij}^{(\sigma)*}}{(\xi\cdot k)^2} \, e^{-i\vec{k}\cdot\vec{b}} \,,
    \end{align}
\end{subequations}
where $k_s^i(k) = k^i - f_2(\xi\cdot k)\,{b^i}/{b}$, obtaining
\begin{equation}\label{eq:wf-free}
    \wf_{\sigma}^{\mathrm{free}}(k) = \frac{2\kappa E_2\,e^{-i\vec{k}\cdot\vec{b}}}{\xi\cdot k} \left(\frac{\pol^{(\sigma)}(k_s,k_s)}{\vec{k}_s^{2} + i0^+} \, e^{if_1 (\xi\cdot k)/2} - \frac{\pol^{(\sigma)}(k,k)}{\vec{k}^{\,2} - i0^+}\right)\,,
\end{equation}
in which we have introduced the notation $\pol^{(\sigma)}(v,w)=\pol^{(\sigma)*}_{ij} v^i w^j$.

As for the free propagating waveform, we again use the decomposition $\bar{T}_{\ml}^{\alpha\beta} = {}^e\bar{T}_{\ml}^{\alpha\beta} + {}^\ell\bar{T}_{\ml}^{\alpha\beta}$ to express the interacting waveform as the sum of an early-sourced contribution from ${}^e\bar{T}_\ml$ and a late-sourced one from ${}^\ell\bar{T}_\ml$.
However, we find that the latter is identically zero
\begin{equation}\label{eq:wf-late-int}
    \wf_{\sigma}^{\mathrm{int}}\big|_{\text{late sourced}} = 0\,,
\end{equation}
as can be guessed from causality principles: a graviton emitted by the secondary body after it has interacted with the primary one cannot reach the shockwave front and interact with it, since both travel at the speed of light.
In other words, a graviton can either be sourced from the deflected part of the trajectory, ${}^\ell \bar{T}_\ml^{\alpha\beta}$, or interact with the background, but never both.
Indeed, using its definition~\eqref{eq:wf-def-2-int} and the absence of light-cone plus components in the vertex, we arrive at the integral
\begin{equation}
    \wf_{\sigma}^{\mathrm{int}}\big|_{\text{late sourced}} \supset \int G^{(0)}(r) \, \psi(r) \,\hat{d}r^{+} \,.
\end{equation}
Both $\psi$ and the Mandelstam-Leibbrandt propagator have poles in the lower-half complex plane, so we close the contour in the upper-half plane, proving~\eqref{eq:wf-late-int}.
Thus the term $\wf_{\sigma}^{\mathrm{int}}$ is entirely given by its early-sourced part
\begin{equation}
    \wf_{\sigma}^{\mathrm{int}}(k) = -\frac{i\kappa}{2} \int_{r} \pol_{\alpha\beta}^{(\sigma)*}(r) \, {}^{e}\bar{T}_\ml^{\alpha\beta}(r) \, i\propp^{(0)}(r) i\mathbf{V}(r,-k)\,.
\end{equation}
The tensor contraction has already been computed in~\eqref{eq:contraction-1}, and since the vertex does not depend on the plus components, the $r^+$ integral can be performed without specifying $\mathbf{V}$ explicitly.
Using the residue theorem we obtain
\begin{equation}\label{eq:wf-early-int}
    \wf_{\sigma}^{\mathrm{int}}(k) = - \kappa E_2 \int_{r_\pperp,r^{-}} \frac{\pol^{(\sigma)}(r,r)}{( r\cdot\xi)^2\,\vec{r}^{\,2}}\,e^{-i\vec{r}\cdot\vec{b}}\,i\mathbf{V}(r,-k)\,.
\end{equation}
Unlike the free waveform $\wf_\sigma^\mathrm{free}$, this integration requires some approximations or systematic expansions.
Given the presence of the Weinberg factor $W=2\gn E_1 (\xi\cdot k)$ in the dressed vertex $\mathbf{V}(r,-k)$, we expand the waveform around two complementary limits:
\begin{enumerate}
    \item [$(i)$] The limit $W\ll1$ can be rephrased as $\mpl^2 \gg E_1 (\xi\cdot k)$, where $\mpl$ is the Planck mass.
        We are also expanding in Self-Force, which requires the perturbation to satisfy $\kappa h_{\alpha\beta} \lesssim \mathcal{O}(E_2/E_1)$, and thus $k^\mu \lesssim \mpl E_2/E_1$.
        Combining the limit $W\ll1$ with the SF expansion then gives
        \begin{equation}\label{eq:W-pm-exp}
            \mpl, E_1\gg E_2\,.
        \end{equation}
        On the other hand, the low-energy expansion ($\gn E_1 E_2 \ll 1$) combined with the Self-Force expansion implies $\mpl\gg E_1 \gg E_2$, so that condition~\eqref{eq:W-pm-exp} is only a slight generalisation of this.
        Thus, expanding in $W\ll1$ coincides with expanding in the small coupling, so we refer to this expansion as the {\it Post-Minkowskian} expansion.
    \item [$(ii)$] The strong-field regime, characterised by a large Weinberg factor $W = 2 \gn E_1 (\xi\cdot k) \gg 1$, which, by arguments similar to those of the previous point, leads to
        \begin{equation}\label{eq:W-strong-exp}
            E_1 \gg E_2 \gg \mpl\,.
        \end{equation}
        It implies that, once a frame is chosen in which the Self-Force expansion is well defined for a GSW background, we are accessing to strong fields.
        Indeed, this regime overlaps with {\it trans-Planckian} scattering, defined by $GE_1E_2\gg1$~\cite{Amati:2007ak,Ciafaloni:2015xsr,Ciafaloni:2018uwe,Amati:1990xe,Amati:1993tb,Verlinde:1991iu}.
\end{enumerate}
In the next two paragraphs we approach the computation of the waveform in these two limits.

\subsection{All-orders Post-Minkowskian series}
Here we compute the interacting waveform in its Post-Minkowskian expanded form
\begin{equation}
    \wf_\sigma^\mathrm{int}(k) = \sum_{n=1}^{\infty} \wf^{(n)}_\sigma(k) \,,
\end{equation}
with $\mathcal{W}^{(n)}_\sigma(k) \propto \kappa^{1+2n}$.
The interaction vertex was obtained order by order in the PM expansion in the previous section in terms of bubble integrals, see equation~\eqref{eq:vert-n-momenutm-space}.
Substituting the series for $\mathbf{V}$ into the definition~\eqref{eq:wf-early-int} of the interacting waveform and performing some manipulations, we obtain a sequence of contributions of the form
\begin{equation}\label{eq:wf-early-int-PM}
    \wf^{(n)}_\sigma(k) = \frac{i\kappa^3E_1 E_2}{2} \, \Big[iW \Gamma(-\epsilon)(4\pi\muir^2 e^{\gamma_\mathrm{E}})^\epsilon\Big]^{n-1} \,\frac{\Gamma(-\epsilon)\Gamma(\alpha_n(\epsilon))}{n!\,\Gamma(-n\epsilon)}\,\pol^{(\sigma)}(\partial_b,\partial_b) \, S_{1,\alpha_n(\epsilon)}(k,b)\,,
\end{equation}
with $\alpha_n(\epsilon)=1+n\epsilon-\epsilon$ and
\begin{equation}
    S_{\alpha_1,\alpha_2}(k,b) = \int \frac{\hat{d}^d\ell_\pperp\,\text{exp}\left({i\ell_\pperp\cdot b_\pperp}\right)}{\left[-\ell_\pperp^2\right]^{\alpha_1}\left[-(\ell-k)_\pperp^2\right]^{\alpha_2}} \,,
\end{equation}
so that the complete 1SF waveform is
\begin{equation}\label{eq:wf-PM-all-series}
    \begin{aligned}
        \wf_\sigma(k)
        & = \frac{2\kappa E_2\,e^{-i\vec{k}\cdot\vec{b}}}{\xi\cdot k} \left(\frac{\pol^{(\sigma)}(k_s,k_s)}{\vec{k}_s^{2} + i0^+} \, e^{if_1 (\xi\cdot k)/2} - \frac{\pol^{(\sigma)}(k,k)}{\vec{k}^{\,2} - i0^+}\right) \\
        & + \frac{i\kappa^3E_1 E_2}{2} \, \pol^{(\sigma)}(\partial_b,\partial_b) \sum_{n=1}^{\infty} \Big[iW \Gamma(-\epsilon)(4\pi\muir^2 e^{\gamma_\mathrm{E}})^\epsilon\Big]^{n-1} \,\frac{\Gamma(-\epsilon)\Gamma(\alpha_n(\epsilon))}{n!\,\Gamma(-n\epsilon)} \, S_{1,\alpha_n(\epsilon)}(k,b) \,.
    \end{aligned}
\end{equation}
Although we have not found a closed form for the integrals $S_{\alpha_1,\alpha_2}$, we provide a parametric solution in Appendix~\ref{app:integrals}, together with their integrated form in the soft series expansion for $k^i \ll 1/b$, valid at all orders in $\epsilon$.
The exactness in $\epsilon$ is required in order to perform the derivatives $\partial_b$, which must be carried out before performing the $\epsilon\to0$ limit.

Before turning to some sanity checks of the expression~\eqref{eq:wf-PM-all-series}, we discuss its infrared (IR) divergences, which appear as poles in $\epsilon$ in $\wf^{\rm int}_\sigma$ and in the PM expansion of $\wf^{\rm free}_\sigma$.
Following~\cite{Caron-Huot:2023vxl}, the infrared poles exponentiate and can be absorbed into a redefinition of the retarded time, which receives two distinct contributions.
One is the time shift associated with the sources of gravitons, {\it i.e.} with the worldlines $x_a^\mu(\tau)$, while the other is the Shapiro time delay of the radiation itself, which coincides with the exponentiation of the IR divergences predicted at the amplitude level by Weinberg.
As we proved at the end of section~\ref{sec:gauge-choice}, the primary source does not deviate, $\delta x_\mh^\mu=0$, while the trajectory~\eqref{eq:deflections-geodesics} of the secondary one has no logarithmic behaviour $\sim\log\tau$ for $|\tau|\to\infty$, since the shockwave field is supported on the null plane $\xi\cdot x=0$.
Its time shift reduces to the instantaneous shift $f_1$ at the crossing of the front.
Therefore, in the massless-massless collision the whole IR divergence is captured by the Weinberg phase, namely the Shapiro delay, predicted at the end of the previous section.
A straightforward calculation shows that the phase is given by
\begin{equation}
    e^{-i\Phi_{\rm IR}} = e^{if_1 (\xi\cdot k)/2}\,,
\end{equation}
where
\begin{equation}
    \Phi_{\rm IR} = -2GE_1 (\xi\cdot k)\,\Gamma(-\epsilon)\left(b^2 \muirt^2\right)^\epsilon = \frac{W}{\epsilon} + W\log\!\left(\pi b^2 \muir^2 e^{2\gamma_{\rm E}}\right) + \mathcal{O}(\epsilon)\,,
\end{equation}
and therefore coincides with the instantaneous shift of $\bar{x}_\ml^+$ in~\eqref{eq:deflections-geodesics}.
Thus we factor out the IR divergences through
\begin{equation}
    \widetilde{\wf}_\sigma (k) = e^{i\Phi_{\rm IR}}\,\wf_\sigma(k)\,.
\end{equation}

We validate the all-order PM interacting waveform in equation~\eqref{eq:wf-PM-all-series} by providing two tests.
First, we derive the leading-order Post-Minkowskian amplitude, given by the $\mathcal{O}(\kappa^3)$ terms of the free propagating waveform $\widetilde{\wf}_\sigma^{\rm free}$ in~\eqref{eq:wf-free} and the $\widetilde{\wf}_\sigma^{(1)}(k)$, which reads
\begin{equation}\label{eq:wf-early-int-1PM}
    \widetilde{\wf}_\sigma^{(1)}(k) =  \frac{i\kappa^3}{2} E_1 E_2 \,\pol^{(\sigma)}(\partial_b,\partial_b)\, S_{1,1}(k,b)\,.
\end{equation}
Combining them together, we obtain\footnote{At tree level there is no retarded-time renormalisation $\widetilde{\wf}^{\lo}_\sigma = \wf^{\lo}_\sigma$.}
\begin{equation}\label{eq:lo-waveform-1sf}
    \begin{aligned}
        \widetilde{\wf}^{\lo}_\sigma(k,b) = & \frac{i\kappa^3E_1 E_2}{2\pi \vec{k}^{\,2} b^2}e^{-i\vec{k}\cdot\vec{b}} \left(i \frac{\vec{k}\cdot\vec{b}}{\vec{k}^{\,2}}\pol^{(\sigma)}(k,k)-i\pol^{(\sigma)}(b,k) - \frac{b^2}{2} \pol^{(\sigma)}(k,k) \log(b\muirt) \right) \\
        & + \frac{i\kappa^3}{2} E_1 E_2 \,\pol^{(\sigma)}(\partial_b,\partial_b)\, S_{1,1}(k,b)\,.
    \end{aligned}
\end{equation}
This expression is in agreement with the result obtained in Appendix~\ref{app:lowf}, equation~\eqref{eq:lo-waveform}, from the KMOC formalism applied to the tree-level S-matrix element~\cite{Kosower:2018adc,Cristofoli:2021vyo}.

As a second test, we match our result against the predictions given by the classical soft-graviton theorems~\cite{Laddha:2019yaj,Saha:2019tub,Sahoo:2021ctw} at 1SF order\footnote{I thank Francesco Alessio for encouraging me to investigate the soft expansion of my results.}.
We express the on-shell momentum of the emitted graviton in terms of its frequency $\omega$ and the null direction $n^\mu$ ($n^2=0$)
\begin{equation}
    k^\mu = \omega n^\mu = \omega (2 c^2_{\theta/2},2 s^2_{\theta/2},s_\theta c_\varphi, s_\theta s_\varphi)\,,
\end{equation}
together with our result~\eqref{eq:double-soft} for the soft expansion of the integrals $S_{1,\alpha_n(\epsilon)}(k,b)$.
The general structure of the expansion of the $n$-th PM term of the waveform, for small frequencies $\omega$, is~\cite{Saha:2019tub,Sahoo:2021ctw,Laddha:2019yaj,Alessio:2024onn}
\begin{equation}\label{eq:structure-soft}
    \widetilde{\wf}_\sigma(k,b) = \frac{w_{\sigma}^{[1/\omega]}}{\omega} + \sum_{r=0}^\infty \left( w_{\sigma}^{[\omega^r\log^{r+1}\omega]}\,\omega^{r}\log^{r+1}\omega + \cdots \right),
\end{equation}
where the ellipsis denotes terms proportional to $\omega^{r}\log^{t}\omega$ with $t<r+1$.
Performing the computation order by order in the PM expansion, we see from~\eqref{eq:wf-early-int-PM}, together with the scalings of the $S_{\alpha_1,\alpha_2}$ integrals in~\eqref{eq:double-soft}, that $\mathcal{W}^{(n)}_\sigma(k)$ starts contributing at order $\omega^{n-1}$.
Thus we only need $\wf_\sigma^{\rm free},\wf_\sigma^{(1)}$ and $\wf_\sigma^{(2)}$ to predict the terms $w_\sigma^{[\log\omega]}$ and $w_\sigma^{[\omega \log^2\omega]}$, in agreement with the truncation predicted in~\cite{Sahoo:2021ctw}.
On the other hand, the free waveform $\mathcal{W}^{\rm free}_\sigma(k)$ is already PM-resummed and contains the entire, non-perturbative linear memory term~$w_\sigma^{[1/\omega]}$ truncated at 1SF order.
We are therefore able, using the replacement $\epsilon^{-1} \to 2\log\omega$ suggested in~\cite{Alessio:2024wmz}\footnote{that is $\muir\to\omega$ in our scheme}, to compute the linear memory and the two tails
\begin{subequations}\label{eq:soft-waveform-memory-tails}
    \begin{align}
        w_{\sigma}^{[1/\omega]} & = - \frac{2 \kappa^3 E_1 E_2 \left(\,\mathcal{N}_1 + \mathcal{N}_2\,\right)}{\left(8\pi b^2 n_\pperp + \kappa^2 E_1 (\xi\cdot n) b_\pperp \right)^2} \,, \label{eq:linear-memory}\\
        w_{\sigma}^{[\log\omega]} & = \frac{i\kappa^3 E_1 E_2}{4\pi} \frac{\pol^{(\sigma)}(n,n)}{n_\pperp^2} \,,\\
        w_{\sigma}^{[\omega\log^2\omega]} & = - \frac{\kappa^5 E_1^2 E_2 (\xi\cdot n)}{64\pi^2} \frac{\pol^{(\sigma)}(n,n)}{n_\pperp^2} = \frac{iW}{\omega} w_{\sigma}^{[\log\omega]} \label{eq:omegalogsquared} \,.
    \end{align}
\end{subequations}
in agreement with the terms of the soft expansion derived in~\cite{Sahoo:2021ctw}.
The two factors $\mathcal{N}_{1,2}$ are defined as
\begin{equation}
    \begin{aligned}
        & \mathcal{N}_1 = 16\pi b^2\left(\pol^{(\sigma)}(n,b)-\frac{n_\pperp\cdot b_\pperp}{n_\pperp^2}\pol^{(\sigma)}(n,n)\right) \,,\\
        & \mathcal{N}_2 = \kappa^2 E_1 (\xi\cdot n)\left( \pol^{(\sigma)}(b,b) + \frac{b^2}{n_\pperp^2} \pol^{(\sigma)}(n,n) \right) \,.
    \end{aligned}
\end{equation}
All the details of the matching with the literature are discussed in Appendix~\ref{app:soft-matching}.
We provide the notebook \texttt{Sab-integrals.nb}, which can be used, in principle, to obtain higher orders of the soft expansion.

\subsection{Trans-Planckian energy limit: convergence of the energy spectrum at high frequencies in the collinear region}\label{sec:trans-planckian}
Starting again from equation~\eqref{eq:wf-early-int}, we compute the PM-resummed waveform using the compact form of the vertex $\mathbf{V}(r,-k)$, computed at the end of the previous section in equation~\eqref{eq:resummed-V-qspace}, namely
\begin{equation}\label{eq:rk-vertex}
    i\mathbf{V}(r,-k) = 2 (\xi\cdot k) \hat{\delta}(\xi\cdot q) \left( \frac{4\pi iW}{\vec{q}^{\,2}} e^{-iW/\epsilon} \frac{\Gamma(1-iW)}{\Gamma(1+iW)}\left(\frac{\vec{q}^{\,2}}{4\pi\muir^2 e^{2\gamma_\mathrm{E}}}\right)^{iW} - \hat{\delta}^{(d)}(q_\pperp)\right),
\end{equation}
with $q=r-k$.
This equation yields two contributions: one, which we call $\wf_\sigma^{\mathrm{int},1}$, from the delta function in the round brackets, and another, named $\wf_\sigma^{\mathrm{int},2}$, from the factor involving the Weinberg infrared phase and the ratio of Gamma functions.
Therefore, the full waveform is the sum of the two contributions mentioned above and of the free waveform in equation~\eqref{eq:wf-free},
\begin{equation}
    \wf_\sigma = \wf_\sigma^{\rm int,1} + \wf_\sigma^{\rm int,2} + \wf_\sigma^{\rm free}\,.
\end{equation}
These three pieces are calculated one by one in the $W\gg1$ asymptotic expansion.
The first piece of the interacting waveform is trivially computed using the delta functions. 
Re-expressing $\xi\cdot k$ by inverting the definition of the Weinberg factor $\xi\cdot k = 16\pi W/\kappa^2 E_1$, we get
\begin{equation}\label{eq:wf-early-int1}
    \wf_\sigma^{\mathrm{int},1} = \frac{\kappa^3 E_1 E_2}{8\pi W} \, \frac{\pol^{(\sigma)}(k,k)}{\vec{k}^{\,2}} \, e^{-i\vec{k}\cdot\vec{b}}\,.
\end{equation}
The second term can be manipulated into the form
\begin{equation}\label{eq:wf-early-int2-not-computed}
    \wf_\sigma^{\mathrm{int},2} = -\frac{i\kappa^3 E_1 E_2}{2} \frac{e^{-iW/\epsilon}}{(4\pi\muir^2e^{2\gamma_\mathrm{E}})^{iW}} \frac{\Gamma(1-iW)}{\Gamma(1+iW)} \, e^{-i\vec{k}\cdot\vec{b}} \, \pol^{(\sigma)*}_{ij}\mathcal{I}^{ij}(k,b)\,,
\end{equation}
with the integral $\mathcal{I}^{ij}(k,b)$ defined in Appendix~\ref{app:saddle-point}, where we solve it for $W\gg1$ using the steepest-descent method.
There, we use Stirling's approximation to asymptotically expand the ratio of Gamma functions for large argument, finally obtaining
\begin{equation}\label{eq:wf-early-int2}
    \wf_\sigma^{\mathrm{int},2} = -\frac{\kappa^3 E_1 E_2}{8 \pi W} \, \frac{e^{-iW/\epsilon}}{\big(\pi \muir^2 b^2 e^{2\gamma_\mathrm{E}}\big)^{iW}} \, \frac{\pol^{(\sigma)}(b,b)}{b^2} \, e^{-i\vec{k}\cdot\vec{b}} + \mathcal{O}(W^{-2}) \,.
\end{equation}
All subleading contributions can be obtained by applying corrections to the leading-order saddle-point result.
We note that the dependence on $W$ enters only through a phase, which makes the amplitude of the waveform finite as $W\to\infty$.
This is unlike the Post-Minkowskian terms~\eqref{eq:wf-early-int-PM}, which diverge proportionally to $W^n$, demonstrating the need to resum the PM expansion in order to obtain finite observable results~\cite{DiVecchia:2022nna}.
Lastly, we also expand the free propagating waveform, obtaining
\begin{equation}
    \wf_\sigma^{\rm free} = \frac{\kappa^3 E_1 E_2}{8\pi W}\,e^{-i\vec{k}\cdot\vec{b}}\left(\frac{\pol^{(\sigma)}(b,b)}{b^2} e^{if_1(\xi\cdot k)/2} - \frac{\pol^{(\sigma)}(k,k)}{\vec{k}^{\,2}}\right)\,,
\end{equation}
where the second term cancels against $\wf_\sigma^{\mathrm{int},1}$ in equation~\eqref{eq:wf-early-int1}.
Moreover, the phase $e^{if_1(\xi\cdot k)/2}$ can be re-expressed using the Weinberg phase $\Phi_{\rm IR}$, obtaining an exact cancellation against $\wf_\sigma^{\rm int,2}$, proving that the full waveform is at least of order $\widetilde{\wf}_\sigma = \mathcal{O}\left(1/W^{2}\right)$.

We now discuss what physical consequence this cancellation produces for the spectrum of radiated energy when $W\gg1$.
Since $\xi\cdot k \sim \omega s^{2}_{\theta/2} \sim W$, this limit can be approached for large graviton frequency $\omega$ and emission angles $\theta\lesssim\pi$, for which the other components are suppressed: $n\cdot\chi \sim (\theta-\pi)^2$ and $\mid\!\!n_\pperp\!\mid\sim\pi-\theta$.
The flux of radiated energy due to the emission of gravitational waves is expressed in terms of the on-shell waveform amplitude~\cite{Weinberg:1965nx,Kosower:2018adc,Cristofoli:2021vyo,Bini:2026vaq}
\begin{equation}\label{eq:energy-spectrum-def}
    \frac{dE_{\rm gw}(\omega,n)}{d\omega\,d\Omega} = \frac{\omega^2}{32\pi^3} \sum_{\sigma=\oplus,\ominus} \big|\wf_\sigma(\omega n)\big|^2\,,
\end{equation}
and the convergence of the total emitted energy $E_{\rm gw}$ at large $\xi\cdot k$ is governed precisely by the asymptotics of $\wf_\sigma$ derived above.
Order by order in the Post-Minkowskian expansion, equation~\eqref{eq:wf-early-int-PM} shows that the $n$-th term $\mathcal{W}^{(n)}_\sigma$ grows as $W^{n-1}$ for large $\xi\cdot k$. If inserted naively into~\eqref{eq:energy-spectrum-def}, any finite truncation of the PM series would therefore produce an integrand that grows without bound in the high-frequency/collinear tail, in contradiction with any physical notion of total radiated energy.
The exact resummation performed above completely cures this pathology.
The cancellation between $\wf_\sigma^{\rm free}$, $\wf_\sigma^{\mathrm{int},1}$ and $\wf_\sigma^{\mathrm{int},2}$ improves the naive $\mathcal{O}(W^{-1})$ falloff of each individual piece to $\wf_\sigma(k) = \mathcal{O}\left(W^{-2}\right)$, so that in the limit $\xi\cdot k \sim \omega \to \infty$ we obtain
\begin{equation}
    \frac{dE_{\rm gw}}{d\Omega} \bigg|_{{\rm high-freq.},\;\chi\text{-collinear}} \, \sim \, \int_{\omega_*}^{\infty} d\omega \, \omega^2 \sum_\sigma\big|\wf_\sigma(k)\big|^2 \, \sim \, \int_{\omega_*}^{\infty} \frac{d\omega}{\omega^{2}} \,<\, \infty \,,
\end{equation}
which is manifestly convergent.
This confirms that, at least at leading Self-Force order, the trans-Planckian resummation of the interaction vertex is not just a technical device for obtaining a finite $E_1\to\infty$ limit of the waveform itself, but a genuine physical requirement for the angular spectrum of energy to converge.
This computation opens the way to a possible solution to the logarithmic divergence of the total radiated energy by massless-particle collisions~\cite{DiVecchia:2022nna,Ciafaloni:2015xsr,RajVenugopalan-universal-features,Gruzinov:2014moa}.
The convergence of the angular spectrum of radiated energy has been proved in the limit $\xi\cdot k = 2\omega s^2_{\theta/2} \to \infty$, which corresponds to high-frequency emission in the collinear direction $\theta\lesssim\pi$.
A similar derivation for $\chi\cdot k \to \infty$ emission should, for completeness, also be carried out, as well as the angular integration which requires expansions in all the emission regions -- $\xi$-collinear, $\chi$-collinear and multi-Regge.
We leave the complete analysis for future work.
{\section{Conclusions}\label{sec:conclusions}}
In this paper we developed and tested a Self-Force Effective Field Theory for the high-energy collision of two spinless bodies, in the regime where the primary particle is massless and sources a gravitational shockwave background.
We showed that the SF-EFT of~\cite{Cheung:2023lnj,Cheung:2024byb}, originally formulated for generic backgrounds and sources, applies successfully to the GSW background up to 1SF order, once extended to the Schwinger-Keldysh (in-in) path integral~\cite{Schwinger:1960qe} to include radiation-reaction effects in the observables.
A key simplification throughout is the choice of the light-cone gauge, whose reference vector coincides with the Killing vector of the Aichelburg-Sexl metric.
This choice makes the recoil operator vanish identically at 1SF and 2SF order, and, as we prove, at any Self-Force order.
It also reduces the tensor structure of propagators and vertices to a remarkably compact form.
This confirms the efficiency of the Self-Force EFT, already tested on the Schwarzschild~\cite{Cheung:2023lnj,Cheung:2024byb} and Kerr~\cite{Akpinar:2025huz} backgrounds, and lets us probe the collision of shockwaves non-perturbatively in Newton's constant, while remaining perturbative in the small parameter controlled by the secondary ``light'' body.

The Self-Force expansion is not covariantly defined for a massless-massless collision.
The expansion parameter $q$ rescales under collinear boosts, so the SF expansion we employ is valid only in frames boosted along one of the two directions of the process.
This asymmetric frame breaks a symmetry relating the two colliding bodies, but it is precisely this asymmetry that enables the resummations in $\gn$.

We computed the graviton propagator on GSW background and we matched it against the independent response-theory (WQFT) computation of~\cite{Bohnenblust:2026ujk}, finding complete agreement, and confirmed that the propagator calculated by Raj and Venugopalan in the eikonal approximation~\cite{Raj:2024xsi} is, in fact, exact.
We further expect our result to match the ultrarelativistic limit of the Compton amplitude in $\mathcal{N}=8$ supergravity, to be presented in forthcoming work~\cite{n8compton}.
Despite arising from entirely different frameworks, the two computations should agree, lending further support to the graviton dominance expected at high energies~\cite{tHooft:1987vrq,DiVecchia:2020ymx}.

Using the graviton propagator, we obtained the leading order waveform in the $E_2/E_1\ll 1$ expansion.
Although we could not resum the resulting Post-Minkowskian series for the so called {\it interacting} piece of the waveform, we computed it to all orders.
The result is expressed in terms of a family of integrals $S_{\alpha_1,\alpha_2}$ which, while admitting no closed-form solution at present, we cast in a parametric integral representation, together with a complete series expansion around the soft limit.
We validated the waveform against two independent checks: $(i)$ at leading order in the Post-Minkowskian expansion, it reproduces the tree-level waveform obtained via the KMOC formalism~\cite{Kosower:2018adc,Cristofoli:2021vyo};
$(ii)$ in the soft expansion, it matches the classical soft-graviton theorems~\cite{Saha:2019tub,Sahoo:2021ctw,Laddha:2019yaj,Alessio:2024onn}.
We find complete agreement for the $1/\omega$ (linear) memory term and the $\log\omega$ and $\omega\log^2\omega$ terms, which truncates exactly at $\mathcal{O}(\kappa^3)$ and $\mathcal{O}(\kappa^5)$, respectively.

Finally, we used a saddle-point approximation to show that the 1SF waveform, in the strong-field limit $W = 2GE_1(\xi\cdot k) \gg 1$, is at most of order $\mathcal{O}(W^{-2})$.
This implies that the angular spectrum of radiated energy, computed in the collinear direction $\theta\lesssim\pi$, converges, approaching the long-standing problem of the logarithmic divergence of the total radiated energy in trans-Planckian scattering of massless particles in GR~\cite{DiVecchia:2022nna,Ciafaloni:2015xsr}.
Since our result is frame-dependent, the radiated energy we obtain differs by a boost from that computed, {\it e.g.}, in~\cite{Ciafaloni:2015xsr}.
This is a genuinely non-perturbative result: as our computation holds at any order in $G$ from the outset, it yields non-analytic information in the coupling, which we believe to be essential for a complete understanding of scattering in the trans-Planckian regime~\cite{Amati:2007ak,Ciafaloni:2018uwe,Amati:1990xe,Amati:1993tb,Verlinde:1991iu}.
A full study of the radiated energy spectrum -- including the complementary collinear limit $\chi\cdot k\to\infty$ ($\theta \sim 0$) and multi-Regge limit ($\theta \sim \pi/2$) -- is left to future work.

\acknowledgments
I am extremely grateful to Riccardo Gonzo for taking part in the initial stages of this project, as well as for many enlightening discussions and suggestions.
I give special thanks to Francesco Alessio and Riccardo Gonzo for reviewing the draft and useful comments and suggestions.
I also thank Lara Bohnenblust, Vittorio Del Duca, Carl Jordan Eriksen, Jitze Hoogeveen, Gustav Jakobsen, Jan Plefka, Ira Rothstein and Michael Saavedra for discussions on this paper and related topics.

\appendix

\section{Geodesic motion on Aichelburg-Sexl\label{app:geodesic-motion}}
We give a review of the solutions to eq.~\eqref{eq:geodesic-equations} for massless $\dot{\bar{x}}_\ml^2=0$ and massive $\dot{\bar{x}}_\ml^2=1$ particles, already computed in~\cite{FerrariVeneziano,Dray:1984ha}.
The non-vanishing components of the curvature symbols are calculated using \texttt{xCoba}~\cite{xAct}
\begin{equation}
    \bar\Gamma^+_{--} = \ag G_d(r)\dot{\delta}(x^-)\,, \qquad \bar\Gamma^i_{--} = \frac{\ag}{2}\delta(x^-)\pd_i G_d(r)\,, \qquad \bar\Gamma^+_{i-} = \ag\delta(x^-)\pd_i G_d(r)\,,
\end{equation}
and, fixing the affine parameter $s= x^-$ and using the $SO(d)$ symmetry of the problem, we get the following equations of motion
\begin{subequations}
    \begin{align}
        & {\bar{x}}_\ml^{-} = s \,, \\
        & \ddot{\bar{x}}_\ml^{+} = -\ag\dot\delta(s)G_d(\bar{r}(s)) - 2\ag\delta(s)G_d'(\bar{r}(s))\dot{\bar{r}}(s) \,, \label{eq:equation-xplus}\\
        & \ddot{\bar{r}}_\ml = -\frac{\ag}{2}\delta(s)\,G_d'(\bar{r}(s)) \,,
    \end{align}
\end{subequations}
in which $G_d^\prime = \partial_r G_d$. 
In formulating the Cauchy conditions we distinguish two cases: $(i)$ the projectile particle is massless, having a light-like initial velocity proportional to $\chi^\mu$
\begin{equation}
    m=0:\quad
    \begin{cases}
        \bar{x}_\ml^{+}(-\infty) = 0\,, & \dot{\bar{x}}_\ml^{+}(-\infty) = 0\,, \\
        \bar{r}_\ml(-\infty) = b \,, & \dot{\bar{r}}_\ml(-\infty) = 0\,,\\
    \end{cases}
\end{equation}
or $(ii)$ it is massive,
\begin{equation}
    m\ne0:\quad
    \begin{cases}
        \bar{x}_\ml^{+}(-\infty) = s\,, & \dot{\bar{x}}_\ml^{+}(-\infty) = 1\,, \\
        \bar{r}_\ml(-\infty) = b \,, & \dot{\bar{r}}_\ml(-\infty) = 0\,.\\
    \end{cases}
\end{equation}
We use the frames in which the massive particle is initially at rest while the massless one hits the shockwave on a collinear configuration with an impact parameter living in the transverse space $b^\alpha=b_\perp^\alpha$.\footnote{Only for the impact parameter we use the convention in which $b=|b_\perp|$, so that $b^2>0$.}

\paragraph{Radial equation}
The radial equation is the same for both massive and massless motion, with solution
\begin{equation}\label{eq:radial-motion}
    \bar{r}_\ml(s) = b - \frac{\ag}{2}s \Theta(s) G_d'(b) = b + s\Theta(s)f_2\,,
\end{equation}
Equation~\eqref{eq:radial-motion} predicts a kick in the radial velocity equal to $f_2\cdot\hat{b}^\alpha = - ({\ag}/{2})G_d'(b) \cdot \hat{b}^\alpha$ after the transition through the wavefront.
We remark the introduction of three constants $f_1,f_2,f_3$ defined in equation~\eqref{eq:f1-f2-f3}, which are derived in this appendix and are related to the coordinate and velocity shifts of the geodesic motion.

\paragraph{Light-front equation}
The equation governing $\bar{x}_\ml^{+}$ is subject to differences between the two cases.
Using the distributional identity $f(s)\dot\delta_s = f(0)\dot\delta_s-\dot{f}(0)\delta_s$, eq.~\eqref{eq:equation-xplus} is re-expressed as
\begin{equation}
    \ddot{\bar{x}}_\ml^{+} = -\ag\dot{\delta}(s) G_d(b) + \frac{\ag^2}{4} \delta(s) \left(G_d'(b)\right)^2 \,.
\end{equation}
For the light-like geodesic, we obtain
\begin{equation}
    \bar{x}_{\ml,m=0}^{+} (s) = -\ag G_d(b)\Theta(s) + \frac{\ag^2}{4}s \Theta(s)\left(G_d'(b)\right)^2 = f_1\Theta(s) + f_3 s\Theta(s)\,,
\end{equation}
which gives an instantaneous shift $f_1 = -\ag G_d(b)$ and a velocity shift $f_3 = (\ag^2/4)\left(G_d'(b)\right)^2$.
We further observe that $f_3 = f_2^2$, which indeed preserves the on-shell condition $\dot{\bar{x}}_\ml^2 = 0$ for $m=0$.

The massive plus component law is instead
\begin{equation}
    \bar{x}_{\ml,m\ne0}^{+} (s) = s + \Theta(s)f_1 + s\Theta(s)f_3\,,
\end{equation}
Thus changing the plus velocity into $f_{3,m} = \Theta(s)f_3 + 1$.
The shift itself therefore remains unchanged with respect to the previous computation, while $\dot{\bar{x}}_\ml^2=1$ is preserved for $m\ne0$.
\section{Light-cone decomposition\label{app:projectors}}
We decompose Minkowski space in light-cone coordinates using the vectors $\xi$ and $\chi$ and an orthogonal projector.
One defines three projectors
\begin{equation}\label{eq:projectors}
    \Pi_{+\,\alpha\beta} = \frac{\chi_\alpha \xi_\beta}{\xi\cdot\chi}\,, \qquad 
    \Pi_{-\,\alpha\beta} = \frac{\xi_\alpha \chi_\beta}{\xi\cdot\chi}\,, \qquad 
    \Pi_{\pperp\,\alpha\beta} = \eta^\pperp_{\alpha\beta}\,,
\end{equation}
such that their sum equals the metric $\eta_{\alpha\beta}$.
Applying the identity four times to the light-cone gauge projector of equation~\eqref{eq:sum-polarisations} and using~\eqref{eq:projectors}
\begin{equation}\label{eq:projected-P-graviton}
    \begin{aligned}
        P_{\alpha\beta\gamma\delta} & = P_{----}\,\xi_\alpha\xi_\beta\xi_\gamma\xi_\delta + P_{---i} \, \xi_\alpha\xi_\beta\xi_\gamma {\eta^\pperp}_\delta^{\;i} \\
            & + P_{--ij} \, \xi_\alpha\xi_\beta {\eta^\pperp}_\gamma^{\;i} {\eta^\pperp}_\delta^{\;j} + P_{-i-j} \, \xi_\alpha\xi_\gamma {\eta^\pperp}_\beta^{\;i} {\eta^\pperp}_\delta^{\;j} \\
            & + P_{-rij} \, \xi_\alpha {\eta^\pperp}_\beta^{\;r} {\eta^\pperp}_\gamma^{\;i}{\eta^\pperp}_\delta^{\;j} + P_{ijkl} \, {\eta^\pperp}_{\alpha}^{\;i} {\eta^\pperp}_{\beta}^{\;j} {\eta^\pperp}_{\gamma}^{\;k} {\eta^\pperp}_{\delta}^{\;l} + \mathrm{perms} \,,
    \end{aligned}
\end{equation}
where the permutations are obtained using trivial symmetries of the tensor $P_{\alpha\beta\gamma\delta}$ and
\begin{equation}\label{eq:pijkl-projected}
    \begin{alignedat}{3}
        & P_{----} = \frac{(D-3)(k\cdot\chi)^2}{(D-2)(k\cdot\xi)^2}\,, &\hspace{50pt} P_{-i-j} = \frac{1}{2}\frac{(D-4)}{(D-2)} \frac{k_{\pperp\,i} k_{\pperp\,j}}{(k\cdot\xi)^2} - \frac{1}{2} \frac{(k\cdot\chi)}{(k\cdot\xi)} \eta^{\pperp}_{ij}\,, \\
        & P_{---i} = \frac{(D-3)(k\cdot\chi)\;}{(D-2)(k\cdot\xi)^2}k_{\pperp\,i}\,,  &\hspace{50pt} P_{-rij} = \frac{\eta^\pperp_{ij} k_{\pperp\,r}}{(D-2)(k\cdot\xi)} - \frac{\eta^\pperp_{ir} k_{\pperp\,j} + \eta^\pperp_{rj} k_{\pperp\,i}}{2 (k\cdot\xi)} \,, \\
        & P_{--ij} = \frac{k_{\pperp\,i}k_{\pperp\,j}}{(k\cdot\xi)^2} + \frac{(k\cdot\chi)}{(k\cdot\xi)}\frac{\eta^\pperp_{ij}}{D-2}\,,  &\hspace{50pt} P_{ijkl} = \frac{1}{2}\eta^\pperp_{ik}\eta^\pperp_{jl} + \frac{1}{2}\eta^\pperp_{il}\eta^\pperp_{jk} - \frac{1}{D-2}\eta^\pperp_{ij}\eta^\pperp_{kl} \,.
    \end{alignedat}
\end{equation}
We usually rename the last symbol as $P_{ijkl} \doteq P^\pperp_{\alpha\beta\gamma\delta}$ with greek indices.
Similarly, we decompose the two-entry projector $\mathcal{P}_{\alpha\beta\gamma\delta}(k_1,k_2)$ in~\eqref{eq:projector-P2}
\begin{equation}\label{eq:pijkl-projected2}
    \begin{alignedat}{3}
        & \mathcal{P}_{----} = \frac{(k_{1\pperp}\cdot k_{2\pperp})^2}{(k_1\cdot\xi)^2 (k_2\cdot\xi)^2} - \frac{1}{D-2}\frac{(k_1\cdot\chi)(k_2\cdot\chi)}{(k_1\cdot\xi)(k_2\cdot\xi)} \,,\\
        & \mathcal{P}_{---i} = -\frac{(k_{1\pperp}\cdot k_{2\pperp}) (k_{1\pperp})_i}{(k_1\cdot\xi)^2 (k_2\cdot\xi)} - \frac{(k_{2\pperp})_i}{D-2}\frac{(k_1\cdot\chi)}{(k_1\cdot\xi) (k_2\cdot\xi)}\,,\\
        & \mathcal{P}_{--ij} = \frac{(k_{1\pperp})_i (k_{1\pperp})_j}{(k_1\cdot\xi)^2} + \frac{\eta^\pperp_{ij}}{D-2}\frac{(k_1\cdot\chi)}{(k_1\cdot\xi)} \,,\\
        & \mathcal{P}_{-i-j} = \frac{1}{2}\frac{(k_{2\pperp})_i (k_{1\pperp})_j}{(k_1\cdot\xi) (k_2\cdot\xi)} - \frac{1}{D-2}\frac{(k_{1\pperp})_i(k_{2\pperp})_j}{(k_1\cdot\xi) (k_2\cdot\xi)} + \frac{1}{2}\frac{\eta^{\pperp}_{ij} k_{1\pperp}\cdot k_{2\pperp}}{(k_1\cdot\xi) (k_2\cdot\xi)}\,,\\
        & \mathcal{P}_{-kij} = -\frac{1}{2}\frac{(k_{1\pperp})_i \eta^\pperp_{jk} + (k_{1\pperp})_j \eta^\pperp_{ik}}{(k_1\cdot\xi)} + \frac{1}{D-2}\frac{\eta^\pperp_{ij} (k_{1\pperp})_k}{(k_1\cdot\xi)} \,,\\
        & \mathcal{P}_{ijkl} = \frac{1}{2}\eta^\pperp_{ik}\eta^\pperp_{jl} + \frac{1}{2}\eta^\pperp_{il}\eta^\pperp_{jk} - \frac{1}{D-2}\eta^\pperp_{ij}\eta^\pperp_{kl} \,.
    \end{alignedat}
\end{equation}
\section{$S_{\alpha_1,\alpha_2}$ integrals\label{app:integrals}}
The 1SF waveform in momentum space~\eqref{eq:wf-early-int} and its Post-Minkowskian expansion~\eqref{eq:wf-early-int-PM} are given in terms of derivatives of the following family of Euclidean integrals
\begin{equation}\label{eq:definition-sa1a2}
    S_{\alpha_1,\alpha_2}(k,b) = \int \hat{d}^d\ell_\pperp \frac{e^{-i\vec{\ell}\cdot\vec{b}}}{\big(\vec{\ell}^{\,2}\big)^{\alpha_1}\big((\vec{\ell}-\vec{k})^2\big)^{\alpha_2}} \,,
\end{equation}
which is the massless counterpart of the family used in the computation of the NLO waveform in the scattering of massive spinless compact bodies~\cite{Brunello:2025eso}.
This appendix furnishes a parametric integral representation for the $S_{\alpha_1,\alpha_2}(k,b)$, as well as a series expansion for their solution.
Afterwards we specialise to the indices $\alpha_1=1$ and $\alpha_2=1+(n-1)\epsilon$ appearing in the Post-Minkowskian series~\eqref{eq:wf-early-int-PM}, obtaining an analytic solution to the integrals as a series of small graviton momentum $k^i \ll 1/b$, which we refer to as {\it soft expansion}.
All the computations related to these integrals and the soft expansion of the waveform~\eqref{eq:soft-waveform-memory-tails} are given in the ancillary notebook~\texttt{Sab-integrals.nb}.

\paragraph{Parametric integral representation}
We express the two propagators using the Schwinger representation as suggested in~\cite{Riva:2021vnj} and, after performing the loop integral, we obtain
\begin{equation}
    \begin{aligned}
            S_{\alpha_1,\alpha_2}(k,b) & = \frac{(4\pi)^{-d/2}}{\Gamma(\alpha_1)\Gamma(\alpha_2)} \\
            & \int_{\mathbb{R}^+\times\,\mathbb{R}^+} dt_1 \, dt_2 \, t_1^{\alpha_1-1} \, t_2^{\alpha_2-1} \, (t_1+t_2)^{-\frac{d}{2}} \exp \left(-\frac{\vec{b}^{\,2} + 4it_2 \vec{k}\cdot\vec{b} + 4t_1 t_2\vec{k}^{\,2}}{4(t_1+t_2)}\right)\,.
    \end{aligned}
\end{equation}
The change of variables $t_1 = \lambda(1-x)$ and $t_2 = \lambda x$, with $x\in [0,1]$ and $\lambda\in(0,\infty)$, brings the integral into
\begin{equation}
    \begin{aligned}
        S_{\alpha_1,\alpha_2}(k,b) & = \frac{(4\pi)^{-d/2}}{\Gamma(\alpha_1)\Gamma(\alpha_2)} \int_0^1 dx \, (1-x)^{\alpha_1-1} x^{\alpha_2-1} \, e^{-ix\vec{k}\cdot\vec{b}} \\
        & \qquad \int_{0}^{\infty} d\lambda\, \lambda^{\alpha_1 + \alpha_2 -\frac{d}{2}-1} \, \exp\left(-\frac{\vec{b}^{\,2}}{4\lambda}-\lambda x(1-x)\vec{k}^{\,2}\right)\,,
    \end{aligned}
\end{equation}
and, using
\begin{equation}\label{eq:bessel-integral}
    \int_0^\infty ds \, e^{-a/s-s} s^{b-1} = 2a^{b/2} \, K_{-b}\left(2\sqrt{a}\right) \,,
\end{equation}
we get
\begin{equation}\label{eq:sa1a2}
    \begin{aligned}
            S_{\alpha_1,\alpha_2}(k,b) & = 2^{1-\alpha_1-\alpha_2}\frac{(2\pi)^{-d/2}}{\Gamma(\alpha_1)\Gamma(\alpha_2)} \left(\frac{k}{b}\right)^{\frac{d}{2}-\alpha_1-\alpha_2} \\
            & \int_0^1 dx\,e^{-ix\vec{k}\cdot\vec{b}}\,x^{d/4+(\alpha_2-\alpha_1)/2-1}(1-x)^{d/4+(\alpha_1-\alpha_2)/2-1}\, K_{d/2-\alpha_1-\alpha_2}\left(kb\sqrt{x(1-x)}\right)\,,
    \end{aligned}
\end{equation}
with $b=|\vec{b}|$ and, only within this appendix, $k=|\vec{k}|$.
This form is a simplification of the integrals already derived in~\cite{Brunello:2025eso}.
An integrated result is given below in series expansion, once the modified Bessel function is expanded in powers of its argument.

There are two special instances of the family of integrals $S_{\alpha_1,\alpha_2}$, namely when either one of the two indices $\alpha_i$ is zero.
These two cases cannot be obtained directly from~\eqref{eq:sa1a2} and require a separate computation.
Using just one Schwinger parameter, we obtain
\begin{subequations}\label{eq:sa1a2-special}
    \begin{align}
        & S_{\alpha_1,0}(k,b) = \frac{\big(\vec{b}^{\,2}/4\big)^{\alpha_1}}{(\pi \vec{b}^{\,2})^{d/2}}\frac{\Gamma\left(d/2-\alpha_1\right)}{\Gamma(\alpha_1)}\,,\\
        & S_{0,\alpha_2}(k,b) = \frac{\big(\vec{b}^{\,2}/4\big)^{\alpha_2}}{(\pi \vec{b}^{\,2})^{d/2}}\frac{\Gamma\left(d/2-\alpha_2\right)}{\Gamma(\alpha_2)} \, e^{-i\vec{b}\cdot\vec{k}}\,.
    \end{align}
\end{subequations}

\paragraph{Series representations}
We calculate an integrated expression for the family of integrals~\eqref{eq:sa1a2} through the series expansion of the modified Bessel function for small argument.
This is physically motivated by considering $bk \ll 1$, which could be either an expansion in the collinear emission limit - for which the transverse component $k_\pperp^\mu$ is small - or a soft expansion.
To perform the last integration in equation~\eqref{eq:sa1a2}, we use the definition of the modified Bessel function
\begin{equation}
    K_\nu(s) = \frac{\pi}{2\sin(\pi\nu)} \sum_{n=0}^{\infty} \frac{1}{n!} \left( \frac{(s/2)^{2n-\nu}}{\Gamma(n-\nu+1)} - \frac{(s/2)^{2n+\nu}}{\Gamma(n+\nu+1)} \right) \,,
\end{equation}
while keeping the exponential $e^{-ix\vec{k}\cdot\vec{b}}$ fixed, although it should be expanded too.
We obtain the series for small argument $z$
\begin{equation}
    \begin{aligned}
        & \int_0^1 dx\,e^{ix\gamma}x^{a_1-1}(1-x)^{a_2-1} K_\nu(z\sqrt{x(1-x)}) = \\
        & \quad \frac{\pi (2z)^{-\nu}}{2\sin(\pi\nu)} \sum_{n=0}^\infty \frac{1}{n!}\left(\frac{z}{2}\right)^{2n} \cdot \\
        & \quad \Bigg(4^{\nu}\frac{\Gamma(a_1+n-\frac{\nu}{2})\Gamma(a_2+n-\frac{\nu}{2})}{\Gamma(1+n-\nu)\Gamma(a_1+a_2+2n-\nu)}{}_1F_1\left(a_1+n-\frac{\nu}{2},a_1+a_2+2n-\nu;i\gamma\right) \\
        & \quad \; -z^{2\nu} \frac{\Gamma(a_1+n+\frac{\nu}{2})\Gamma(a_2+n+\frac{\nu}{2})}{\Gamma(1+n+\nu)\Gamma(a_1+a_2+2n+\nu)} {}_1F_1\left(a_1+n+\frac{\nu}{2},a_1+a_2+2n+\nu;i\gamma\right) \Bigg)\,.
    \end{aligned}
\end{equation}
The convergence conditions are $\mathsf{Re}(a_1\pm\nu/2)>0$ and $\mathsf{Re}(a_2\pm\nu/2)>0$.
Using the integral above, we get the series expansion of the members of the family of integrals $S_{\alpha_1,\alpha_2}(k,b)$, namely
\begin{equation}\label{eq:sab-1f1}
    \begin{aligned}
        S_{\alpha_1,\alpha_2}(k,b) = &\frac{(4\pi)^{-d/2}}{\Gamma(\alpha_1)\Gamma(\alpha_2)}\,\frac{\pi}{\sin\left(\pi\alpha\right)}\left(\frac{b}{k}\right)^{\alpha} \sum_{n=0}^{\infty}\frac{(bk)^{2 n}}{n!4^n} \\
        & \Bigg( \big(bk\big)^{-\alpha} \frac{\Gamma\left(-\alpha_1+\frac{d}{2}+n\right)\Gamma\left(-\alpha_2+\frac{d}{2}+n\right)}{\Gamma\left(1-\alpha+n\right)\Gamma\left(-\alpha+\frac{d}{2}+2n\right)} {}_1F_1\left(-\alpha _1+\frac{d}{2}+n,-\alpha+\frac{d}{2}+2n;i\gamma\right) \\
        & \,- \left(\frac{bk}{4}\right)^{\alpha} \frac{\Gamma\left(\alpha_1+n\right)\Gamma\left(\alpha_2+n\right)}{\Gamma\left(1+\alpha+n\right)\Gamma\left(\alpha_1+\alpha_2+2n\right)} {}_1F_1\left(\alpha_2+n,\alpha_1+\alpha_2+2n;i\gamma\right) \Bigg) \,, \\
    \end{aligned}
\end{equation}
where $\alpha = \alpha_1 + \alpha_2 - \dfrac{d}{2}$ and $\gamma =-\vec{k}\cdot\vec{b}$.
This form can equivalently be computed using a Mellin-Barnes representation of the integral~\eqref{eq:definition-sa1a2} and closing the contour on the UV poles~\cite{Weinzierl:2022eaz}.

\paragraph{Soft expansion}
If one requires the momentum $k$ to be soft, equation~\eqref{eq:sab-1f1} can be further expanded for small $\gamma$ through the definition of the ${}_1F_1$ function
\begin{equation}
    {}_1F_1(a_1,a_2;z) = \frac{\Gamma(a_2)}{\Gamma(a_1)} \sum_{m=0}^\infty \frac{\Gamma(a_1+m)}{\Gamma(a_2+m)} \frac{z^m}{m!}\,,
\end{equation}
obtaining, with $d=2-2\epsilon$ and $\alpha=\alpha_1+\alpha_2-1+\epsilon$, the expression
\begin{equation}\label{eq:sab-soft}
    \begin{aligned}
        S_{\alpha_1,\alpha_2}(k,b) = &\frac{(4\pi)^{-1+\epsilon}}{\Gamma(\alpha_1)\Gamma(\alpha_2)}\,\frac{\pi}{\sin\left(\pi\alpha\right)}\left(\frac{b}{k}\right)^{\alpha} \sum_{n=0}^{\infty} \sum_{m=0}^{\infty} \frac{(bk)^{2 n}}{n!4^n} \frac{(i\gamma)^m}{m!} \\
        & \Bigg( \big(bk\big)^{-\alpha_1-\alpha_2-\epsilon+1} \frac{\Gamma(n+m-\alpha_1-\epsilon+1)\Gamma(n-\alpha_2-\epsilon+1)}{\Gamma(n-\alpha_1-\alpha_2-\epsilon+2)\Gamma(2n+m-\alpha_1-\alpha_2+2-2\epsilon)} \\
        & \,- \left(\frac{bk}{4}\right)^{\alpha_1+\alpha_2+\epsilon-1} \frac{\Gamma\left(n+\alpha_1\right)\Gamma\left(m+n+\alpha_2\right)}{\Gamma\left(2n+m+\alpha_1+\alpha_2\right)\Gamma\left(n+\alpha_1+\alpha_2+\epsilon\right)} \Bigg) \,. \\
    \end{aligned}
\end{equation}
Afterwards one can specialise to the case $\alpha_1=1$ and $\alpha_2=1+p\epsilon-\epsilon\equiv\alpha_p$, which is useful for expressing the Post-Minkowskian terms of the 1SF waveform in the soft limit.
We change the summation indices $m+n\to n$ and $m\to m$ in equation~\eqref{eq:sab-soft} and resum the $m$ contributions, obtaining
\begin{equation}\label{eq:double-soft}
    \begin{aligned}
        S_{1,\alpha_p}(k,b) = &\frac{k^{-2p\epsilon}}{4^{\alpha_p+1}k^2 \Gamma(\alpha_p)} \frac{\pi^\epsilon}{\sin(\pi p\epsilon)}\sum_{n=0}^{\infty} \frac{(i\gamma)^n}{n!} \\
        & \Bigg( (bk)^{2p\epsilon+2}\Gamma(\alpha_p+n) {}_2\tilde{F}_2\left(1,-n;2+p\epsilon,\alpha_p+n+1;-\frac{k^2 b^2}{4i\gamma}\right) \\
        & \qquad -4^{1+p\epsilon}\Gamma(n-\epsilon){}_1\tilde{F}_1\left(-n,n-(p+1)\epsilon;-\frac{k^2 b^2}{4i\gamma}\right) \Bigg)\,.
    \end{aligned}
\end{equation}
It is straightforward to apply the derivatives $\pol_{ij}^{(\sigma)*} \partial_b^i \partial_b^j$ to this expression and afterwards to take the $\epsilon\to0$ and soft $bk\sim\gamma \ll 1$ limits.
\section{Leading order waveform from the KMOC method\label{app:lowf}}
The gravitational waveform at leading order in $\gn$ for massless particles can be computed from the tree-level classical amplitude, which is derived by independent computation.
Using the conventions for the momenta $p_1,\; p_2 \rightarrow p_1+q-k,\; p_2-q,\; k^\ominus$, with $p_1^\mu=E_1\xi^\mu$ and $p_2^\mu=E_2\chi^\mu$, we obtain
\begin{equation}\label{eq:tree-amplitude}
    \frac{i\mathcal{M}_\ominus^{\rm cl.\,tree}(k,q)}{2s} = \frac{i\kappa^3}{4} E_1 E_2 \left(-\frac{2 k_\pperp\cdot q_\pperp \zeta_k^2}{(k\cdot\xi)^2(k\cdot\chi)^2(q-k)_\pperp^2} - \frac{2\zeta_q\zeta_k + \zeta_k^2}{(k\cdot\xi)(k\cdot\chi)(q-k)_\pperp^2} - \frac{\zeta_q^2}{q_\pperp^2(q-k)_\pperp^2}\right) ,
\end{equation}
which satisfy gauge invariance and the leading soft theorem~\cite{Weinberg:1965nx}.
In the equation above we used the notation $\zeta_v = v^x+iv^y$.
We use the KMOC approach~\cite{Kosower:2018adc,Cristofoli:2021vyo}, which gives the LO waveform in terms of the tree-level amplitude
\begin{equation}
    \widetilde{\wf}^{\lo}_\ominus(k,b) = \int_{q_\pperp} \frac{i\mathcal{M}_\ominus^{\rm cl.}(k,q)}{2s}\, e^{-i\vec{q}\cdot\vec{b}}\,.
\end{equation}
We find that all the integrals can be performed using the formulae~\eqref{eq:sa1a2-special} for $S_{\alpha_1,0}$ and $S_{0,\alpha_2}$ and their derivatives, in particular
\begin{equation}
    \partial_b^i S_{1,0}(k,b) = 2\epsilon \frac{b^i}{b^2}S_{1,0}(k,b)\,, \qquad \delta_{ij} \partial_b^i \partial_b^j S_{1,0}(k,b) = -\delta^{(d)}(b)\,,
\end{equation}
together with the integral $S_{1,1}(k,b)$ coming from the last term in equation~\eqref{eq:tree-amplitude}.
All the local terms $\delta^{(d)}(b)$ appearing in the above equation cancel from the final result.
The result is
\begin{equation}
    \widetilde{\wf}^{\lo}_\ominus(k,b) = \frac{i\kappa^3}{4} E_1 E_2 \Bigg\{ \frac{e^{-i\vec{k}\cdot\vec{b}}}{\vec{k}^{\,2}} \left[\left(1-4i\epsilon \frac{\vec{k}\cdot\vec{b}}{\vec{k}^{\,2}b^2}\right)\zeta_k^2 + \frac{4i\epsilon}{b^2}\zeta_k\zeta_b\right] S_{1,0}(k,b) + 2\pol_{ij}^{\ominus*} \partial_b^i \partial_b^j S_{1,1}(k,b) \Bigg\}.
\end{equation}
When expanding for $\epsilon\to0$ we get a $1/\epsilon$ pole that we associate with a logarithm of an infrared scale $\muir$ through the replacement $1/\epsilon\to2\log\muir$, obtaining
\begin{equation}\label{eq:lo-waveform}
    \begin{aligned}
        \widetilde{\wf}^{\lo}_\sigma(k,b) = & \frac{i\kappa^3E_1 E_2}{4\pi \vec{k}^{\,2}}e^{-i\vec{k}\cdot\vec{b}} \left(2i \frac{\vec{k}\cdot\vec{b}}{\vec{k}^{\,2}b^2}\pol^{(\sigma)}(k,k)-2i\frac{\pol^{(\sigma)}(b,k)}{b^2} - \pol^{(\sigma)}(k,k) \log(b\muirt) \right) \\
        & + \frac{i\kappa^3}{2} E_1 E_2 \,\pol^{(\sigma)}(\partial_b,\partial_b)\, S_{1,1}(k,b)\,,
    \end{aligned}
\end{equation}
where we adopted the notation $\pol^{(\sigma)}(v,w) = \pol_{ij}^{(\sigma)*} v^i w^j$ so that $\zeta_v \zeta_w= 2\pol^{\ominus}(v,w)$.
The waveform above agrees with the 1SF waveform at the leading order in PM expansion, calculated from the SF-EFT in equation~\eqref{eq:lo-waveform-1sf}.

We also give the soft expansion of the LO waveform, obtained with the help of the series~\eqref{eq:double-soft} for the integrals $S_{1,\alpha_p}(k,b)$.
The general structure of the soft expansion is given in equation~\eqref{eq:structure-soft} and for the LO waveform reads~\cite{Saha:2019tub,Sahoo:2021ctw,Laddha:2019yaj,Alessio:2024onn}
\begin{equation}
    \widetilde{\wf}^{\lo}_\sigma(k,b) = \frac{w^{\lo}_{\sigma\,[1/\omega]}}{\omega} + \sum_{r=0}^\infty w^{\lo}_{\sigma\,[\omega^r\log^{r+1}\omega]}\,\omega^{r}\log^{r+1}\omega + \cdots
\end{equation}
where the ellipsis denotes lower powers of $\log\omega$ associated to the same power of $\omega$, {\it i.e.} subleading logarithmic contributions.
Expressing the graviton momentum in terms of its frequency and its direction $k^\mu = \omega n^\mu$ we get
\begin{subequations}\label{eq:soft-factors-lo}
    \begin{align}
        w^{\lo}_{\sigma\,[1/\omega]}    & = - \frac{\kappa^3 E_1 E_2}{2\pi b^2 n_\pperp^2} \left(\pol^{(\sigma)}(n,b) - \frac{n_\pperp \cdot b_\pperp}{n_\pperp^2}\pol^{(\sigma)}(n,n)\right) \,, \label{eq:soft-factors-lo-1}\\
        w^{\lo}_{\sigma\,[\log\omega]}  & = \frac{i\kappa^3E_1 E_2}{4\pi n_\pperp^2} \pol^{(\sigma)}(n,n) \,,\\
        w^{\lo}_{\sigma\,[\omega\log^2\omega]} & = 0 \,.
    \end{align}
\end{subequations}
The $\omega^{-1}$ term is only the leading order in $\kappa$ of the full 1SF linear memory term~\eqref{eq:linear-memory}, while the second term is exact in the coupling, which is exhausted at order $\mathcal{O}(\kappa^3)$, as predicted in~\cite{Sahoo:2021ctw}.
\section{Predictions for the soft behaviour of the massless-massless scattering waveform\label{app:soft-matching}}
This appendix gives an overview of the results obtained in~\cite{Saha:2019tub,Laddha:2019yaj,Sahoo:2021ctw,Alessio:2024onn} and their matching with our expression~\eqref{eq:soft-waveform-memory-tails} for the soft expansion of the 1SF waveform emitted in a massless-massless scattering.
Specifically all the equations are taken from~\cite{Sahoo:2021ctw}, in which the authors obtain the memory term and the two tails as
\begin{equation}
    {\rm FT}[h_{\mu\nu}](\omega,n^\mu,R) = \frac{i}{\omega}A_{\mu\nu} - \left(B_{\mu\nu}-C_{\mu\nu}\right)\log\omega + i \frac{F_{\mu\nu}-G_{\mu\nu}}{2}\omega\log^2\omega + \cdots\,,
\end{equation}
where $R$ is the distance between the massless sources and the observer and the expression is truncated at the first order in the expansion for large $R$.
The coefficients are given in equations~$(4.3-4.4-4.5)$ of~\cite{Sahoo:2021ctw}\footnote{The change of signs is due to the change of metric convention, which is mostly plus in the reference.}
\begin{subequations}
    \begin{align}
        & A^{\mu\nu} = - \frac{2\gn}{R}\left[\sum_{i\in{\rm in}} \frac{p_i^\mu p_i^\nu}{n\cdot p_i} - \sum_{i\in{\rm out}} \frac{{p'_i}^\mu {p'_i}^\nu}{n\cdot p'_i}\right] \,, \\
        & B^{\mu\nu} = - C^{\mu\nu} = - \frac{4\gn^2}{R} \left[(n\cdot P)\sum_{i\in{\rm in}} \frac{p_i^\mu p_i^\nu}{n\cdot p_i} - P^\mu P^\nu\right] \,, \\
        & F^{\mu\nu} = - G^{\mu\nu} = \frac{16\gn^3}{R} (n\cdot P) \left[(n\cdot P)\sum_{i\in{\rm in}} \frac{p_i^\mu p_i^\nu}{n\cdot p_i} - P^\mu P^\nu\right]
    \end{align}
\end{subequations}
where $P^\mu = p_1^\mu + p_2^\mu = E_1\xi^\mu + E_2\chi^\mu$, and $A^{\mu\nu}$ arises from eikonal emission off the external legs~\cite{Weinberg:1964ew}, predicting the linear memory~\cite{Strominger:2014pwa}.
We draw the reader's attention to the fact that the coefficients $B,C,F,G$ are built using ingoing momenta only, so that their expressions are exact in $\gn$: $B$ and $C$ are tree-level-exact, while $F$ and $G$ come from the loop-level computation.
Therefore, to predict the terms above, we need the conservative motion to all-loop order, while it is sufficient the tree-level and one-loop waveform.

To match the waveform amplitude, we use the saddle-point approximation for the gravitational field at large distances $R$, which reads~\cite{Brandhuber:2023hhy}
\begin{equation}
    {\rm FT}\big[h^{\mu\nu}\pol_{\mu\nu}^{(\sigma)*}\big](\omega,n^\mu,R)  \simeq \frac{i\kappa}{8\pi R} \wf_\sigma (\omega n^\mu) + \mathcal{O}(R^{-2})\,,
\end{equation}
from which we obtain
\begin{subequations}
    \begin{align}
        w_{\sigma}^{[1/\omega]} & = -\frac{\kappa}{2}\left[E_2 \frac{(\chi\cdot\pol^{(\sigma)})^2}{\chi\cdot n} - \sum_{i\in{\rm out}} \frac{({p'_i}\cdot\pol^{(\sigma)})^2}{n\cdot p'_i}\right] \,,\\
        w_{\sigma}^{[\log\omega]} & = \frac{i\kappa^3E_1 E_2}{4\pi n_\pperp^2} \pol^{(\sigma)}(n,n) \,,\\
        w_{\sigma}^{[\omega\log^2\omega]} & = - \frac{\kappa^5 E_1 E_2}{64\pi^2 n_\pperp^2 } \Big[E_1 (n\cdot\xi) + E_2 (n\cdot\chi)\Big] \pol^{(\sigma)}(n,n) \label{eq:prediction-o-lo-sq}\,.
    \end{align}
\end{subequations}
The $\log\omega$ term is in perfect agreement with the result obtained from the SF-EFT in the main text, while the term $\omega(\log\omega)^2$ can be compared with~\eqref{eq:omegalogsquared} only at the leading order in $E_2/E_1\ll1$, finding agreement.
Focusing on the Weinberg term $A^{\mu\nu}$, its determination requires knowledge of the full dynamics of the two bodies, including radiation effects.
Fortunately, we only need to match the 1SF waveform, we therefore expand the asymptotic outgoing momenta in powers of $q=E_2/E_1$, namely
\begin{equation}
    \begin{aligned}
        & {p'_1}^\mu = E_1 \xi^\mu + q E_1 \Delta v_1^\mu + \mathcal{O}(q^2)\,,\\
        & {p'_2}^\mu = E_2 \left( \chi^\mu + \Delta v_2^\mu \right) + \mathcal{O}(q^2)\,,
    \end{aligned}
\end{equation}
where $\Delta v_2^\mu = \dot{\bar{x}}_\ml^\mu - \chi^\mu = \dfrac{d{\bar{x}}_\ml^\mu}{d\tau} - \chi^\mu $ comes from the geodesic motion~\eqref{eq:deflections-geodesics}.
After applying the light-cone gauge property $p_1\cdot\pol^{(\sigma)} = 0$, we get
\begin{equation}
    \begin{aligned}
        w_{\sigma}^{[1/\omega]} & \simeq -\frac{\kappa}{2} \left[ E_2 \frac{(\chi\cdot\pol^{(\sigma)})^2}{\chi\cdot n} - E_2 \frac{(\chi\cdot\pol^{(\sigma)} + \Delta v_2\cdot\pol^{(\sigma)})^2}{\chi\cdot n + \Delta v_2\cdot n} - \underbrace{q^2E_1\frac{(\Delta v_1\cdot \pol^{(\sigma)})^2}{\xi\cdot n + q\Delta v_1\cdot n}}_{\simeq\,0} \right] \\
            & = - \frac{2 \kappa^3 E_1 E_2 \left(\,\mathcal{N}_1 + \mathcal{N}_2 \,\right)}{\left(8\pi b^2 n_\pperp + \kappa^2 E_1 (\xi\cdot n) b_\pperp \right)^2}\,,
    \end{aligned}
\end{equation}
with $\mathcal{N}_{1,2}$ given in the main text. The equation above matches the result~\eqref{eq:soft-waveform-memory-tails} obtained from the SF-EFT and its leading order PM expansion is in agreement with the KMOC computation in App.~\ref{app:lowf}, equation~\eqref{eq:soft-factors-lo-1}.
Moreover, we checked that these are also correctly reproduced by the ultrarelativistic limit of the all-order ansatz $a_\ell^{\mu\nu}$ in reference~\cite{Alessio:2024onn}.

\section{Waveform computation in the strong-field approximation\label{app:saddle-point}}
In this appendix we compute the asymptotic expansion of the piece $\wf_\sigma^{\rm int,2}$ of the 1SF waveform for $W\gg1$, which we have introduced in equation~\eqref{eq:wf-early-int2-not-computed}
\begin{equation}\label{eq:apppog}
    \wf_\sigma^{\mathrm{int},2} = -\frac{i\kappa^3 E_1 E_2}{2} \frac{e^{-iW/\epsilon}}{(4\pi\muir^2e^{2\gamma_\mathrm{E}})^{iW}} \frac{\Gamma(1-iW)}{\Gamma(1+iW)} \, e^{-i\vec{k}\cdot\vec{b}} \, \pol^{(\sigma)*}_{ij}\mathcal{I}^{ij}(k,b)\,.
\end{equation}
To do so, we need the asymptotic behaviour of the ratio of Gamma functions $\Gamma(1-iW)/\Gamma(1+iW)$ and of the integral $\mathcal{I}^{ij}$.
We compute the expansion of the Gamma functions using Stirling's approximation for the logarithm of $\Gamma(z)$ at large absolute value of its argument, $|z|\gg1$
\begin{equation}
    \log\left(\frac{e^z\,\Gamma(z)}{z^{z-1/2}\sqrt{2\pi}}\right) \sim \sum_{n=1}^{\infty} \frac{B_{2n}}{2n(2n-1)z^{2n-1}}\,,
\end{equation}
so that
\begin{equation}
    \frac{\Gamma(1-iW)}{\Gamma(1+iW)} = -i W^{-2iW} \text{exp}\left\{2iW + \frac{i}{6W} + \frac{i}{180W^3} + \mathcal{O}(W^{-5}) \right\} \,.
\end{equation}
The integral is instead
\begin{equation}\label{eq:integral-I-ij}
    \mathcal{I}^{ij}(k,b) = \int_{r_\pperp} F^{ij}(r)\, e^{iW \varphi(r)}\,,
\end{equation}
with
\begin{equation}\label{eq:integral-I-ij-2}
    F^{ij}(r) = \frac{(r^i+k^i)\,(r^j+k^j)}{r_\pperp^2 (r_\pperp+k_\pperp)^2}\,,
    \hspace{25pt}
    \varphi(r) = \log(-r_\pperp^2) + \frac{r_\pperp\cdot{b}_\pperp}{W}\,.
\end{equation}
We solve it using the steepest-descent method at leading order of accuracy for large $W$, for which the saddle point is $\vec{r}_0 = 2W\vec{b}/b^2$, and the saddle-point formula gives
\begin{equation}
    \mathcal{I}^{ij}(k,b) \simeq F^{ij}(r_0) \frac{e^{iW\varphi(r_0)}}{\sqrt{ (2\pi W)^{d} \, \|\varphi''(r_0)\|}} = \frac{1}{4\pi W} \, \frac{b^i b^j}{b^2} \left(\frac{4W^2}{b^2e^2}\right)^{iW} + \mathcal{O}(W^{-2})\,,
\end{equation}
in which we set $d=2$.
Substituting this result into~\eqref{eq:apppog} gives the formula~\eqref{eq:wf-early-int2} in the main text.

\bibliographystyle{JHEP}
\bibliography{../bibliography}

\end{document}